%% file: main.tex
\documentclass[manuscript, screen]{acmart}
\usepackage{epsfig,endnotes,url,color,subfigure}
\usepackage{booktabs} 

\usepackage{balance}
\usepackage[linesnumbered,lined,ruled,commentsnumbered]{algorithm2e}
\usepackage{colortbl}
\usepackage{amsmath,amssymb}

\usepackage{natbib}

\usepackage{multirow}
\usepackage{listings}
\usepackage{tikz}
\newcommand{\tabitem}{~~\llap{\textbullet}~~}

\usepackage{graphicx}

\DeclareMathOperator*{\argmax}{argmax} 
\DeclareMathOperator*{\argmin}{argmin} 

\setcopyright{rightsretained}

\input{mac}

\begin{document}
\title{Orchestrating the Development Lifecycle of Machine Learning-Based IoT Applications: A Taxonomy and Survey}

\author{Bin Qian}
\author{Jie Su}
\author{Zhenyu Wen}
\authornote{Zhenyu is the corresponding author}
\author{Devki Nandan Jha}


\author{Yinhao Li}
\author{Yu Guan}
\author{Deepak Puthal}
\author{Philip James}
\affiliation{%
  \institution{Newcastle University, UK}
}

\author{Renyu Yang}
\affiliation{%
  \institution{University of Leeds, UK}
}

\author{Albert Y. Zomaya}
\affiliation{%
  \institution{The University of Sydney, Australia}
}

\author{Omer Rana}
\affiliation{%
  \institution{Cardiff University, UK}
}

\author{LIZHE WANG}
\affiliation{%
  \institution{China University of Geoscience (Wuhan), China}
}

\author{Maciej Koutny}
\author{Rajiv Ranjan}
\affiliation{%
  \institution{Newcastle University, UK}
}


\input{abstract}

\maketitle


\input{introduction}
\input{model_develop}

\input{deployment}

\input{action}
\input{collection}
\input{conclusion}
\bibliographystyle{ACM-Reference-Format-num}
\bibliography{main}  
\appendix
\input{appendix}

\end{document}

%% file: mac.tex
\newcommand{\myparagraph}[1]{\smallskip \noindent{\bf {#1}.}}

\newcommand{\out}[1] {}

\newcounter{codeLineCntr}

\reversemarginpar
\newif\ifnotes
\notestrue

\newcommand{\punt}[1]{}

\renewcommand{\eqref}[1]{Equation~(\ref{#1})}

\newcommand{\proc}[1]{\ifmmode\mbox{\textsc{#1}}\else\textsc{#1}\fi}

  \newcommand{\func}[1]{\ifmmode\mathrm{#1}\else\textrm{#1}fi} %

\newcounter{remark}[section]



%% file: abstract.tex
\begin{abstract}
Machine Learning (ML) and Internet of Things (IoT) are complementary advances: ML techniques unlock the potential of IoT with intelligence, and IoT applications increasingly feed data collected by sensors into ML models, thereby employing results to improve their business processes and services. Hence, orchestrating ML pipelines that encompass model training and implication involved in the holistic development lifecycle of an IoT application often leads to complex system integration. This paper provides a comprehensive and systematic survey of the development lifecycle of ML-based IoT applications. We outline the core roadmap and taxonomy, and subsequently assess and compare existing standard techniques used at individual stages.
\end{abstract}

%% file: introduction.tex
\section{Introduction}
\label{sec:introduction}

Rapid development of hardware, software and communication technologies boosts the speed of connection of the physical world to the Internet via Internet of Things (IoT). A report \footnote{https://www.statista.com/statistics/471264/iot-number-of-connected-devices-worldwide/} shows that about 75.44 billion IoT devices will be connected to the Internet by 2025. These devices generate a massive amount of data with various modalities. 
Processing and analyzing such big data is essential for developing smart IoT applications. Machine Learning (ML) plays a vital role in data intelligence which aims to understand and explore the real world. ML $+$ IoT type applications thus are experiencing explosive growth. However, there are unfilled gaps between current solutions and the demands of orchestrating the development lifecycle of ML-based IoT applications.
Existing orchestration frameworks for example \emph{Ubuntu Juju}, \emph{Puppet} and \emph{Chef} are flexible in providing solutions for deploying and running 
applications over public or private clouds. These frameworks, however, neglect the heterogeneity of IoT environments that encompasses various hardwares, communication protocols and operating systems. More importantly, none of them are able to completely orchestrate a holistic development lifecycle of ML-based IoT applications. 
The development lifecycle must cover the following factors: 1) \emph{how} the target application is specified and developed, 2) \emph{where} the target application is deployed, (3) \emph{what} kind of information the target application is being audited. Application specification defines the requirements including the ML tasks, performance, accuracy and execution workflow. Based on the specification and the available computing resources, the ML models are developed to meet the specified requirements while optimizing the training processes in terms of the cost of time and computing resources. Next, the model deployment considers the difficulty of the heterogeneity of the IoT environment for running a set of composed ML models. Finally, ML-based IoT applications closely connect with people's lives and some applications such as autopilot require high reliability. Therefore, essential monitoring information has to be collected to improve the performance of the application in the next iteration of the lifecycle.

In this survey, we present comprehensive research on orchestrating the development lifecycle of ML-based IoT applications. We first present the core roadmap and taxonomy, and subsequently summarize, compare, and assess the variety of techniques used in each step of the lifecycle.  
Previous efforts provided broad knowledge that can drive us to build the taxonomy. For instance, \cite{stoica2017berkeley} discussed encountered challenges of developing the next generation of AI systems. \cite{zhang2019deep,mohammadi2018deep} gave comprehensive reviews of  available deep learning architectures and algorithms in IoT domain.  To the best of our knowledge, this is the first work that presents a comprehensive survey to illustrate the whole development lifecycle of ML-based IoT application, which paves the way for developing an agile, robust and reliable smart IoT application. Before introducing the roadmap and taxonomy, we provide a smart city example in the next subsection that illustrates  ML-based IoT applications in real-world.


\subsection{Smart City Applications}\label{sec:smartcity}      

%
\input{figuresTex/fig-smart-city}
Smart city uses modern communication and information techniques to monitor, integrate and analyze the data collected from core systems running across cities. Meanwhile, smart city makes intelligent responses to various use cases, such as traffic control, weather forecasting, industrial and commercial activities. Fig.~\ref{fig:smart_city} represents a smart city which consists of various IoT applications with many of them using Machine Learning (ML) techniques. 
For example, a \emph{smart traffic routing} system consists of a large number of cameras monitoring the road traffic and a smart algorithm running on the cloud recommending the optimal routes for users~\cite{zhou2010adaptive}.
On the other hand, a \emph{smart car navigation system} \cite{ivanecky2012car} allows the passengers to set and change destinations via built-in car audio devices.
The two systems work together to provide real-time interactive routing services.
More specifically, the user's voice commands are translated in the car edge side and sent to the cloud where the \emph{smart traffic routing} system works. The best route is translated back to voice guiding the users to their destinations.
The above-mentioned applications involve various computing resources (e.g., cloud, edge, and IoT devices) and ML techniques, making the development of these ML-based IoT applications very challenging both for the ML models and the IoT system. To fill this gap, we orchestrate the development lifecycle of an ML-based IoT application. In the next subsection, we present a roadmap for the development lifecycle along with a comprehensive taxonomy that surveys the techniques relevant for developing the application.


\subsection{Roadmap and Taxonomy}

\input{figuresTex/fig-architecture}

\myparagraph{Roadmap} Fig~\ref{fig:architecture} shows the roadmap of developing an ML-based IoT application.
The roadmap starts with the requirements specification where the required computing resources (hardware and software) and ML models are specified. 
Based on the specification, we carefully design the infrastructure protocol, data acquisition approach and machine learning model development pipeline. 
Next, we implement and train the model with various ML algorithms. We also evaluate and optimize the models to achieve high efficiency without sacrificing too much accuracy.  
After the model development, an optimized deployment plan is generated based on the specified ML models and infrastructures.   
The deployed application must be audited while it is running on real IoT environments; the audit aims to explore the performance issues in terms of security, reliability and other QoS metrics.
Finally, the audited issues will guide the corrections of orchestration details in the next iteration of the application development.

\myparagraph{Taxonomy} Fig.~\ref{taxonomy1} depicts our taxonomy which systematically analyzes the core components in the orchestration of the development lifecycle of a ML-based IoT application.
Note that the survey in \cite{weerasiri2017taxonomy} has reviewed cloud resource orchestration techniques. It outlines the key infrastructure orchestration challenges for cloud-based application as well as being extendable for IoT applications. Thus, in this survey, we focus more on the challenges of implementing ML models and orchestrating their IoT application development lifecycle. 
To this end, we extract the core building blocks of the development lifecycle relevant to ML
and identify \emph{four} main categories based on their specific functionality during the development process. The outline of the paper follows the structure of the taxonomy as well. 
\begin{enumerate}
	\item \textbf{Model Development.} We propose a general pipeline for developing a ready-to-deploy ML model. We investigate the ML techniques to build each block of the pipeline (refer to \S \ref{sec:training}).
	\item \textbf{Model Deployment.} In our work, we review the software deployment techniques and analyze the challenges of applying such techniques to deploy the ML models in IoT environments (refer to \S \ref{sec:deployment}).  
	\item \textbf{Model Audit.} Audit is one of the important dimensions in building a robust application. We survey the main security, reliability and performance issues in ML-based IoT applications (refer to \S \ref{sec:audit}).  
	\item \textbf{Data Acquisition.} Data quality is important in building ML models. We identify three dimensions that are important throughout the data acquisition pipeline: data collection, data fusion and data preprocessing (refer to \S \ref{sec:collection}).
\end{enumerate}


\input{taxonomy}

%% file: figuresTex/fig-smart-city.tex
\begin{figure}[ht]
    \centering
    \includegraphics[width=3.5in]{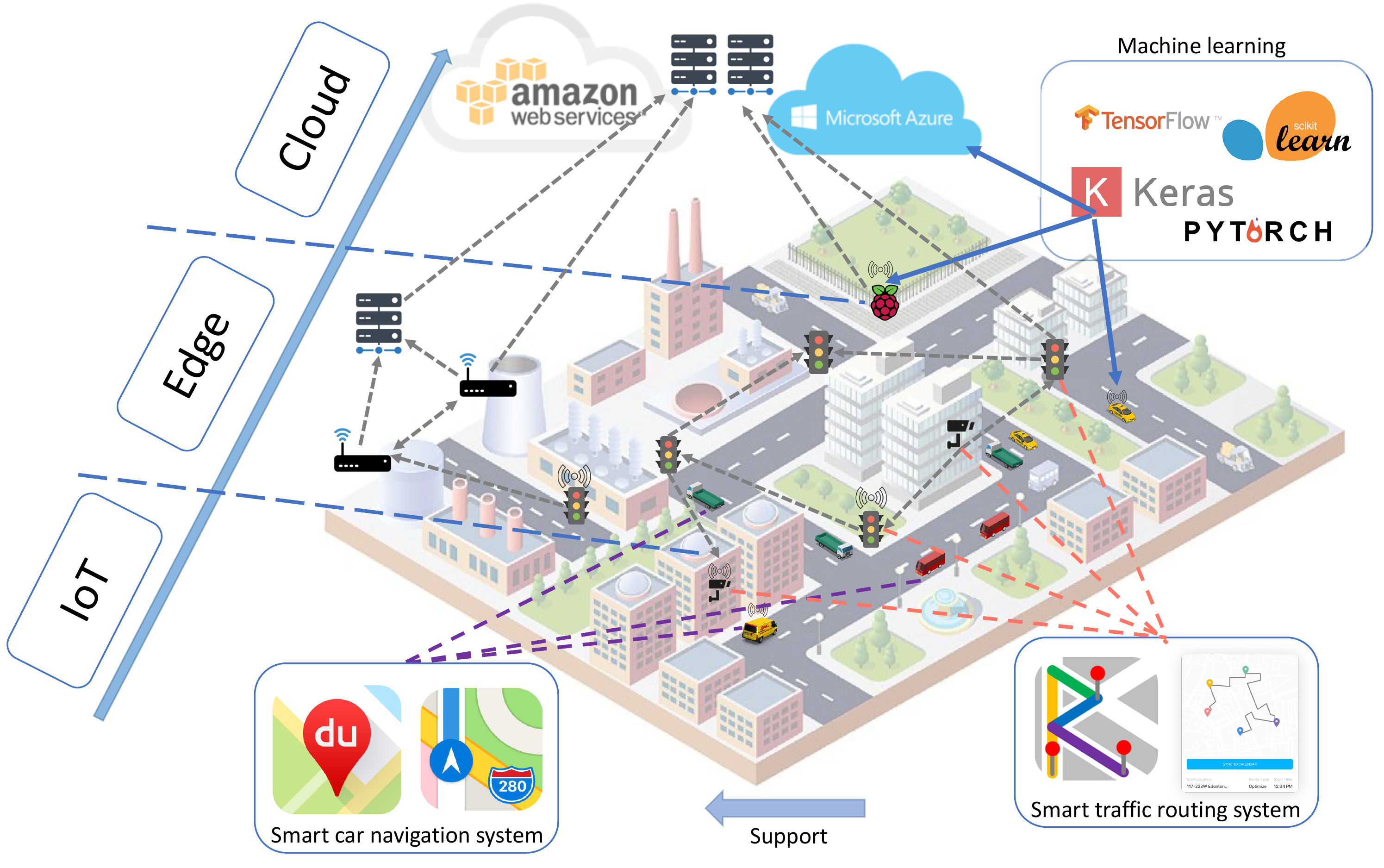}
     \vspace{-5mm}
     \caption{Smart City}
    \label{fig:smart_city}
     \vspace{-5mm}
    \end{figure}

%% file: figuresTex/fig-architecture.tex
\begin{figure}[h]
\centering
\includegraphics[width=5.5in]{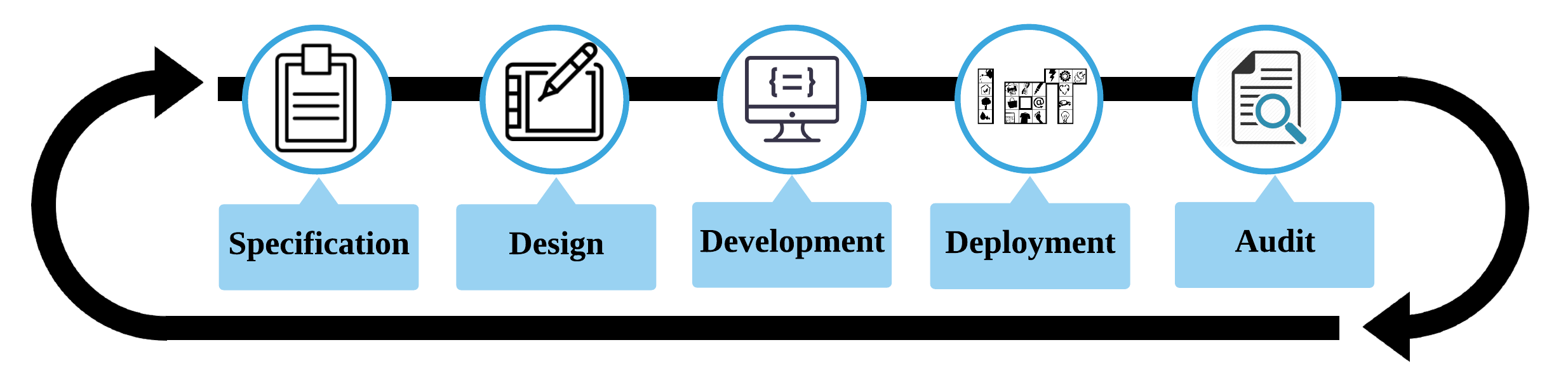}
 \vspace{-5mm}
 \caption{The development lifecycle of an ML-based IoT application}
\label{fig:architecture}
 \vspace{-4mm}
\end{figure}

%% file: taxonomy.tex
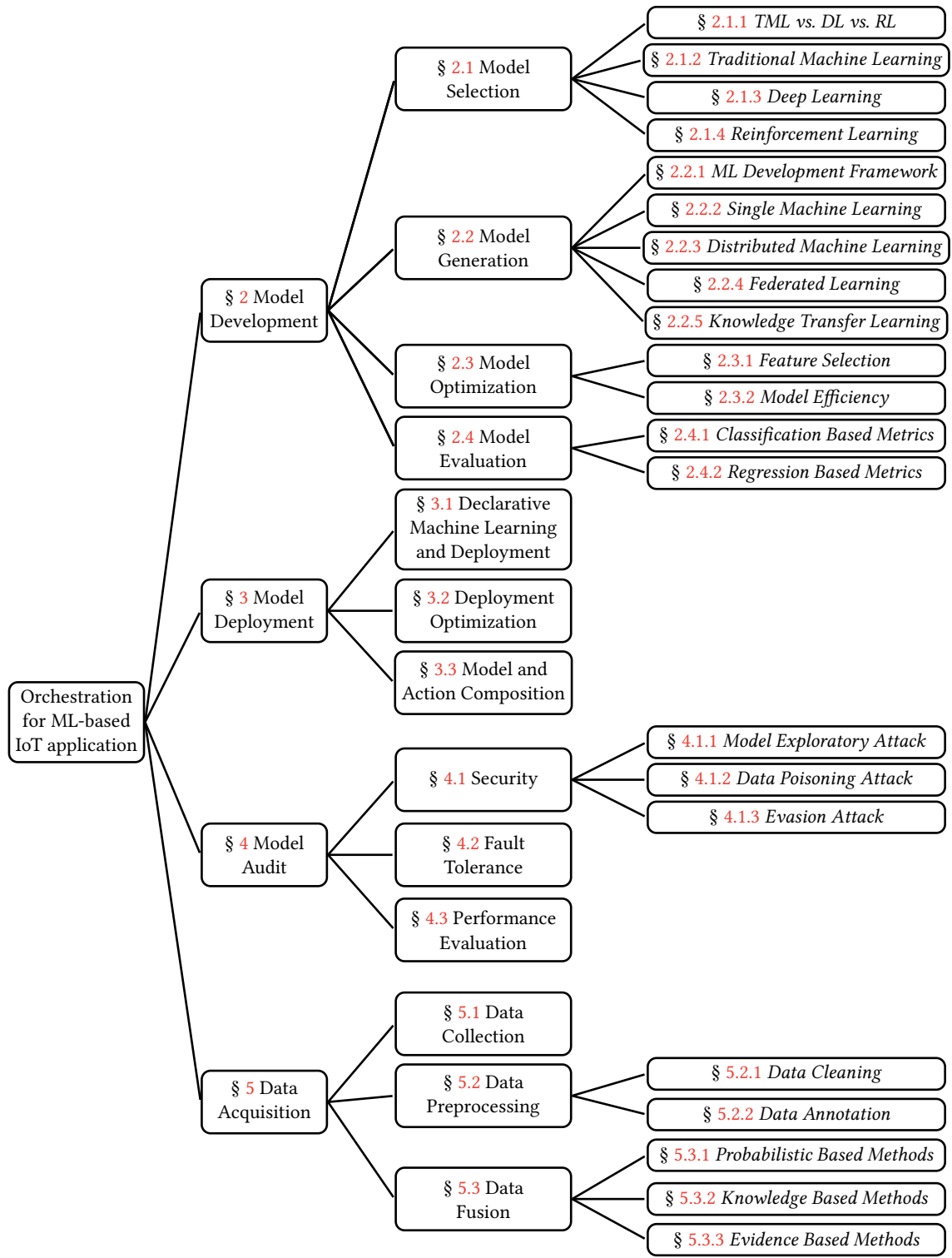
\begin{figure}
\begin{tikzpicture}[every node/.style = {shape          = rectangle,
                                         rounded corners,
                                         draw=black,
                                         minimum width  = 3cm,
                                         minimum height = 1.5cm,
                                         align          = center,
                                         text           = black},
                                         line width=1pt,
                  blue edge/.style  = { -,
                                         line width=1pt,
                                         black,
                                         shorten >= 1pt}]

\node[minimum width=2cm, minimum height=1.0cm,](0;0) at (0,0.42) {Orchestration \\for ML-based \\IoT application};

  \node[minimum width=2cm, minimum height=1.0cm,](1;1)  at (3, 7) {$\S$ \ref{sec:training} Model\\Development};
       \node[minimum width=2.8cm, minimum height=1.0cm,](2;6)  at (6.5,10.7) {$\S$ \ref{selection} Model\\Selection};
            \node[minimum width=4.75cm, minimum height=0.5cm,](3;12)  at (11.5,11.6) {$\S$ \ref{big-categories comparison} \textit{TML vs. DL vs. RL}};
            \node[minimum width=4.75cm, minimum height=0.5cm,](3;11)  at (11.5,11) {$\S$ \ref{sec:tml}  \textit{Traditional Machine Learning}}; 
            \node[align=left, minimum width=4.75cm, minimum height=0.5cm](3;10)  at (11.5,10.4) {$\S$ \ref{sec:deep learning} \textit{Deep Learning}};
            \node[minimum width=4.75cm, minimum height=0.5cm,](3;9)  at (11.5,9.8) {$\S$ \ref{rl} \textit{Reinforcement Learning}}; 
            
       \node[minimum width=2.8cm, minimum height=1.0cm,](2;5)  at (6.5,8.0) {$\S$ \ref{generation} Model\\Generation};
            \node[minimum width=4.75cm, minimum height=0.5cm,](3;18)  at (11.5,9.2) {$\S$ \ref{mlframework} \textit{ML Development Framework}}; 
            \node[minimum width=4.75cm, minimum height=0.5cm,](3;17)  at (11.5,8.6) {$\S$ \ref{sec:singlemachine} \textit{Single Machine Learning}};
            \node[minimum width=4.75cm, minimum height=0.5cm,](3;16)  at (11.5,8.0) {$\S$ \ref{sec:distributed} \textit{Distributed Machine Learning}};
            \node[minimum width=4.75cm, minimum height=0.5cm,](3;15)  at (11.5,7.4) {$\S$ \ref{sec:federatedlearning} \textit{Federated Learning}};
            \node[minimum width=4.75cm, minimum height=0.5cm,](3;14)  at (11.5,6.8) {$\S$ \ref{sec:transferlearning} \textit{Knowledge Transfer Learning}};
            
       \node[minimum width=2.8cm, minimum height=1.0cm,](2;4)  at (6.5,5.95) {$\S$ \ref{optimisation} Model\\Optimization}; 
             \node[minimum width=4.75cm, minimum height=0.5cm,](3;6)  at (11.5,6.2) {$\S$ \ref{fs} \textit{Feature Selection}};  
             \node[minimum width=4.75cm, minimum height=0.5cm,](3;5)  at (11.5,5.6) {$\S$ \ref{sec:model_eff} \textit{Model Efficiency}};  
             
       \node[minimum width=2.8cm, minimum height=1.0cm,](2;3)  at (6.5,4.8) {$\S$ \ref{evaluation} Model\\Evaluation};
            \node[minimum width=4.75cm, minimum height=0.5cm,](3;4)  at (11.5,5.0) {$\S$ \ref{classificationeval} \textit{Classification Based Metrics}};  
             \node[minimum width=4.75cm, minimum height=0.5cm,](3;3)  at (11.5,4.4) {$\S$ \ref{regreeval} \textit{Regression Based Metrics}};

  \node[minimum width=2cm, minimum height=1.0cm,](1;0)  at (3, 2.2) {$\S$ \ref{sec:deployment} Model\\Deployment}; 
         \node[minimum width=2.8cm, minimum height=1.0cm,](2;2)  at (6.5,3.5) {$\S$ \ref{sec:declarative} Declarative \\Machine Learning\\ and Deployment};
         
          \node[minimum width=2.8cm, minimum height=1.0cm,](2;1)  at (6.5,2.2) {$\S$ \ref{sec:optimization} Deployment\\Optimization};
                    \node[minimum width=2.8cm, minimum height=1.0cm,](2;0)  at (6.5,1.05) {$\S$ \ref{sec:compose} Model and\\ Action Composition};
  
  \node[minimum width=2cm, minimum height=1.0cm,](1;-1) at (3,-1.7) {$\S$ \ref{sec:audit} Model\\Audit}; 
            \node[minimum width=2.8cm, minimum height=1.0cm,](2;-1)  at (6.5,-0.5) {$\S$ \ref{security} Security};
                \node[minimum width=4.75cm, minimum height=0.5cm,](3;-2)  at (11.5,0.1) {$\S$ \ref{modelexattack} \textit{Model Exploratory Attack}}; 
                \node[minimum width=4.75cm, minimum height=0.5cm,](3;-3)  at (11.5,-0.5) {$\S$ \ref{sec:data_poison} \textit{Data Poisoning Attack}}; 
                \node[minimum width=4.75cm, minimum height=0.5cm,](3;-4)  at (11.5,-1.1) {$\S$ \ref{evasion} \textit{Evasion Attack}}; 
                
             \node[minimum width=2.8cm, minimum height=1.0cm,](2;-2)  at (6.5,-1.70) {$\S$ \ref{sec:fault} Fault\\Tolerance};
             \node[minimum width=2.8cm, minimum height=1.0cm,](2;-3)  at (6.5,-2.9) {$\S$ \ref{performanceeva} Performance\\Evaluation};

  \node[minimum width=2cm, minimum height=1.0cm,](1;-2) at (3,-5.65) {$\S$ \ref{sec:collection} Data\\Acquisition};
        \node[minimum width=2.8cm, minimum height=1.0cm,](2;-4)  at (6.5,-4.40) {$\S$ \ref{datacollection} Data\\ Collection};
       
                  \node[minimum width=2.8cm, minimum height=1.0cm,](2;-5)  at (6.5,-5.55) {$\S$ \ref{datapreproc} Data\\Preprocessing};
          		\node[minimum width=4.75cm, minimum height=0.5cm,](3;-5)  at (11.5,-5.2) {$\S$ \ref{dataclean} \textit{Data  Cleaning}};
          		\node[minimum width=4.75cm, minimum height=0.5cm,](3;-6)  at (11.5,-5.85) {$\S$ \ref{dataanno} \textit{Data Annotation}};

         \node[minimum width=2.8cm, minimum height=1.0cm,](2;-6)  at (6.5,-7.2) {$\S$ \ref{datafusion} Data\\Fusion};
         		\node[minimum width=4.75cm, minimum height=0.5cm,](3;-7)  at (11.5,-6.5) {$\S$ \ref{probdatafusion} \textit{Probabilistic  Based Methods}};
         		\node[minimum width=4.75cm, minimum height=0.5cm,](3;-8)  at ( 11.5,-7.2) {$\S$ \ref{knowledagedatafusion} \textit{Knowledge Based Methods}};
         		\node[minimum width=4.75cm, minimum height=0.5cm,](3;-9)  at ( 11.5,-7.85) {$\S$ \ref{evidancedatafusion} \textit{Evidence Based Methods}};       

\foreach \j in {-2,...,1}
  { \draw[blue edge] (0;0.east) -- (1;\j.west); }
          
  \foreach \j in {3,...,6}
  { \draw[blue edge] (1;1.east) -- (2;\j.west);} 
  \foreach \j in {3,...,6}
  { \draw[blue edge] (1;1.east) -- (2;\j.west);} 
    \foreach \j in {0,...,2}
  { \draw[blue edge] (1;0.east) -- (2;\j.west);} 
      \foreach \j in {-3,...,-1}
  { \draw[blue edge] (1;-1.east) -- (2;\j.west);} 
        \foreach \j in {-6,...,-4}
  { \draw[blue edge] (1;-2.east) -- (2;\j.west);}

   \foreach \j in {9,...,12}
 { \draw[blue edge] (2;6.east) -- (3;\j.west);} 
    \foreach \j in {14,...,18}
 { \draw[blue edge] (2;5.east) -- (3;\j.west);}
     \foreach \j in {5,...,6}
 { \draw[blue edge] (2;4.east) -- (3;\j.west);}
      \foreach \j in {3,...,4}
 { \draw[blue edge] (2;3.east) -- (3;\j.west);}
        \foreach \j in {-4,...,-2}
 { \draw[blue edge] (2;-1.east) -- (3;\j.west);}
         \foreach \j in {-6,...,-5}
 { \draw[blue edge] (2;-5.east) -- (3;\j.west);}
          \foreach \j in {-9,...,-7}
 { \draw[blue edge] (2;-6.east) -- (3;\j.west);}

\end{tikzpicture}
 \caption{A taxonomy for orchestrating ML-based IoT application development lifecycle}
 \label{taxonomy1}
\end{figure}

%% file: model_develop.tex
\input{model_selection}

\input{model_generation}

\input{model_opt}

\input{model_eva}

%% file: model_selection.tex
\section{Model development}
\label{sec:training}
\vspace{-2mm}

\input{figuresTex/fig-training-cycle}
One of the core components in this paper is machine learning (ML) models, which may be roughly divided into three categories: Traditional Machine Learning (TML), Deep Learning (DL) and Reinforcement Learning (RL).
To develop ML models in the IoT environment, we propose a generic pipeline (see Fig. \ref{fig:train_cycle}) including \textit{model selection}, \textit{model generation}, \textit{model optimization} and \textit{model evaluation}. We initially introduce the generic pipeline by presenting adaptive video streaming~\cite{mao2017neural} as an example. 


\myparagraph{Adaptive video streaming}   
Video transmission between server and mobile devices employs http-based adaptive streaming techniques. 
In a typical video server (e.g., DASH \footnote{https://github.com/Dash-Industry-Forum/dash.js}), videos are encoded and stored as multiple chunks at different bitrates. 
One video usually consists of several chunks with each containing seconds of content.
To maximize video quality, the video player in a client (e.g., mobile device) usually employs adaptive bitrate (ABR) algorithms aiming to pull high-bitrate chunks from the server without compromising the latency. 
As shown in Fig. \ref{fig:pensieve}, ABR algorithms use simple heuristics to make bitrate decisions based on various observations such as the estimated network throughput and playback buffer occupancy. 
ABR algorithms require fine-grained tuning and can be hardly generalized to handle various network conditions that fluctuate across time and different environments. Thus we are seeking to solve the problem using modern ML technologies.

\input{figuresTex/fig-pensieve}
To this end, we first need to perform \textbf{model selection} (\S \ref{selection}) to find a subset of suitable models. 
In this scenario, the server must give a bitrate decision so that the client can return feedback that conveys whether the decision is satisfactory.
Such interaction problems necessitate further use of RL and 
we will present proper choice of RL algorithms based on different selection criteria (\S \ref{rl}).  
%
Next, we will choose a suitable development framework to implement the model and utilize different acceleration techniques to reduce the latency of \textbf{model generation} (\S \ref{generation}).  
In this example, Tensorflow and  A3C algorithm \cite{mnih2016asynchronous} are used as the development framework and  distributed training protocol respectively for faster convergence.
Once generated, the model has to be adapted into the real environment. Considering heterogeneity of IoT infrastructure, models need to be optimized according to the computing resources. This procedure is called \textbf{model optimization} (\S \ref{optimisation}).
In \textbf{model evaluation} (\S \ref{evaluation}), model performance is measured to validate whether the model meets expected results. Particularly in this case, performance is evaluated by the total reward obtained from the simulated environment. The following subsections will discuss the the pipeline in detail.  

%

\subsection{Model Selection}\label{selection}


Model selection aims to find the \textit{optimal} ML model to perform a user's specified tasks, whilst adapting to the complexity of IoT environments. In this section, we first discuss the model selection from three main categories i.e., TML, DL and RL, followed by a survey of well-known models (or algorithms) in each category and their corresponding criteria for model selection.

\subsubsection{TML vs. DL vs. RL} \label{big-categories comparison}
In this work we roughly divide the ML approaches/concepts into TML, DL and RL. 
Compared with the most popular DL , TML is relatively lightweight. 
It is a set of algorithms that directly transform the input data (to output), according to certain criteria. 
For supervised cases when a class label is available for training, TML aims to map the input data to the labels by optimising a model, which can be used to infer unseen data at the test stage.
However, since the relationship between raw data and label might be highly non-linear, feature engineering--- a heuristic trial-and-error process --- is normally required to construct the appropriate input feature. 
The TML model is relatively simple, the interpretability (e.g., the relationship between the engineered features and the labels) tends to be high. 

DL has become popular in recent years. Consisting of multiple layers,
DL is powerful for modeling complex non-linear relationships (between the input and output) and thus does not require the aforementioned heuristic (and expensive) feature engineering process, making it a popular modelling approach in many fields such as computer vision and natural language processing. 
Compared with TML, DL models tend to have more parameters (to be estimated) and generally they require more data for reliable representation learning. 
However, it is crucial to guarantee the data quality and 
a recent empirical study\cite{nakkiran2019deep} suggested the increasing number of noisy/less-representative training samples may harm DL's performance, making it less generalizable to unseen test data.  
 Moreover, DL's multilayer structures make it difficult to interpret the complex relationship between input (i.e., raw features) and output. However, more and more visualisation techniques (e.g., attention map \cite{zhang2018visual}) were used, which play an important role in understanding DL's decision-making process. 
 

RL has become increasingly popular due to its success in addressing challenging sequential decision-making problems \cite{sutton2018reinforcement}. 
Some of these achievements are based on the combination of DL and RL,  i.e., Deep Reinforcement Learning.
It has shown its considerable performance in natural language processing~\cite{xiong2017dcn,li2016deep}, computer vision~\cite{arulkumaran2017brief,ren2018deep,yun2017action,supancic2017tracking,chen2018part}, robotics~\cite{quillen2018deep} and  IoT systems~\cite{mao2017neural,mao2016resource,zhao2016deep} and  related applications like video games~\cite{arulkumaran2017brief}, visual tracking~\cite{ren2018deep,yun2017action,supancic2017tracking}, action prediction~\cite{chen2018part}, robotic grasping~\cite{quillen2018deep}, question answering~\cite{xiong2017dcn}, dialogue generation~\cite{li2016deep}, etc.
In RL, there is usually one or more agent(s) interacting with the outside environment, where optimal control policies are learnt through experience. 
Fig.~\ref{fig:rl} illustrates the iterative interaction circle, where the agent starts without knowing anything about environment or task. 
Each time the agent takes action based on the environment states, and it receives a reward from the environment.
RL optimises this process  such that it learns to make decisions with higher rewards received.

\vspace{-3mm}
\input{figuresTex/fig-reinforcement-learning}

\myparagraph{Discussion} In IoT environments, a variety of problems can be modelled by using the aforementioned three approaches.
The applications range from system and networking \cite{mao2017neural} \cite{mao2016resource}, smart city \cite{zhao2016deep} \cite{li2016traffic}, to smart grid \cite{wen2015optimal} \cite{ruelens2016residential}, etc.
To begin with modeling, it is essential for users to choose a suitable learning concept at the first stage.
The main selection criteria can be divided into two categories: \textit{Function-based selection} and \textit{Power Consumption-based selection}. 

\textit{Function-based selection} aims to choose an appropriate concept based on their functional difference. 
For example, 
RL benefits from its iterative environment $\leftrightarrow$ agent interaction property, and can be applied to various applications which need interaction with environment or system such as smart temperature control systems, or recommendation systems (with cold start problem). 
On the other hand, TML algorithms are more suitable for modelling structured data (with high-level semantic attributes), especially when interpretability is required.  
DL models are typically used to model complex unstructured data,  e.g., images, audios, time-series data, etc. and are an ideal choice especially  with high amount of training data and low requirement on interpretability.

\textit{Power Consumption-based selection} aims to choose an appropriate model given constraints in computational power or latency.
In contrast to TML, the powerful RL/DL models are normally computationally expensive with high overhead.
Recently, model compression techniques were developed, which may provide a relatively efficient solution for using RL/DL models for some IoT applications.
However, on some mobile platforms with very limited hardware resources (e.g., power, memory, storage), it is still challenging to employ compressed RL/DL models, especially when there are some performance requirements (e.g., accuracy, or real-time inference) \cite{cheng2017survey}.  
On the other hand, lightweight TML may be more efficient, yet
reasonable accuracy can only be achieved with appropriate features (e.g., high level attributes derived from the time-consuming feature engineering). 
 \vspace{-3mm}

\input{figuresTex/fig-new}

\subsubsection{Traditional Machine Learning}\label{sec:tml} 
Herein we demonstrate several popular TML algorithms (algorithm details are available in TML method \textbf{Appendix B}),
and discuss the criteria for choosing the TML algorithms. 
Given different tasks, TML can be further divided into \emph{Supervised Learning} and \emph{Unsupervised Learning}.

\myparagraph{Supervised Learning} Supervised learning algorithm~(i.e., Fig. \ref{fig:supandunsup}) can be used when both the input data $X$ and the corresponding labels $Y$ are provided (for training), and it aims to learn a mapping function such that $Y: \leftarrow f(X)$. 
Supervised learning algorithms have been widely used in IoT applications, we introduce the  most representative classifiers below.

\emph{Perceptron} and \emph{Logistic Regression (LR)} are probably the simplest linear classifiers. 
For both models, the model (i.e., weights and bias) is basically a simple linear transformation. 
Perceptron can perform binary classification simply based on the sign of the (linearly) transformed input data, while LR will further scale the transformed value into probability (via \emph{sigmoid} function), before a thresholding function is applied for the binary classification decision.
LR can also be extended to process multi-class classification scenarios by using \emph{softmax} as the scaling function, with class-wise probabilities as output.

\emph{Artificial Neural Networks (ANN)} is a general extension of the aforementioned linear classifiers. 
Compared with \emph{Perceptron} or \emph{LR} which linearly project input data to the output,  \emph{ANN} has an additional ``hidden layer''  (with a non-linear activation function), 
which enables \emph{ANN} to model non-linearity. 
However, in contrast to linear classifiers, this additional hidden layer makes it more difficult to see the relationship between the input and output data (i.e., low interpretability). 
Although in theory, with one hidden layer \emph{ANN} can model any complex non-linear functions, in practice it has limited generalization capabilities when facing unseen data.
\emph{ANN} with more layers, also referred to as deep neural networks, tend to have better modelling capability, which will be introduced in the next subsection.


\emph{Decision Tree~(DT)}~\cite{quinlan1986induction} and \emph{Random Forest~(RF)}~\cite{breiman2001random} are two tree-structure based non-linear classifiers. 
Based on certain attribute-splitting criteria (e.g., \emph{Information Gain} or \emph{Gini Impurity}), \emph{DT} can analyse the most informative attributes sequentially (i.e., splitting) until the final decision can be made.
The tree structure makes it interpretable and it has reasonable accuracy with low-dimensional semantic attributes. 
However, it faces ``the curse of dimensionality'' problem and does not generalize well when the input feature quality is low.
\emph{RF}, on the other hand, can effectively address this overfitting issue. 
\emph{RF} is an ensemble approach on aggregating different small-scale DTs, which are derived based on random sampling of the features/datasets.
The random sampling mechanism can effectively reduce the dimensionality (for each individual \emph{DT}) while the aggregation function can smooth the uncertainty of individual \emph{DT}s, making \emph{RF} a powerful model with great generalisation capabilities. 
However, the interpretability of \emph{RF} tends to be less obvious than that of \emph{DT}, owing to the random sampling and aggregation mechanisms.

\emph{Support Vector Machine~(SVM)}~\cite{cortes1995support} is another popular supervised learning method. 
It is also called large margin classifier as it aims at finding a hyperplane that is capable of separating the data points (belonging to different classes) with the largest margin. 
For non-linearly separable datasets, various kernels (e.g., \emph{RBF (Radial Basis Function)}) can be applied into the \emph{SVM} framework with good generalization ability.  
Yet the time complexity for training this algorithm can be very high (i.e., $O(N^3)$ \cite{abdiansah2015time}, where $N$ represents the dataset size), making it less suitable for big datasets.
On the other hand, \emph{K-Nearest Neighbour~(KNN)}\cite{cover1967nearest}, which does not require a training process (also referred to as lazy learning), is another powerful non-linear classifier. 
The classification is performed by distance calculation (between query and all the training examples), distance ranking, and 
majority voting among the (\emph{K}) nearest neighbours. 
So selecting suitable distance functions/metrics (for different tasks) is one of the key issues in \emph{KNN}. 
Since for any query sample, the distance calculation has to be performed for every sample in the whole training set, it can be time-consuming and thus less scalable to large datasets.
Different from the aforementioned methods, \emph{Naive Bayesian~(NB)} algorithm \cite{friedman1997bayesian} takes the prior knowledge of the class distribution into account. 
Based on the assumption that the features are conditionally independent, the likelihood of each feature can be calculated independently, before being combined with the prior probability according to the Bayes' rule.
If the feature-independence assumption is not significantly violated (e.g., low-dimensional structured data), it can be a very effective and efficient tool.

\myparagraph{Unsupervised Learning} The unsupervised learning algorithm~(see Fig. \ref{fig:supandunsup} right) aims at learning the inherent relationship between the data when only input data $X$ exists (without class label $Y$). 
For example, the clustering algorithm can be used to find the potential patterns of some unlabelled data and the obtained results can be used for future analysis. 
\emph{K-Means}\cite{hartigan1979algorithm} and \emph{Principal Component Analysis~(PCA)}~\cite{shlens2014tutorial} are the two most popular unsupervised learning algorithms. \emph{K-means} aims to find $K$ group patterns from data by iteratively assigning each sample to different clusters based on the distance between the sample and the centroid of each cluster. 
\emph{PCA} is normally used for dimensionality reduction, which can de-correlate the raw features before selecting the most informative ones.

\myparagraph{Discussion} 
For IoT applications, a common principle is to select the algorithm with the highest performance in terms of effectiveness and efficiency. 
One can run all related algorithms (e.g., supervised, or unsupervised), before selecting the most appropriate one.
For effectiveness, one has to define the most suitable evaluation metrics, which can be  
task-dependent, e.g., accuracy or mean-f1 score for classification tasks, or mean squared errors for regression, etc. 
Before model selection,
a number of factors should be taken into account: data structure (structured data, or unstructured data which may need additional preprocessing), data size (small or large), prior knowledge (e.g., class distribution), data separability (linearly, or non-linearly separable which may require additional feature engineering), dimensionality (low, or high which may require dimensionality reduction), etc.
There may also exist additional requirements from the users/stakeholders, e.g., interpretability for health diagnosis.
Additionally, it is necessary to understand the efficiency requirement specific to an IoT application and one has to consider how the training/testing time grows with respect to data size. 
Time complexity shown in Table 2 in \textbf{Appendix B} provides more insights.
Take \emph{KNN} as an example: although no training time is taken,  \emph{KNN}'s inference time can be very high (especially with a large training set), and 
thus presumably unsuitable for certain time-critical IoT applications. Also, the deployment environment is another non-negligible factor when developing IoT applications since many applications run (or partially run) on low power computing resources. 
\subsubsection{Deep Learning} \label{sec:deep learning}

In this section, we primarily introduce three classical deep models (i.e.,  \emph{Deep Neural Networks} (DNN)/ \emph{Multilayer Perceptron} (MLP),  \emph{Convolutional Neural Networks} (CNN) and  \emph{Recurrent Neural Networks} (RNN)) for supervised learning tasks on unstructured data such as  image, video, text, time-series data, etc.
We also brief two popular unsupervised models: \emph{Autoencoder} (AE), and  \emph{Generative Adversarial Networks} (GAN). 

\myparagraph{Supervised DL} Next, we will introduce three basic supervised DL models: DNN, CNN and RNN, which require both the data and label for training.

\emph{Deep Neural Networks (DNN).} As previously mentioned, a deep neural network (\emph{DNN}) is an  \emph{ANN} with more than one hidden layer, and hence it is also called multilayer perceptron (\emph{MLP}). 
Compared with  \emph{ANN} with a single hidden layer,  \emph{DNN} has more powerful modelling capabilities and its deep structure makes it easier for it to learn higher-level semantic features, which is crucial for classification tasks on complex data.
However, for high-dimensional unstructured input data (such as images), there may be many model parameters to be estimated, and in this case, overfitting may occur if there is not enough labelled data.
Nevertheless, generally  \emph{DNN} has decent performance when input dimensionality is not extremely high, and it has been successfully applied  
to various applications, for example human action recognition \cite{vacher2015speech}, traffic congestion prediction\cite{devi2017machine} and healthcare\cite{7821702}.

\emph{Convolutional Neural Network (CNN).}  When it comes to high-dimensional unstructured data such as images, in visual recognition tasks it is hard to directly map the raw image pixels into target labels due to the complex non-linear relationship. 
The traditional way is to perform feature engineering, which is normally a trial-and-error process, and may require domain knowledge in certain circumstances, before TML is applied.
This heuristic approach is normally time-consuming, and there exist substantial recognition errors even in simple tasks since it is very challenging to hand-engineer the high-level semantic features.
 \emph{CNN}, a deep neural network with convolutional layers and pooling layers, can address this issue effectively. 
The convolution operation can extract the higher level features while the pooling operation can keep the most informative responses and reduce the dimensionality.  
Compared with  \emph{DNN}, the weight sharing concept (of the convolution operation) enables  \emph{CNN} to capture the local pattern without suffering from the ``curse of high-dimensionality'' from the input. 
These operations and the hierarchical nature make \emph{CNN} a powerful tool for extracting high-level semantic representations from raw image pixels directly, and successfully applied to various recognition tasks such as object recognition, image segmentation~\cite{Everingham10} and object detection~\cite{He2017MaskR}. Because of the decent performance on various visual analysis tasks,  \emph{CNN} is usually considered as the first choice for some camera-based IoT applications, for example  traffic sign detection~\cite{shustanov2017cnn}.


\emph{Recurrent Neural Networks (RNN).} Nowadays, with the increasing amount of generated stream and sequential data from various sensors, time series analysis has become popular among the machine learning (ML) community. 
 \emph{RNN} is a sequential modelling technique that can effectively combine the temporal information and current signal into the hidden units for time-series classification/prediction. 
An improved  \emph{RNN} named Long Short Term Memory (LSTM) \cite{hochreiter1997long}, including complex gates and memory cells within the hidden units for ``better memories'', became popular in various applications such as speech recognition~\cite{Graves2013SpeechRW}, video analysis~\cite{Ullah2018ActionRI}, language translation~\cite{luong-etal-2015-effective}, activity recognition \cite{guan2017ensembles} etc.  
Since data streaming is most common in the IoT environment, RNN (LSTM) is deemed as one of the most powerful modelling techniques, and there are various IoT applications such as  
smart assistant~\cite{frewat2016android,vacher2015speech}, \textit{smart car navigator system}~\cite{ivanecky2012car}, malware threat hunting~\cite{haddadpajouh2018deep}, network traffic forecasting~\cite{ramakrishnan2018network}, equipment condition forecasting~\cite{zhang2018lstm}, energy demand prediction system~\cite{munir2017rnn}, load forecasting~\cite{kong2017short}, etc.

\myparagraph{Unsupervised DL} We will also introduce two unsupervised DL models: 
\textit{Autoencoder~(AE)}~\cite{baldi2012autoencoders} and \textit{Generative Adversarial Network~(GAN)}~\cite{goodfellow2014generative}. 

Without requiring any label information, \emph{AE} can extract compact features and reconstruct the original (high-dimensional) data with the extracted features. 
It is normally used for dimensionality reduction, latent distribution analysis or outlier detection.
\emph{GAN}, on the other hand, applies an adversarial process to learn the ``real'' distribution from the input data. 
More precisely, \emph{GAN} consists of two parts, namely generator and discriminator. 
The generator aims at generating indistinguishable samples compared to the real data. While the discriminator works adversarially to distinguish the generated fake samples from the real data. 
It is an iterative competition process that will eventually lead to a state where the generated samples are indistinguishable from the real data. 
With the learnt "real" distribution, one can generate various samples for different purposes.
 \emph{AE} and  \emph{GAN} are both powerful tools in the computer vision field, and their properties make them promising approaches for IoT applications.
 \emph{AE} can be used for diagnosis/fault detection tasks~\cite{chopra2015fault, oh2018residual} or simply as a preprocessing tool (i.e., feature extraction/dimensionality reduction). 
\emph{GAN} has been used for studies on generating rare category samples, and this upsampling approach may further improve the model performance~\cite{zhao2018application, zhao2018research}. 

\myparagraph{Discussion} 
The aforementioned DL models can be effective tools for processing different unstructured data types. 
The way of applying them is generally very flexible, and they can be used jointly to process the complex data from various sources in the IoT environments.
For example, although  \emph{CNN}/\emph{RNN} could be used in an end-to-end manner (e.g., as image/time-series classifiers), they could also be used as feature extractors, based on which one can easily aggregate features extracted from different sources (e.g., audio, images, sensor data). 
With high-dimensional video data, one can either model by training  \emph{CNN}+ \emph{LSTM} jointly  \cite{venugopalan2014translating}, or use  \emph{CNN}/\emph{AE} as feature extractors, before the sequential modelling (e.g., using  \emph{LSTM}).
However, when modelling the data with limited labels (e.g., rare event), one needs to consider the potential overfitting effect when using DL directly.
One may go back to the TML approaches or use some upsampling techniques (e.g.,  \emph{GAN}) to alleviate this effect.

\subsubsection{Reinforcement Learning (RL)} \label{rl}
In this section, we first introduce the strategies used to formulate the aforementioned video streaming example (see \S \ref{sec:training}) with Reinforcement Learning (RL). 
As mentioned earlier, in RL an \textit{agent} interacts with the \textit{environment}, learning an optimal control policy through experience. 
It requires three key elements, \textit{observation}, \textit{action}, and \textit{reward}. 
Based on these, we can formulate the adaptive bitrate streaming problem.
Specifically, \textit{observation} can be the buffer occupancy, network throughput, etc. 
At each step, the \textit{agent} decides the bitrate of the next chunk. 
A \textit{reward} (for example the quality of service feedback from the user) is received after the \textit{agent} takes \textit{action} (chunk bitrate).
The algorithm proposed in \cite{mao2017neural} collects and generalizes the results of performing the past decisions and optimizes its policy from different network conditions. 
This RL-based algorithm can also make the system robust to various environmental noises such as unseen network conditions, video properties, etc.

As shown in Fig. \ref{fig:rl-categorisation}, there is a plethora of algorithms in the whole reinforcement learning family. 
More details of these RL algorithms can be found in the  \textbf{RL methods in Appendix B}, and here  
we focus on selecting appropriate RL algorithms based on different selection criteria. 
\vspace{-5mm}

\input{figuresTex/fig-rl-categorisation}

\textbf{Environment Modelling Cost}
In RL modelling, sample efficiency is one of the major challenges. 
Normally the RL agent can interact either with the real world or a simulated environment during training. 
However, it can be difficult to simulate the heterogeneous IoT environments and complex IoT devices. 
RL models can also be trained directly in real world IoT environments, yet one major limitation is the heavy training cost, which may range from
 seconds to minutes for each step.  
The model-based RL method, a method that can reduce the sample complexity, can decrease the training time significantly.
It first learns a predictive model of the real world, based on which the decisions can be made. 
When compared with model-free approaches, model-based methods are still in their infancy, and because of the efficiency property, they may attract more attention in the near future.


\textbf{Action Space:}
The action space of RL algorithms can be either continuous or discrete.
 For those RL algorithms with discrete action space, they choose from a finite number of actions at runtime. 
 Take the video streaming task for example, the action space is different bitrates for each chunk. 
 Another task formulated in discrete action space can be found in \cite{mao2016resource}, where the action space is the ``schedule of the job at $i$-th slot''.
Available algorithms for discrete action space tasks most reside in the policy gradient group, for example DQN, DDQN.
The continuous action space, on the other hand, is infinite for all possible actions. 
Relationships exist between the actions that are usually sampled from certain distributions such as Gaussian distribution. 
For example, in an energy-harvesting management system, PPO algorithm \cite{schulman2017proximal} is used to control IoT nodes for power allocation. 
The action space, as stated in \cite{murad2019autonomous}, is sampled from a Gaussian distribution to denote the load of each node ranging from 0\% to 100\%. 
Similarly, in another work \cite{aoudia2018rlman} that studied energy harvesting WSNs, the Actor-Critic \cite{konda2000actor} algorithm is implemented to control the packet rate during transmission.
One advantage of continuous action space lies in its ability to accurately control the system, thus a higher QoE is expected.

%% file: figuresTex/fig-training-cycle.tex
\begin{figure}[ht]
\centering
\includegraphics[width=4.5in]{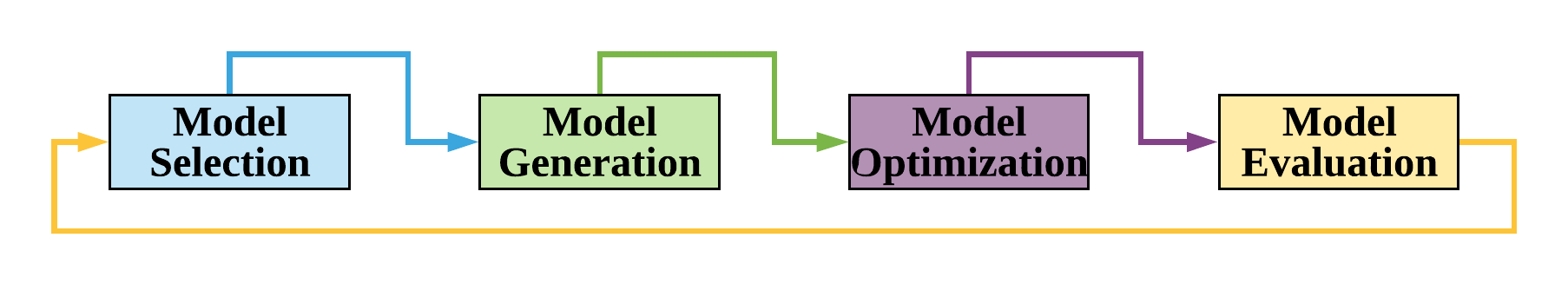}
\vspace{-6mm}
 \caption{A general pipeline of model development}
\label{fig:train_cycle}
\vspace{-5mm}
\end{figure}

%% file: figuresTex/fig-pensieve.tex
\begin{figure}[ht]
\centering
\includegraphics[width=3.5in]{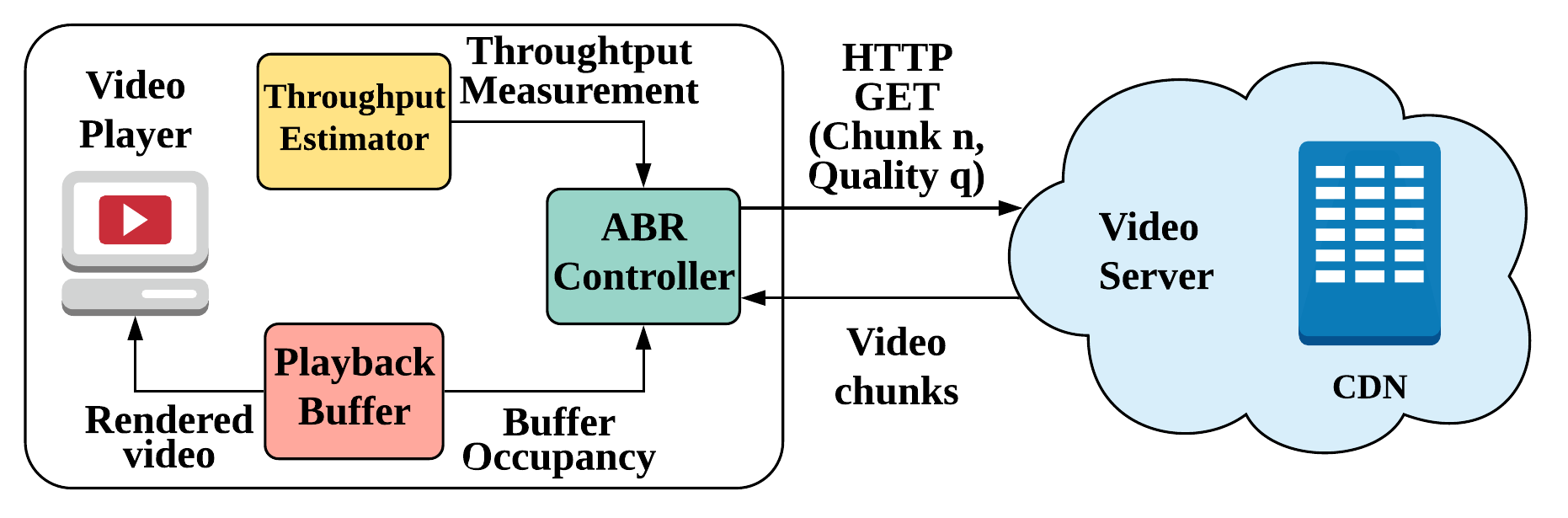}
\vspace{-5mm}
 \caption{Adaptive video streaming}
\label{fig:pensieve}
\vspace{-5mm}
\end{figure}

%% file: figuresTex/fig-reinforcement-learning.tex
\begin{figure}[ht]
\centering
\includegraphics[width=3.0in]{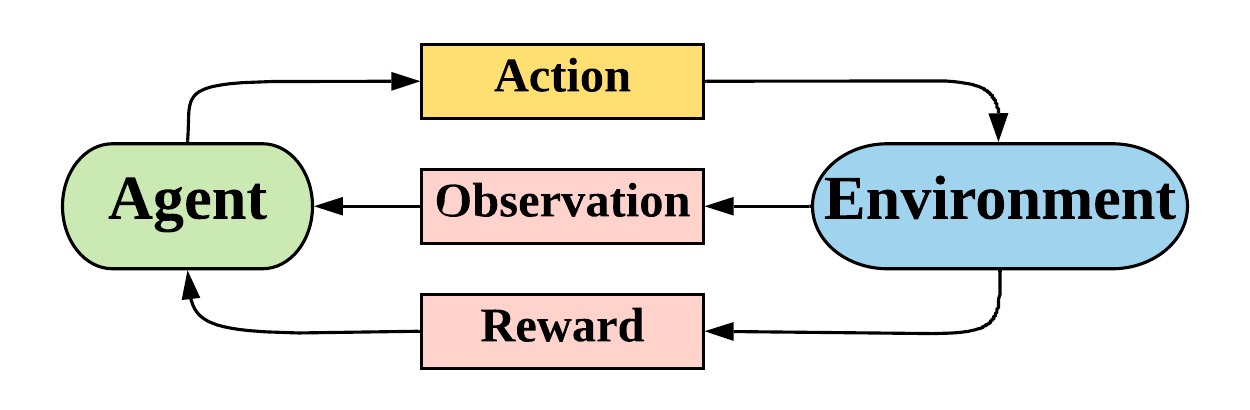}
\vspace{-5mm}
 \caption{Reinforcement Learning Paradigm}
\label{fig:rl}
\vspace{-5mm}
\end{figure}

%% file: figuresTex/fig-new.tex
\begin{figure}[ht]
\centering
\includegraphics[width=4in]{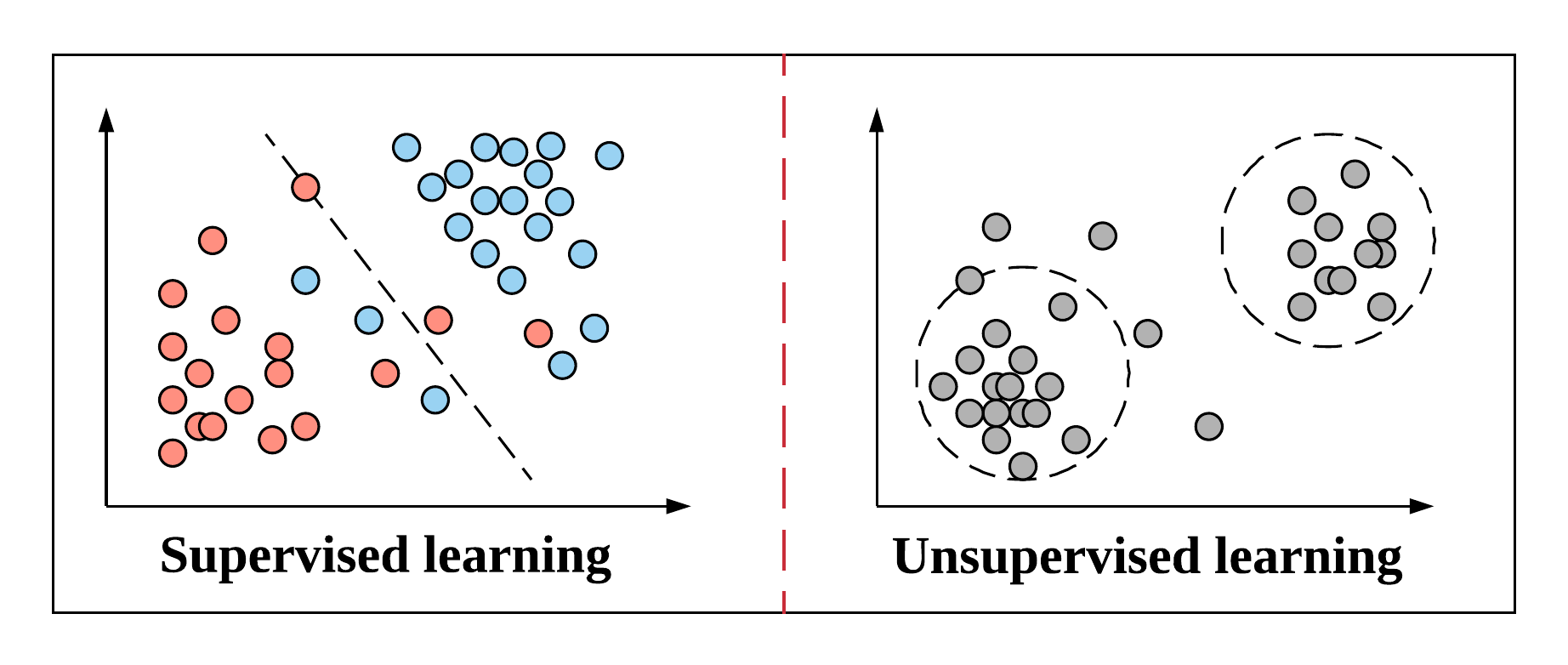}
\vspace{-5mm}
 \caption{Examples of Supervised Learning (Linear Regression) and Unsupervised Learning (Clustering) }
\label{fig:supandunsup}
\vspace{-5mm}
\end{figure}

%% file: figuresTex/fig-rl-categorisation.tex
\begin{figure}[ht]
\centering
\includegraphics[width=3.0in]{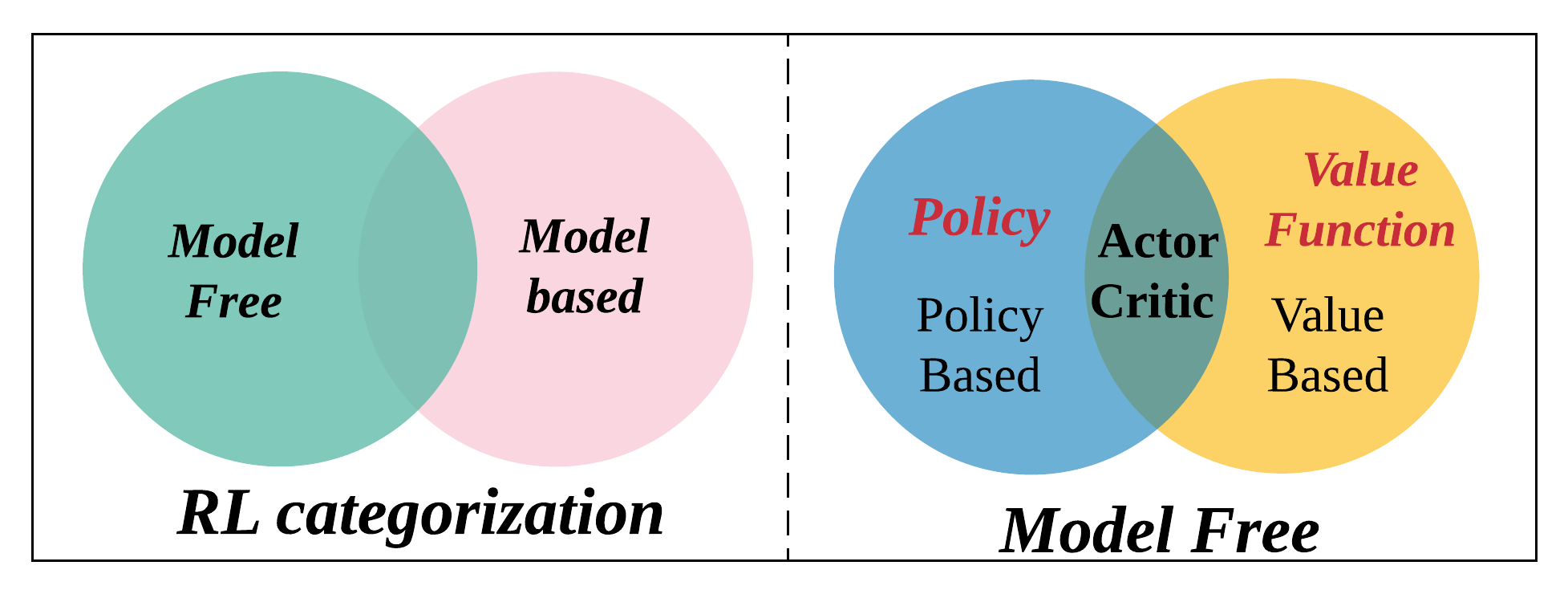}
\vspace{-4mm}
 \caption{Reinforcement Learning Categorization}
\label{fig:rl-categorisation}
\vspace{-4mm}
\end{figure}

%% file: model_generation.tex
\subsection{Model Generation} \label{generation}
Based on the user requirement and task specification, we have selected a variety of models. 
Next, the models need to be developed and implemented.
In this section, we will introduce the available tools for the development. 
We will also present the approaches that can be utilized to accelerate the training process.

\subsubsection{Machine Learning Development Framework} \label{mlframework}
The training and execution of ML models can be tricky and it may require numerous engineering efforts. 
Efforts have been devoted to developing frameworks to support the model development. 
These frameworks have their own strengths and weaknesses in terms of the supported models, usability, scalability, etc. In this section, we will review several development frameworks.   

For TML, the most famous development framework is Sci-kit learn. 
It is a free ML library with Python interface. 
Sci-kit learn supports almost all main-stream machine learning models and is a popular tool for fast prototyping. 
For DL, we list some of the most popular DL frameworks and discuss their pros and cons in Table \ref{DLLibrary}. Uusers can choose the most suitable frameworks based on their needs.

\begin{table}[t]
\footnotesize
\resizebox{\textwidth}{!}{%
\begin{tabular}{|c|c|c|l|l|}
\hline
\textbf{\begin{tabular}[c]{@{}c@{}}DL \\ frameworks\end{tabular}} & \textbf{\begin{tabular}[c]{@{}c@{}}Core\\  language\end{tabular}} & \textbf{Interface}                                                                    & \multicolumn{1}{c|}{\textbf{Pros}}                                                                                                                                                                                                      & \multicolumn{1}{c|}{\textbf{Cons}}                                                                                      \\ \hline
Tensorflow (2)                                                    & C++                                                               & \begin{tabular}[c]{@{}c@{}}Python, Javascript,\\ C++, Java, Go\end{tabular}           & \begin{tabular}[c]{@{}l@{}}- Effective data visualization\\ - Distributed learning\\ - Efficient model serving\\ - On-device inference with low latency for\\   mobile devices\\ - Eager Execution with TF2, easy to debug\end{tabular} & \begin{tabular}[c]{@{}l@{}}- Steep learning curve (migration from TF 1 to \\   TF 2)\\ - Poor results for speed\end{tabular} \\ \hline
Pytorch                                                           & C/C++                                                             & Python, C++                                                                           & \begin{tabular}[c]{@{}l@{}}- Simple and transparent modeling\\ - Eager execution\end{tabular}                                                                                                                                           & - Hard to serve even with ONNX support                                                                                  \\ \hline
Caffe (2)                                                         & C++                                                               & Python, C++                                                                           & \begin{tabular}[c]{@{}l@{}}- Fast, scalable, and lightweight\\ - Server optimized inference\end{tabular}                                                                                                                                & \begin{tabular}[c]{@{}l@{}}- Limited community support\\ - Limited in implementing complex networks\end{tabular}        \\ \hline
Mxnet                                                             & C++                                                               & \begin{tabular}[c]{@{}c@{}}Python, C++, Java, \\ Julia, R, Perl, Clojure\end{tabular} & \begin{tabular}[c]{@{}l@{}}- Fast, flexible, and efficient in terms\\   of running DL algorithms\\ - Run on any device\\ - Easy model serving\\ - Highly scalable\end{tabular}                                                          & \begin{tabular}[c]{@{}l@{}}- Smaller community compared with Tensorflow\\   or Pytorch\end{tabular}                     \\ \hline
DL4J                                                              & Java                                                              & Java, Clojure, Kotlin                                                                 & \begin{tabular}[c]{@{}l@{}}- Robust, flexible and effective\\ - Works with Apache Hadoop and Spark\end{tabular}                                                                                                                         & \begin{tabular}[c]{@{}l@{}}- Robust, flexible and effective\\ - Works with Apache Hadoop and Spark\end{tabular}         \\ \hline
\end{tabular}%
}
\caption{Comparison of Deep Learning Frameworks}
\label{DLLibrary}
\vspace{-8mm}
\end{table}

When IoT comes into context, more challenges arise with edge computing as it is trying to move the computation close where the data is generated \cite{shi2016edge}. The device heterogeneity of edge computing has made the development of the DL models more complicated. 
There are many portable Edge computing devices, each optimized with different inference engines.
For example Nvidia Jetson series GPU computing unit compiles models with TensorRT\footnote{https://developer.nvidia.com/tensorrt} inference engine while TensorFlow Lite\footnote{https://www.tensorflow.org/lite} is specially optimized for Google coral TPU.
These inference engines optimize the model graph and quantize the model parameters to lower precision, thus delivering low latency and high-throughput for on-device inference. 
Some attempts \cite{tensorflow_2019} have been made to integrate both inference engines but the compatibility issue still exists.
TVM \cite{chen2018tvm} breaks the boundaries among diverse hardware, aiming at cross-framework and cross-device end-to-end optimization of DL models.

\subsubsection{Single Machine Learning (Centralized)} \label{sec:singlemachine}
Model training via single machine is a common strategy for ML model generation. By placing the learning-related computation in the same place, the model learns from the data and updates its parameters. In this subsection, we highlight two approaches that leverage hardware for the training process acceleration: \textit{Computation Optimization}, \textit{Algorithm Optmization}.

\emph{Computation Optimization} The basic computation unit of neural networks consists of vector-vector, vector-matrix and matrix-matrix operations. 
Efficient implementation of computations can accelerate the training and inference process. 
The Basic Linear Algebra Subprogram (BLAS) 
\footnote{http://www.netlib.org/blas}  
standardizes the building blocks for basic vector, matrix operations. 
A higher level linear algebra library such as cuBLAS 
\footnote{https://docs.nvidia.com/cuda/cublas/} 
implements BLAS on top of NVIDIA CUDA and is efficient in utilizing the GPU computation resource. 
Intel Math Kernel Library (MKL) \footnote{https://software.intel.com/en-us/mkl}, on the other hand, maximizes performance on Intel processors and is compatible with BLAS without the change of code.

Different DL architectures (e.g., DNNs, CNNs and RNNs) may require different optimizations in terms of basic computations. 
The DNN computation is usually basic matrix-matrix multiplication and the aforementioned BLAS libraries can efficiently accelerate the computations with GPU resources. The CNNs and RNNs are different in their convolution and recurrent computations. 
Convolutions can not fully utilize the multi-core processors and the acceleration can be achieved by unrolling the convolution \cite{chellapilla2006high} to matrix-matrix computation or computing convolutions as point-wise product \cite{mathieu2013fast}. 
For RNN (LSTM), the complex gate structures and consecutive recurrent layers differ from the DNNs and CNNs in that these computation units can not be split and deployed directly at different devices. This has  made parallel computation difficult to apply.
Optimization is possible though, with implementations on top of NVIDIA cuDNN \cite{chetlur2014cudnn}. Computations among the same gates can be grouped into larger matrix operations \cite{appleyard2016optimizing} and save intermediate steps. 
We can also accelerate by caching RNN units' weights with the GPU's inverted memory hierarchy \cite{diamos2016persistent}. The weights are reusable between time steps, making a maximum $30\times$ speed up on a TitanX GPU.

\emph{Algorithm Optimization}
Apart from the resource utilization optimization, the algorithmic level optimization is another important research direction for efficient model training and faster convergence.
Optimization algorithms aim at minimizing/maximizing a loss function that varies for different machine learning tasks.
They can be divided into two categories: \textit{First Order Optimization} and \textit{Second Order Optimization}.

\textit{First Order Optimization} methods minimizing/maximizing the loss function with the gradient values with respect to the model parameters. Gradient Descent  is one of the most important algorithms for neural networks.
After back-propagation from the loss function, the model parameters are updated towards the opposite direction of the gradient.
Gradient descent approaches fall into local optima when the absolute value is either too big or too small.
Also it updates the gradient of the whole data set at one time, memory limitation is always a big problem.
Variants have been proposed to address the aforementioned problems, including Stochastic gradient descent \cite{bottou2012stochastic}, mini-batch gradient descent \cite{dekel2012optimal}. 
Also, much famous research enables faster model convergence: Momentum \cite{qian1999momentum}, AdaGrad \cite{duchi2011adaptive}, RMSProp \cite{RMSprop}, ADAM \cite{kingma2014adam}. 
\textit{Second Order Optimization} methods take second order derivative for minimizing/maximizing loss function. Compared to the \textit{First Order Optimization}, it consumes more computation power and is less popular for machine learning model training. However, \textit{Second Order Optimization} considers the surface curvature performance and is less likely to get stuck on saddle points. Thus it sometimes outperforms the \textit{First Order Optimization}. Famous \textit{Second Order Optimization} methods include \cite{ba2016distributed, byrd2016stochastic, he2017distributed, moritz2016linearly, osawa2018second}.
For more systematic survey on the optimization methods for machine learning training, one can refer to \cite{bottou2018optimization}.

\subsubsection{Distributed Machine Learning}\label{sec:distributed}
Modern ML models such as neural networks require a substantial amount of data for the training process.
These data are usually aggregated and stored in the cloud server where training happens.  
However, when the training process of large volume data outpaces the computing power of a single machine, we need to leverage multiple machines available in the server cluster. This requires the development of novel distributed ML systems and parallel training mechanisms which distribute and accelerate the machine learning workload.

\input{figuresTex/fig-distributed_ml}

Fig \ref{fig:distributed_ml} shows the schematic diagram of a distributed ML pipeline. It has multiple components which are engaged in \textit{Training Concurrency}, \textit{Single Machine Optimization}, and \textit{Distributed System}. 
In \textit{Training Concurrency}, either the models or the data are split into small chunks and placed on different devices. With \textit{Single Machine Optimization} (which shares similar techniques as conventional ML, see 
\S~\ref{sec:singlemachine}) that accelerates the training process, we get all local gradient updates. Finally, \textit{Distributed System} discusses strategies that efficiently aggregate the gradient updates.


\myparagraph{Training Concurrency in Distributed ML}
In the distributed machine learning, the selection of parallel strategy depends on two factors: data size and model size. When either the datasets or the model parameters are too big for single-machine processing, it is straightforward to consider partitioning them into smaller chunks for processing at different places. Here we first introduce two basic methods \textit{data parallel}, \textit{model parallel}. We also introduce \textit{pipeline parallel} and other hybrid approaches that take advantage of both approaches.

\emph{Data Parallel.} In a multi-core system where a single core can not store all the data, data parallel is considered by either splitting the data samples or the features.
Data parallel has been successfully applied to numerous machine learning algorithms \cite{chu2007map} with each core working independently on a subset of data. 
It can be used for training ML algorithm for example decision trees and other linear models where the features are relatively independent.
Parallel with the split of data features, though, it can not be used directly with neural networks because different dimensions of the features are highly correlated. 

In deep learning, data parallel works by distributing the training dataset across different GPU units. 
The dominant data parallel approach is \textit{batch parallelism} where mini-batch SGD is employed to compute local gradient updates on a subset of the data. 
A central server is responsible for aggregating all local updates to global parameter and pushing new models back to the working units.
One of the earliest works trained with GPUs can be found in \cite{raina2009large} where the authors implemented distributed mini-batch SGD unsupervised learning concurrently with thousands of threads in a single GPU. 
By varying the batch size \cite{goyal2017accurate, smith2017don, you2017large}, this method is effective in reducing the communication cost without too much accuracy loss.
In the next paragraph, we will discuss more about the parallel SGD algorithms \cite{mcdonald2010distributed, zinkevich2010parallelized, zhang2016parallel, stich2019local, yu2019parallel} for improving the communication efficiency, which can be seen as one way of improving the performance of data parallelism.
Another type of data parallel that addresses the memory limit on single GPU is \textit{spatial parallelism} \cite{jin2018spatially}. \textit{Spatial parallelism} considers partitioning spatial tensors into smaller subdivisions and allocating them to separate processing units. It thus differs from \textit{batch parallelism} in that the latter puts the groups of data in the same process. \textit{Spatial parallelism} approach has proven to show near linear speedup on modern multi-GPU systems.

\emph{Model Parallel.} 
Data parallel suffers from the infeasibility of dealing with very large models especially when it exceeds the capacity of a single node. Model parallel addresses this problem by splitting the model with only a subset of the whole model running on each node \cite{bradley2011parallel, dean2012large, lee2014model, kim2016strads}. The computation graphs can be divided within the layers (horizontal) or across the layers (vertical). Mesh-tensorflow \cite{shazeer2018mesh} allows linear within-layer scaling of model parameters across multiple devices after compiling a computation graph into a SPMD program. However, this approach requires high communication cost as it needs to split and combine model updates across a large number of units. \cite{huo2018decoupled} introduced decoupled parallel backpropagation to break the sequential limitation of the back-propagation between the nodes, greatly increasing the training speed without much accuracy loss. 
For CNN, as each layer can be specified as five dimensions including: samples, height, width, channels, and filters, existing literature \cite{dryden2019improving, dryden2019channel} studies the split of models among dimensions.
Another research direction that optimizes the communication overhead is by searching the optimal partition and device placement of computation graphs via reinforcement learning \cite{mirhoseini2018hierarchical}. The literature \cite{jia2018beyond, wang2019supporting} followed this idea and shows interest in automatic search of optimal parallel strategies.

\emph{Pipeline Parallel.}
Although model parallel has proven successful in training extremely large models, the implementation is complicated due to the complexity of the neural network structure. This is especially true for CNNs since the convolution operators are highly correlated. 
Also, GPU utilization is low for model parallel. Due to the gradient interdependence between different partitions, usually only one GPU is in use each time. 
To solve the aforementioned problems, pipelining has been studied \cite{petrowski1993performance, wu2016google} for speeding up the model training. 
With pipeline parallel, models are partitioned and displayed across different GPUs. Then mini-batches of training data are injected to the pipeline for concurrent processing of different inputs at the same time.
Fewer worker GPUs are idle in the pipeline parallel setting as each node is allocated jobs, without waiting for other nodes to finish their work. 
According to the synchronization strategy we discussed earlier, gradients are aggregated by either synchronous pipeline model (GPipe \cite{huang2019gpipe}) or asynchronous pipeline model (PipeDream \cite{harlap2018pipedream}, SpecTrain \cite{chen2018efficient}, XPipe \cite{guan2019xpipe}). 
Theoretical analysis of pipeline parallel optimzation has also been studied and with Pipeline
Parallel Random Smoothing (PPRS) \cite{colin2019theoretical}, convergence rates can be further accelerated.

\emph{Hybrid.} 
Data and model parallel are not mutually exclusive. Hybrid approaches that combine the benefits of both methods are effective in further accelerating the training process.
Pipeline parallel \cite{harlap2018pipedream, huang2019gpipe} can be seen as an approach built on top of data parallel and model parallel. Apart from that, \cite{krizhevsky2014one} proposed combining data parallel and model parallel for different types of operators. With data parallel for CNN layers and model parallel for DNN layers, it achieved a $6.25\times$ speed up with only $1\%$ of accuracy loss on eight GPUs. 
Another implementation MAPS-Multi \cite{ben2015memory} borrows the idea of \cite{krizhevsky2014one} and automates the partitioning of workload among multiple GPUs, achieving $3.12\times$ speed up on four GTX 780 GPUs.
Other forms of data parallel and model parallel hybrids exist in the literature \cite{dean2012large, chilimbi2014project, gholami2018integrated, gaunt2017ampnet} that reduce the overall communication and computation overhead.

\myparagraph{Distributed ML System}
When we have acquired a local model update with partial data slice, multi-node and multi-thread collaboration are important for effectively updating the model. 
Network communication plays an important role in sharing the information across the nodes. 
In this section, we present the three most important features in network communication: 1) network topology, 2) synchronization strategy and 3) communication efficiency.

\noindent \emph{Network Topology.} The network topology defines the node connection approach in the distributed machine learning system. 
When the data and models are relatively simple, it is common to utilize existing Message Passing Interface (MPI) or MapReduce infrastructure for the training. 
Later when the systems are becoming more and more complex, new topologies should be designed to facilitate the parameter update.

The Iterative MapReduce (IMR) or AllReduce approaches are commonly used for synchronous data parallel training. Typical IMR engines (for example the Spark MLlib \cite{meng2016mllib}) generalizes MapReduce and enables the iterative training required by most ML algorithms. 
Synchronous training can also be implemented by AllReduce topology.
MPI\footnote{https://computing.llnl.gov/tutorials/mpi/} (Message Passing Interface) supports AllReduce and is efficient for CPU-CPU communication. Many researchers implement their own version of AllReduce for example Caffe2 Gloo\footnote{https://github.com/facebookincubator/gloo}, Baidu Ring AllReduce\footnote{https://github.com/baidu-research/baidu-allreduce}.
In the ring-Allreduce topology, all nodes connect to each other without a central server, just like a ring. 
The training gradients are aggregated through their neighbors on the ring.
To provide more efficient communication for DL workload in the GPU cluster, libraries such as Nvidia NCCL \cite{nccl} are developed and support the AllReduce topology. In NCCL2 \cite{nccl2}, the multi-node distribution feature is also introduced. 
Horovod \cite{sergeev2018horovod} replaces the Baidu ring-Allreduce backend with NCCL2 for  efficient distribution.

A Parameter Server (PS) infrastructure \cite{li2014scaling} is usually composed of a set of worker nodes and a server node which gathers and distributes computation from worker nodes.
As asynchronous training of PS neglects stragglers, it provides better fault tolerance capability when some of the nodes break down.
Parameter server also features high scalability and flexibility.
Users can add nodes to the cluster without restarting the cluster. 

Famous projects such as DMTK Microsoft Multiverso \cite{dmtk}, Petuum \cite{xing2015petuum} and DistBelief \cite{dean2012large} enable training of even larger networks. 

\noindent \emph{Synchronization Strategy.}
In distributed ML, model parameter synchronization between worker nodes is cost-extensive. The trade-off between the communication and the fresher updates has great impact on the parallelism efficiency.

Bulk Synchronous Parallel (BSP) \cite{bsp1995} is the simplest strategy for ensuring model consistency of all worker nodes. For each training iteration, all nodes wait for the last (slowest) node to finish the computation and the next iteration does not start before the all the model updates are aggregated. Total Asynchronous Parallel (TAP) \cite{dean2012large} approaches are proposed to address the problem of the stragglers within the network. With TAP, all worker nodes access the global model via a shared memory. They can pull and update global model parameters any time when the training is finished. As there is no update barrier for this approach, the system fault tolerance is greatly improved. However, stale model updates can not guarantee convergence to global optimum. Many famous frameworks use the TAP strategy, including Hogwild! \cite{recht2011hogwild} and Cyclades \cite{pan2016cyclades}.

`Stale Synchronous Parallel (SSP) \cite{ho2013more} compromises between fully-synchronous and asynchronous schemes. It allows a maximum staleness by allowing faster working nodes to read global parameters without waiting for slower nodes. As a result,  the workers spend more time doing valuable computation, thereby improving the training speed greatly. But when there is too much staleness within the system, the convergence speed can be significantly reduced. Many state-of-the-art distributed training systems implement BSP and SSP for efficient parallelism, for example tensorflow \cite{abadi2016tensorflow}, Geeps \cite{cui2016geeps}, Petuum \cite{xing2015petuum}.

In contrast to the SSP which limits the staleness of the model update, the Approximate Synchronous Parallel \cite{hsieh2017gaia} (ASP) limits the correctness. In Gaia \cite{hsieh2017gaia}, for each local model updates, the global parameter is aggregated only if the parameter change exceeds a predefined threshold. This ``significance" only strategy eliminates unnecessary model update and is efficient in utilizing the limited bandwidth. However, the empirical determination of threshold only considers the network traffic and is insufficient for dealing with dynamics in the IoT environment. \cite{wang2018edge} has addressed this problem by also considering resource constraints for efficient parallelism.

\noindent \emph{Communication Efficiency.} 
Communication overhead is the key and often the bottleneck in distributed machine learning \cite{li2014communication}. 
The sequential optimization algorithms implemented in the worker nodes require frequent read and write from the global shared parameters which poses great challenge on balancing network bandwidth and communication frequency.
To increase the communication efficiency, we can either reduce the size of the model gradient (communication content) or the communication frequency.

\emph{Communication content.}
The gradient size between working nodes is correlated to both the model size itself and the gradient compression rate. We have reviewed four types of \textit{model compression} techniques in \S \ref{sec:model_eff} which are effective in reducing the overall gradient size. Hereby we focus on the techniques that compress the gradient before transmission, discusses the gradient \textit{quantization} and \textit{sparsification}.

Gradient quantization differs from the weight quantization (\S \ref{sec:model_eff}) as the former  compresses the gradient transmission between worker nodes while the latter focuses on faster inference via smaller model size. Works that reduce the gradient precision \cite{de2015taming} have been proposed and 1-bit quantization \cite{seide20141, strom2015scalable} is effective in greatly reducing the computation overhead. Based on the idea, QSGD \cite{alistarh2017qsgd} and Terngrad \cite{wen2017terngrad} consider stochastic quantization where gradients are randomly rounded to lower precision. Additionally, weight quantization and gradient quantization can also be combined  \cite{zhou2016dorefa, hubara2017quantized, zhang2017zipml, wu2018training, hou2018analysis} for efficient on device acceleration.

The weights of the DNNs are usually sparse and due to the large number of unchanged weights in each iteration, the gradient updates are even more sparse. This sparsification nature of the gradient transmission has been utilized for more efficient communication. Gradient sparsification works by sending only important gradients when exceeding a fixed threshold \cite{strom2015scalable} or adaptive threshold \cite{dryden2016communication}. Gradient Dropping \cite{aji2017sparse} uses layer normalization to keep the convergence speed. DGC \cite{lin2017deep} uses local gradient clipping for sending important gradients first while the less important ones are aggregated with momentum correction for later transmission. 

\emph{Communication Frequency.} 
Local (Parallel) SGD \cite{mcdonald2010distributed, zinkevich2010parallelized, zhang2016parallel, stich2019local, yu2019parallel} entails performing local updates several times before parameter aggregation. Motivated by reducing the inter-node communication, this approach is also called model averaging. 
One-shot averaging \cite{zinkevich2010parallelized, mcdonald2010distributed} considers only one aggregation during the whole training process.
While \cite{zhang2016parallel} argues that one-shot averaging can cause inaccuracy and proposes more frequent communications, many works \cite{povey_zhang_khudanpur_2017, lin2018don, yu2018parallel, zhang2015deep} prove the applicability of the model averaging approach in various deep learning applications.  
In an asynchronous setting, the communication frequency can also be maneuvered through the push and pull operations in the worker nodes. 
DistBelief \cite{dean2012large} has adopted this approach with a larger push interval compared to the pull interval.

\subsubsection{Federated Learning} \label{sec:federatedlearning}

In traditional distributed machine learning, the training usually happens on the cloud data center with aggregated training data generated by collecting, labelling and shuffling  raw data. 
The training data is thus considered \textit{identical and independent distributed (IID)} and balanced.
This facilitates the training process as one only needs to consider distributing the training task across various computation units and updating the model by aggregating all local gradient updates.
However, this is not the case when IoT comes into play. 
The ML-based IoT applications differ from the traditional ML applications in that they usually generate data from heterogeneous geo-distributed devices (e.g., user behavior data from mobile phones). 
These data can be privacy-sensitive as users usually prefer not to leak personal information, making conventional distributed ML algorithms infeasible for solving such problems.
Thus novel optimziation techniques are required to enable training in such scenarios.

Federated learning (FL) \cite{konevcny2016federated} is a type of distributed machine learning research that moves the training close to the distributed IoT devices. 
It learns a global model by aggregating local gradient updates and does not require the movement of the raw data to the cloud center.
FederatedAveraging (FedAvg) \cite{mcmahan2016communication} is a decentralized learning algorithm specifically designed for the FL. It implements synchronous local SGD \cite{chen2016revisiting} on each device with a global server averaging over a fraction of all the model updates per iteration. 
FedAvg is capable of training high-accuracy models on various datasets  with many fewer communication rounds.
Following this work, \cite{konevcny2016federated} proposed two approaches: \textit{Structured updates} and \textit{sketched updates} for reducing the communication cost, achieving higher communication efficiency.
Further research addresses the privacy limitation of FL by Differential Privacy \cite{mcmahan2017learning} and Secure Aggregation \cite{bonawitz2019federated}. 
Finally, \cite{bonawitz2019towards} delivers system-level implementation of FL based on previously mentioned techniques. It is able to train deep learning models with local data stored on mobile phones.

FL is still developing rapidly with many challenges remaining to be solved. 
On the one hand, FL shares similar challenges as in conventional distributed machine learning methods in terms of more efficient communication protocol, synchronization strategy as well as parallel optimziation algorithms. 
On the other hand, the distinct setting of FL requires more research preserving the privacy of training data, ensuring the fairness and addressing bias in the data.
For a more thorough survey on details of FL, one can refer to \cite{kairouz2019advances}.

\subsubsection{Knowledge Transfer Learning} \label{sec:transferlearning}

The knowledge learnt from trained models can be transferred and adapt to new tasks, 
and it is especially helpful when limited data/labels are available. 
In this section we introduce four types of knowledge transfer learning (KTL) approaches: Transfer learning, Meta learning, Online learning and Continual learning.

\textbf{Transfer Learning}~\cite{tan2018survey} --- transferring knowledge across datasets--- is the most popular KTL approach.
It trains a model in the source domain (with adequate data/labels, e.g., on ImageNet \cite{deng2009imagenet} for general visual recognition tasks), and fine-tunes the model parameters in the target domain to accommodate the new tasks (e.g., medical imaging analysis on rare diseases).
The rationale behind is that low-level and mid-level features can be representative enough and thus shared across different domains. In this case, only the parameters related to high-level feature extraction need to be updated.
This mechanism does not require a large amount of data annotation for learning reliable representation in the new tasks, which could be useful in cases when annotations are expensive (e.g., medical applications). 
\textbf{Meta Learning}~\cite{vanschoren2018meta} is another popular KTL approach; instead of transferring knowledge across datasets, it focuses on knowledge transfer across tasks. 
Meta learning means learning knowledge or patterns from a large number of tasks, then transfer this knowledge for more efficient learning of new tasks.
When with continuous data streaming, it is also desirable to update the model with incoming data, and in this case \textbf{Online Learning}~\cite{hoi2018online} can be used. 
However, it is difficult to model when the incoming data is from a different distribution or a different task. 
Most recently, \textbf{Continual Learning}\cite{parisi2019continual} was proposed to address this issue. 
Not only can it accommodate the new tasks or data with unknown distribution, it can also maintain the performance on the old/historical tasks (i.e., no forgetting \cite{kemker2018measuring}), making it a practical tool for real-world IoT applications.
These four KTL approaches are similar in concept yet have different use cases. 
Transfer/Meta learning are focused on knowledge transfer across datasets or tasks (irrespective of data types),  while online/continuous learning are more suitable for data streaming and can transfer the knowledge continuously to the new incoming data or tasks.

\subsubsection{Discussion}
Effort has been devoted to implementing the distributed machine learning on top of modern deep learning frameworks. 
Remarkable results have been achieved where with proper implementation \cite{goyal2017accurate} with Tensorflow, the training time of the state-of-art ImageNet can be reduced from days to one hour. 
Compared with Tensorflow, more efficient implementation such as Horovod 
can increase the GPU utilization for even more acceleration. 
Horovod has already been incorporated in various deep learning framework ecosystems (e.g., Pytorch, Mxnet).

Deep learning ecosystems free the researcher from heavy implementation effort. 
There are however, challenges for model generation in a distributed setting: 
(1) \textit{The choice of hardware.} The same implementation can have different performance on different devices. 
One would have to be aware of the device features for efficient acceleration. 
(2) \textit{Parallel hyperparameter tuning strategy.} Compared with single machine training, the distributed system is more complex and it is thus more difficult to find an optimal structure. 
(3) \textit{Effective work of DL frameworks with other big data application like Hadoop/Spark.} 
Existing big data frameworks (e.g., Spark/Hadoop) can also be applied for effectively distributing the DL training pipeline, and a deeper integration of both frameworks is urgently required.

%% file: figuresTex/fig-distributed_ml.tex
\begin{figure}[ht]
\centering
\includegraphics[width=3in]{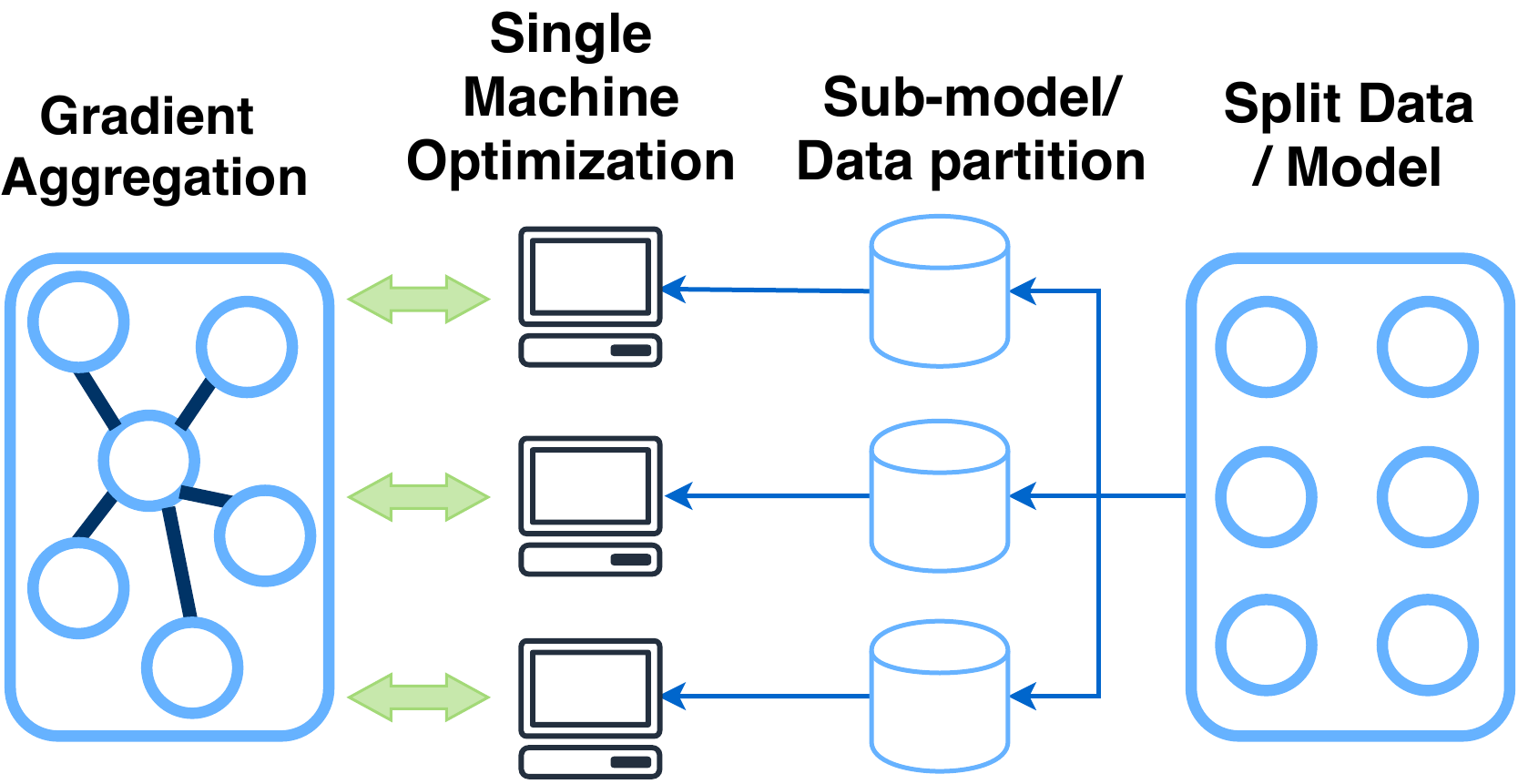}
\vspace{-4mm}
 \caption{Distributed Machine Learning Pipeline}
\label{fig:distributed_ml}
\vspace{-3mm}
\end{figure}

%% file: model_opt.tex
\subsection{Model Optimization} \label{optimisation}
We have discussed the model selection and model generation where a model is generated catering to the specific needs of IoT applications. 
There are, however, still things to be considered before model deployment. 
The IoT application differs significantly from other areas in terms of deployment devices and data sources. 
The limited computational budget of edge devices requires smaller models for small-scale computational workload to ensure low inference latency. Also, heterogeneous data sources in IoT environments usually contain redundant information that can even mislead the decision of the ML models. 
It is important to select only relevant and informative features or to perform model compression for performance optimization.  
In this section, we discuss these two topics. 

\subsubsection{Feature Selection} \label{fs}
The amount of high-dimensional data in IoT environments poses challenges on the training of the ML algorithms. 
Noisy and redundant signals exist and they may consume substantial computational power. 
For unstructured data (e.g., images, audio, etc.), a common way to reduce the data redundancy is to perform feature extraction, which can be realised in many ways, e.g., via feature engineering, supervised/unsupervised learning, etc.
The extracted features can be deemed as structured data, and its redundancy can be further reduced by feature selection.
Feature selection can help reducing the computational complexity, which may improve the performance in terms of both effectiveness and efficiency---crucial factors in the limited-resource IoT environments.
Briefly, feature selection is the process of preserving relevant features while discarding irrelevant/redundant features. 
There are generally three categories of feature selection approach, namely the \textit{Filter} approach \cite{dash2003consistency, hall2000correlation, battiti1994using}, the \textit{Wrapper} approach \cite{marcano2010feature, gutierrez2002pattern, Goldberg:1989:GAS:534133, jouan1995genetic} and the \textit{Embedded} approach \cite{quinlan2014c4, breiman2017classification}. 
The application of feature selection in the IoT literature is emerging especially when data quality plays a significant role in the performance of the learning application.

In \emph{Smart Farming}, early detection and treatment of plant disease or nutrient deficiency would increase the health of the crops. 
An  IoT based \textit{Smart Farming} system can give real-time decisions with optimal fertilizer volume. 
\cite{kale2019iot} \cite{meunkaewjinda2008grape} show that when there are many sensors with a large volume of IoT streaming data, feature selection can remove the redundant information and maximize the prediction accuracy. 
Another application is \emph{Intrusion Detection}, which aims to detect online behaviour that compromises the confidentiality of the network. 
In this case, feature selection can be used to extract the most informative patterns \cite{lee2000adaptive, lappas2007data}, serving as the basis of an effective intrusion detection system. 

 For future research, one could extend the single-object optimization to multi-object optimization. For example in IoT systems, optimal feature selection algorithms assist the machine learning models to optimize the execution time. 
 We can explore the modifications of the feature selection algorithm to minimize the energy consumption of routing decisions as well \cite{egea2017intelligent}.
One can also study how to detect the dynamics within the data flow and then adaptively apply  the search algorithms accordingly to further improve the performance of the feature selection algorithms. 


\subsubsection{Model efficiency} \label{sec:model_eff} 
The state-of-the-art DL models often require high computational resources beyond the capabilities of IoT devices. 
Those models that perform well on large CPU and GPU clusters may suffer from unacceptable inference latency or even be unable to run on edge devices (e.g. Raspberry Pi). 
Tuning the deep neural network architectures to increase the efficiency without sacrificing much accuracy has been an active research area.
In this section, we will cover three main optimization directions: \textit{Efficient architecture design}, \textit{Neural architecture search} and \textit{Model compression}.

\myparagraph{Efficient architecture design}
There exist neural networks that can specifically match the resource and application constraints. They aim to explore highly efficient basic architecture specially designed for platforms such as mobiles, robots as well as other IoT devices. MobileNets \cite{howard2017mobilenets} is among the most famous works and proposed to use depth-wise separable convolutions \cite{sifre2014rigid} to build CNN models. 
By controlling the network hyper-parameters, MobileNets can strike an optimal balance between the accuracy and the constraints (e.g., computing resources). Later in MobileNetv2 \cite{sandler2018mobilenetv2}, the inverted residual with linear bottleneck architecture was introduced to significantly reduce the operations and memory usage. 
Other important works include Xception \cite{chollet2017xception} ShuffleNet \cite{zhang2018shufflenet}, ShuffleNetv2 \cite{ma2018shufflenet}, CondenseNet \cite{huang2018condensenet}. 
These neural networks optimize on-device inference performance via efficient design of building blocks, achieving much less computational complexity while keeping or even raising accuracy on various computer vision datasets. Some work even outperforms the neural architectures generated through exhaustive automatic model search. Also, different building blocks can be combined together for even lighter models.

\myparagraph{Neural architecture search (NAS)}
 Another research direction named neural architecture search aims at searching an optimal network structure in a predefined search space. There are usually three types of algorithms: reinforcement learning approach 
 \cite{liu2018progressive,pham2018efficient},
 Genetic Algorithm (GA) based \cite{liu2017hierarchical,real2019regularized}, 
 and other algorithms \cite{baker2017accelerating,brock2017smash}. 
 
 The models generated by these methods are normally constrained to smaller model sizes.
 Model size and operation quality are the two most common metrics to be optimized, over other metrics such as inference time or power consumption.
Representative works including  MONAS \cite{hsu2018monas}, DPP-Net \cite{dong2018dpp}, RENA \cite{zhou2018resource}, Pareto-NASH \cite{elsken2018multi} and MnasNet \cite{tan2019mnasnet} are interested in finding the best model architectures to meet these constraints. 
These approaches are more straightforward as they optimize directly over real world performance.
However, one drawback of NAS is the extensive computing power required for finding the optimal neural architectures. Thus, the already generated architectures can be utilized as guidance for future design for more efficient neural network architecture.
 
  
\myparagraph{Model Compression}
As modern state-of-art DL models can be very large, reducing the model computation cost is crucial for deploying the models on IoT devices, especially for those latency-sensitive real-time applications. 
Model compression methods can be divided into four categories, \textit{Parameter pruning and sharing}, \textit{Low-rank factorization}, \textit{Transferred/compact convolutional filters} and \textit{Knowledge distillation}. 
Hereby we briefly summaries the categories of model compression techniques and list several important works. 

\textit{Parameter pruning and sharing} method aims to find and remove the redundant parameters (of DL models) for higher efficiency and generalization. 
One direction is to apply quantization techniques. 
The DL's parameters/weights are usually stored in memory with 32-bits, and quantization techniques can compress them into 16-bits \cite{gupta2015deep}, 8-bits \cite{vanhoucke2011improving} or even 1-bit \cite{courbariaux2015binaryconnect,courbariaux2016binarized,rastegari2016xnor}.  
On the other hand, weight pruning and sharing (in pre-trained DL models) has also attracted interest among the community. 
Some popular methods imposed L1 or L2 regularization constraints \cite{lebedev2016fast,wen2016learning}, 
which can penalize models with more parameters, yielding the effect of pruning unnecessary parameters.


\textit{Low-rank factorization} method decomposes the CNN or DNN tensors to lower ranks. 
The tensor matrix decomposition is implemented for each layer of the DL model. 
That is, once the decomposition for a certain layer is completed, the parameter size will be fixed (for this layer) and the decomposition will proceed to the next layer. 
Interesting work can be found in \cite{lebedev2014speeding}. 
However, there are two major drawbacks. 
For very large DL models, it may be very expensive to perform decomposition owing to large parameter matrices. 
On the other hand, its layer-wise nature may also yield cumulative error, diverting the compression results to be far from optimal.

\textit{Transferred/compact convolutional filters} method reduces the memory consumption by implementing special structural convolutional filters. Motivated by the equivariant group theory \cite{cohen2016group}, the transferred convolutional filter transforms the network layers to a more compact structure, thus reducing the overall parameter space.
The family of transformation functions \cite{zhai2016doubly,shang2016understanding,li2016multi} 
operates in the spatial domain of the convolutional filters to compress the whole network parameters. 
Compared to other model compression methods, transferred convolutional filters methods are less stable due to the strong transfer assumptions. 
However, when the assumption holds, the compact convolutional filter can have very good performance.
In  \cite{szegedy2017inception,wu2017squeezedet}, 
the filters were decomposed from $3 \times 3$ or bigger to $1 \times 1$ convolutions---ideal operations for  IoT devices.   

\textit{Knowledge distillation} method learns a new, more compact model that mimics the function presented by the original complex DL model. 
The idea came from the work in \cite{bucilua2006model}, where a neural network was applied to mimic the behavior of a large classifier ensemble system. 
Later this idea has been extended to the complex DL methods \cite{hinton2015distilling}, 
more details can be found in \cite{romero2014fitnets,balan2015bayesian,zagoruyko2016paying}. 
However, currently the knowledge distillation methods are limited to classification tasks and further development is required.

\emph{Discussion.} 

Types of model compression techniques have their own strengths and weaknesses and thus optimal choice is based on specific user requirements.
\textit{Parameter pruning and sharing} methods are the most commonly applied techniques for compression models from original models. It is stable as with proper tuning, this approach usually delivers no or few accuracy losses.
On the other hand, \textit{Transferred/compact convolutional filters} methods address the compression from scratch. This end-to-end efficient design for improving the CNN performance approach shares similar insights to the efficient \textit{neural architecture design} approach as we discussed earlier. 
\textit{Knowledge distillation} methods are promising when working with relatively small datasets as the student model can benefit from the teacher model with less data.
All these methods are not mutually exclusive, we can make combinations based on specific use cases to optimize the models that are more suitable for low-resource IoT devices.

%% file: model_eva.tex
\subsection{Model Evaluation} \label{evaluation}

After the models have been trained, based on suitable metrics their performance should be evaluated  before deployment. 
Accuracy is one of the most popular evaluation metrics in classification tasks, yet it faces several problems in different scenarios.
For example, it is an overall measure without indicating the recognition capability for each class, which may be heavily biased if there exists a significant class imbalance problem. 
There are various evaluation metrics and it is key to select the most appropriate one. 
For the rest of this section we investigate several widely used metrics for classification and regression tasks. 
For classification/regression tasks, one aims to construct a model~(i.e. $f(\cdot)$) that can predict the value of dependent variable $Y$ from independent variable $X$. The difference between these two tasks is the fact that the dependent variable $Y$ is numerical for regression and categorical for classification. 

\subsubsection{Classification Problem based metric}\label{classificationeval}
In classification tasks, one of the most effective evaluation metrics is a confusion matrix  \cite{townsend1971theoretical}. 
As demonstrated in Table \ref{confusion} for a binary classification task, in a confusion matrix the row represents the predicted class and the column represents the ground truth (actual class).
The entries True Positive~($tp$) and True Negative~($tn$) represent the correctly classified positive and negative samples, while the entries False Negative~($fn$) and False Positive~($fp$) denote the misclassified positive and negative samples, respectively.  
\begin{table}[ht]
\footnotesize
\centering
\begin{tabular}{|c|c|c|}
\hline
						 & Actual Positive Class & Actual Negative Class \\ \hline
Predicted Positive Class & True Positive~($tp$)     & False Negative~($fn$)    \\ \hline
Predicted Negative Class & False Positive~($fp$)    & True Negative~($tn$)     \\ \hline
\end{tabular}
\caption{Confusion Matrix for Classification}
\label{confusion}

\vspace{-5mm}
\end{table}

Based on the confusion matrix, several evaluation metrics can be derived.
The \emph{accuracy} (i.e., $\frac{tp+tn}{tp+tn+fp+fn}$) and \emph{error rate} (i.e., $\frac{fp+fn}{tp+tn+fp+fn}$ or $1-accuracy$) are the most commonly used metrics because it is more understandable and intuitive for humans. 
However, these two metrics are powerless in terms of class-wise informativeness~\cite{mackay2003information}, which may neglect the minority class~\cite{chawla2004special} (if there is a class imbalance problem).

The metrics \emph{precision} and \emph{recall} can be used to measure the performance irrespective of the class imbalance problem (more definitions of classification evaluation metrics can be found in \textbf{Table 4 in  Appendix C}). 
In binary classification problems, as mentioned earlier 
$tp, fn, fp$ are defined as the the number of 
``positives correctly classified as positives'',  
``positives incorrectly classified as negatives'' ,
``negatives incorrectly classified as positives'' respectively. 
Then we can see 
\emph{recall} (i.e., $\frac{tp}{tp+fn}$) indicates the ability of a classifier to detect (true) positives out of all positive instances, while \emph{precision} (i.e., $\frac{tp}{tp+fp}$) is the percentage of detected (true) positives out of all the detected ones. 
Since the binary classification's decision may highly depend on the threshold, there is a trade-off between precision and recall.
For example, if a high threshold has been chosen---the similarity scores (the model outputs) have to be higher to give positive decisions---the classifier tends to have high $fn$ and low $fp$, yielding low recall and high precision.
Similarly, reducing the value of the threshold may increase recall and decrease precision accordingly. 
For different applications, one needs to consider the optimal threshold for their requirements.
For example, forensic applications may prioritize high precision (i.e., low in $fp$) while a medical diagnosis may prioritize high recall (i.e., low in $fn$). 

In some tasks when both recall and precision are important, the \textit{F1-score} (i.e., $2 \frac{ precision \cdot recall}{precision+recall}$)---a measure that can balance the precision/recall trade-off--- is normally used. 
It is worth noting that for multi-class cases, the multi-class confusion matrix can be calculated, and the aforementioned precision/recall/F1-score can be extended to measure the class-wise performance. 
Depending on the data/applications, the overall performance can be measured by aggregating all the class-wise metrics. 
Two popular aggregation operations are averaging, and weighted averaging, e.g., mean F1-score, or weighted F1-score (over all the class-wise F1-scores).


\subsubsection{Regression Problem based metric}\label{regreeval}
For regression problems, the evaluation metrics are different from the classification ones. 
Popular evaluation metrics include Mean Squared Error (MSE), Mean Absolute Error (MAE), Mean Percentage Error (MPE), etc. Details and formulas of these metrics can be found in  \textbf{Table 5 Appendix C}.

\emph{Mean Absolute Error (MAE)} and \emph{Mean Squared Error (MSE)} are the simplest metrics for regression evaluation.  
They denote the expected model errors defined in terms of absolute difference and squared difference (between the predicted value and the ground truth), respectively. 
Alternatively, the \emph{Mean Absolute Percentage Error (MAPE)} and \emph{Mean Percentage Error (MPE)} can also be applied to regression problems. 
The MAPE is similar to MAE but more intuitive as it shows percentage. 
The MPE lacks the absolute term on MAPE, which means the positive and negative errors will cancel out. 
In this case, the MPE can not be directly used to measure the performance of a model. 
However, it can be used to check whether the model systematically underestimates (more negative errors) or overestimates (more positive errors).

%
%
%

All of the above metrics can be applied to the regression problem, but it is important to consider the property of the dataset beforehand. 
For example, some fields may (or may not) be more prone to outliers, and the corresponding (effective) evaluation metrics  may be different.


%% file: deployment.tex
\section{Model deployment}
\label{sec:deployment}
When the ML model development process is finished, the developed models are to be deployed and composed as an application in the complex IoT environments. 
To simplify the deployment, the ML models and underlying infrastructure need to be specified (\S \ref{sec:declarative}). Next, the optimization techniques can be applied to generate the deployment plans that select the suitable ML models for the deployment, optimizs the resource utilization of the model deployment and improve the reusability of the deployed models (\S \ref{sec:optimization}). Once the deployment plans are generated, the models will be deployed over the specified infrastructure and the deployed models will be composed as defined in the plan (\S \ref{sec:compose}).

\subsection{Declarative Machine Learning and Deployment}\label{sec:declarative}

\myparagraph{Declarative ML} Declarative ML aims to use high-level language to specify ML tasks by separating the applications from the underlying data representation, model training and computing resources. There are \emph{three} general properties of declarative ML.  
 First, the high-level specification only considers data types of input, intermediate results and output. They are exposed as abstract data types without considering the physical representation of the data or how the data is processed by the underlying ML models. 
 Second, the ML tasks are specified as high-level operations through well-defined semantics. The basic operation primitives and their expected accuracy levels (or confidence interval) are defined accordingly. 
 Based on the operation semantics, declarative ML systems select the features and underlying ML models  automatically or semi-automatically, optimize the model performance and accuracy for varying data characteristics and runtime environments. Notably, the selection is based on the available models, provided as services.    
 Finally, the correctness of the ML models must be satisfied when a given model produces the equivalent results in any computing resources with the same input data and configurations. 
As a result, the declarative ML enables execution of the ML models over various hardware and computation platforms (such as Apache Spark) without any changes. Besides, these specification languages also bring flexibility and usability in the ML model deployment stage.
 
 SystemML \cite{boehm2016systemml} is an implementation of declarative ML on Apache Spark. Through domain-specific languages, it specifies the ML models as abstract data types and operations, independent of their implementation. The system is able to specify the majority of ML models:  matrix factorizations, dimension reduction, survival models for training and scoring, classification, descriptive statistics, clustering and regression. 
There is also other state-of-the-art research on declarative ML, including TUPAQ \cite{sparks2015automating} and Columbus \cite{zhang2016materialization}. They utilize language specification and modelling technologies to describe the ML models for automatic model and feature selection, performance and resource optimization, model and data reuse.  


\myparagraph{Declarative Deployment}
Hardware in the IoT environment consists of three basic types of device: \emph{data generating} devices, \emph{data processing} devices and \emph{data transferring} devices. 
\emph{Data generating} devices are also called ``Things'' (e.g., sensors, CCTV) and are used to collect environmental data. \emph{Data transferring} devices such as router, IoT gateway, base station are used to transfer the generated data to the \emph{data processing} devices. \emph{Data processing} devices are used to run the analytic jobs. They can be GPU, CPU and TPU servers running in cloud or ARM based edge device such as Raspberry Pi and Arduino. 
An ML-based IoT application is usually running across a fully distributed environment, such that it requires correct specification of the component devices as well as the precise interoperation between these devices. 
\cite{singh2017create} lists fundamental aspects that may simplify the hardware specification, i.e., processor, clock rate, general purpose input/output (GPIO), connectivity methods (Wi-Fi, Bluetooth, wired connection) and communication protocols (serial peripheral interface), universal asynchronous receiver-transmitter (UART). 

Regarding the software, it is often categorized into three groups based on operation levels: \emph{operating system (OS)}, \emph{programming language} and \emph{platform}. 
IoT \textit{OS} allows users to achieve the basic behavior of a computer within internet-connected devices. The choice of OS in different layers of the IoT environment depends on the hardware properties such as memory and CPU.
The \textit{programming language} helps the developers to build various applications in different working environments with diverse constraints. The choice depends on the capability of devices and the purpose of the application~\cite{cabe2018iot}. 
The IoT software \textit{platform} is a system that simplifies the development and deployment of the ML-based IoT application. 
It is an essential element of a huge IoT ecosystem which can be leveraged to connect new elements to the system. For more details of the most popular OSs, programming languages and platforms in IoT domain, one can refer to \textbf{Appendix A}. 
The ML development \textit{platforms} have been discussed in 
\S~\ref{mlframework}. 
  
The heterogeneity of IoT infrastructures makes the deployment very complicated and difficult to  automate. To overcome this issue, the infrastructure must be described and specified by machine understandable languages. Then, the declarative deployment systems are able to automatically map the ML models to the infrastructures and generate the deployment plans that optimize the performance and the accuracy. 

The declarative TOSCA model \cite{da2017internet} is able to specify the common infrastructures such as Raspberry Pis and cloud VM (hardware), MQTT and XMPP (communication protocol). The deployment logic can be defined through \emph{TOSCA Lifecycle Interface} that allows users to customize the deployment steps. However, this declarative  model is still very basic and can not handle complex deployments such as specifying the details of ML based application.  Moreover, the  IoT applications consist of installing devices and sensors which require \emph{human tasks}. These tasks are not natively supported by any available declarative deployment  \cite{breitenbucher2017declarative}.  
The imperative tool (e.g., kubectl commands) allows the technical experts with diverse knowledge of different deployment systems and APIs to interact with a deployment system and decide what actions should be taken. However, current imperative frameworks such as Juju, Kubernetes  still do not support  interactions such as sensor installation. 
In future,  declarative deployment systems should interact with declarative ML systems to deploy a complex application over the heterogeneity of IoT infrastructure while supporting the \emph{human} tasks through a more human centered imperative deployment model.

\subsection{Deployment Optimization}\label{sec:optimization}
When the infrastructures and deployment workflow of the ML models are specified, the deployment optimization problem can be formed as a mathematical expression subject to a set of system constraints. Then, resource allocation algorithms can be used to efficiently and precisely find the best solution for the given mathematical expressions. 
Moreover, the optimization objectives are a set of QoS parameters including storage and memory space, budget, task execution time and communication delay etc,. These algorithms can be divided into \emph{two} classes based on whether an optimal solution can be guaranteed: \emph{meta-heuristic method}  and \emph{iterative method} (or \emph{mathematical optimization}). Nowadays, ML methods are becoming popular and being applied to solve these resource allocation problems by learning ``good'' solutions from the data. We investigate the representative works in resource allocation based on these \emph{three} classes.

\myparagraph{Iterative-based method} This class of algorithm generates a sequence of improved approximate solutions where each solution is driven by previous solutions. Eventually, the solutions will converge to an optimal point proved by a rigorous mathematical analysis. The heuristic-based iterative methods are also very common, but we categorize this type of algorithms into the \emph {meta-heuristic based method}.
The most popular algorithms of this class include newton's method \cite{madan2010fast, madan2010cell}, gradient method \cite{beck20141} and ellipsoid method \cite{lu2010utility}. To apply and adapt iterative-based algorithms to optimize resource allocation requires strong mathematical background, which can be an obstruction for software developers to utilize these algorithms to optimize their deployment. 
Furthermore, the algorithms have the variety of performance for different problems in terms of efficiency and accuracy. 
As a result, more algorithms from iterative-based methods need to be studied and simplified by the system researchers, providing toolkits (or solvers) to tackle different optimization problems in IoT application deployment. 

\myparagraph{Meta-heuristic based method}
The optimization problems in IoT applications can have large search spaces or be time-sensitive. The \emph{meta-heuristic} based method is faster than \emph{iterative-based method} in finding a near-optimal solution. This type of method consists of two subclasses:  \emph{trajectory}-based method and  \emph{population}-based method. The \emph{trajectory}-based method finds a suitable solution with a trajectory defined in the search space. First, the resource allocation problems are mapped into a set of search problems such as variable neighborhood search, iterated local search, simulated annealing and tabu search. Then, the \emph{meta-heuristic} algorithms are used to find the solutions. Many survey papers~\cite{singh2016survey, hansen2010variable, liu2014survey} have reviewed the algorithms applied for resource allocation in IoT, cloud computing, mobile computing. 
Additionally,  \emph{population}-based methods aim to find a suitable solution in the search space that is described as the evolution of a population of solutions. This method is also called evolutionary computation and the most well-known algorithm is the genetic algorithm.  \cite{zhan2015cloud} investigates the resource allocation problems solved by evolutionary approaches in cloud computing.

\myparagraph{Machine learning based method}
ML based method is inspired by the ability of data to represent the performance and utilization of the contemporary systems. The ML based methods are used to build data-driven models that allows the target systems to learn and generate an optimized deployment plan. The proposed algorithms have been used to optimize various QoS parameters such as latency \cite{yadwadkar2016multi, mao2017neural}, resource utilization \cite{mirhoseini2017device}, energy consumption \cite{berral2010towards} and many others. Zhang et al. \cite{zhang2019deep} have given a comprehensive survey of the ML based methods used for resource allocation in mobile and wireless networking. 

Deployment (or resource allocation) optimization problems have been studied for decades, and remain a huge legacy for overcoming the optimization problems in deploying ML-based IoT applications. Instead of developing new optimization algorithms, more efforts are required to model the complex optimization problems, in which the system scale, conditions and diversity have been amplified significantly.

\subsection{Action and Model Composition} \label{sec:compose}
Deployment of ML models in a pipeline requires proper model composition to maximize the user QoS. As shown in \S \ref{sec:smartcity}, a smart car navigator system comprises multiple ML models, including speech recognition, text classification, text generation and text-to-speech (TTS) model.

\textit{Action composition} is defined by composing a set of basic actions for complex decisions. In a self-driving car operating system, actions can be accelerating, braking, turning left and right, etc. The combination of various action spaces increases the difficulties of learning optimal decisions in such complex systems.  Hierarchical abstract machines (HAM) \cite{singh1992reinforcement} are well studied in the context of reinforcement learning \cite{parr1998reinforcement, thrun1995finding} by allowing agents to select from a constrained list of action spaces, speeding up the learning and adaptation to the new environment. 

 Model composition aims to create a ML-based IoT application by using reusable, portable, self-contained modules via inserting new components or removing  existing components.
 Apache Airflow \footnote{https://airflow.apache.org/} is an open-source platform for creating, scheduling and monitoring workflows in Python. The Valohai\footnote{https://valohai.com/} operator is an extension of Airflow that utilizes the docker container to build self-contained modules for each model while providing the flexibility for users to define the steps to execute. \cite{kim2018nsml} reported the following challenges for chaining the ML models: 
 \begin{itemize}
 	\item How to allocate computing resources automatically for different models. An application is chained by various ML models require different computing resources across heterogeneous infrastructures. It is challenging to provision the computing resources efficiently for the chained ML models while meeting their performance requirements.
 	\item How to chain the dependent models. Each individual ML model has its own specification and data format of the inputs and outputs. The challenge is to design a data messaging system to orchestrate the data flow across different models while considering their required specification and data format. 

 	\item How to meet the security requirements. 
 	The inferences are performed through various components, with each deployed across different computing resources. This introduces a set of challenges including privacy, verification of outputs of each model, changing the security policies of components, etc.
 	\item How to monitor  failure. The composed application consists of a set of ML models that needs to be monitored, ensuring that everything is streamlined and executed as anticipated.
 \end{itemize}

Apart from challenges mentioned above, the literature discusses the techniques to improve the performance of individual models via system configuration, including \emph{model batch size}, \emph{model replica} and \emph{system buffering}.

\textit{Per-Model Batch Size.} Batching the received user queries optimizes throughput by fully utilizing the features of the pre-trained models, which is faster than processing one query at a time. However, batching query can potentially increase latency because the the model will wait for a whole batch of queries to come before it starts to proceed. The first query is not returned until the final query is processed \cite{crankshaw2018inferline}. The choice of the per-model batch is challenging due to the sequential composition between the models. 

\textit{Model Replica.}
In heavy or bursting loads, a system must quickly respond to the query fluctuations to meet the latency requirements. To alleviate the system congestion and achieve high throughput, it is critical to identify the bottleneck, which can be challenging due to the system dynamics. 
The bottleneck models can be resolved by replicating the model instances across multiple devices \cite{crankshaw2017clipper}, therefore balancing the workload. However, distributing the queries across more model replica in a parallel setting  \cite{crankshaw2017clipper} is also challenging since the optimal placement depends on the model performance and the device capacity.

\textit{System Buffering.}
Serving system as a stream processing system comprises components across multiple devices. These devices usually process at different speeds, making system buffering across nodes necessary. Message queues are usually implemented to ensure smooth running within the system. However, buffering mechanism would increase the latency based on various system configurations \cite{crankshaw2017clipper}. It is thus challenging to design proper strategies to balance the message queue overhead and the system latency.

%% file: action.tex
\section{Model audit}
\label{sec:audit}


Audit aims to evaluate whether the application is operating effectively, safely and reliably with the collected evidence.
To this end, we must know what we should audit. Most work focuses on monitoring or debugging the issues caused by infrastructure failures \cite{krunic2007nodemd}, implementation bugs \cite{kothari2008deriving, dong2013d2} and deployment errors \cite{sultana2014kinesis}.
In this section, we investigate the security, reliability and performance issues caused by ML models, especially DL models.

\subsection{Security}\label{security}
There are many surveys regarding IoT security issues and challenges.
The security of IoT standardized communication protocols were evaluated in \cite{granjal2010secure} based on their proposed model.
\cite{sicari2015security} categorized the security issues of IoT into \emph{eight} domains including authentication, access control, confidentiality, privacy, trust, secure middleware, mobile security and policy enforcement.
\cite{roman2013features} studied the main challenges and solutions of designing and deploying security mechanisms in centralized and distributed IoT architectures. \cite{lin2017survey} discussed the security features of IoT and categorized the attacks into \emph{four} layers. i.e., the perception layer, the network layer, and the application layer.

In this subsection, we discuss security issues for deep learning based IoT applications: \emph{Model exploratory attack}, \emph{Data poisoning attacks} and \emph{Evasion attacks}.
\emph{Model exploratory attacks} do not happen during training, instead the attacker tries to discover information from the trained model including the model itself and training data.  
\emph{Data poisoning attacks} happen during the training phase, where the attacker attempts to shift the boundary of DL models in their favor by polluting the training data.
Finally, \emph{evasion attacks} maliciously craft the inputs for the deep learning based IoT application to trigger abnormal model behavior. Interestingly, the development of the research on adversarial learning has started an arms race between adversaries and defenders.

The following subsection summarizes the most popular attacks and defenses of these attacks.
We also propose research directions for development of robust IoT applications.


\subsubsection{Model exploratory attack}\label{modelexattack}
 This type of attack is usually performed on open-source frameworks such as PredictionIO and cloud-based machine learning service.
This ML-as-a-service may allow users to input partial feature vectors while still being able to receive confidence values in addition to prediction results. Thus, the attacker can leverage this feature to either extract the model or the sensitive information underlying the model.
\emph{Model stealing} and \emph{Membership leakage} are \emph{two} main types of model exploratory attack.
\emph{Model stealing} attack aims to duplicate the functionality of the model that allows the attacker to evade detection by the stolen model \cite{ateniese2013hacking, nelson2012query}. \cite{tramer2016stealing} proposed a method that learns the target models via a prediction API. Evaluations show that this method successfully extracts the models including logistic regression, SVM, neural network and decision tree from BigML and Amazon Web Services. More attack methods can be created based on the extensive literature on learning theory, e.g., PAC learning \cite{valiant1984theory} and its variants \cite{benedek1991learnability}.
\emph{Membership leakage} attacks are interested in stealing the information from the training data which may not be publicly available and may contain some sensitive information such as trade secrets, medical records etc. In this type of attack, an attacker is able to infer the members of the population or the members of the training dataset.
Attacking the members of the population means that the types of data are used to create the model. Therefore, the target model has not been generalized for the adversary, because he/she has the sample of the entire population of the training dataset. The attacks were successfully performed in the dataset including voice, handwritten images, network traffic, online shopping, record of hospital stays etc \cite{shokri2017membership, ateniese2013hacking}.
 The members of the training dataset attack aim to identify the individuals whose data are used for training the model, which causes a serious privacy issue. For example, if an attacker knows that a patient's medical record was used to train a disease detection model, it also reveals that the patient has this disease.  The experiments in \cite{tramer2016stealing} show that the attacks are able to extract the training dataset when the model is based on kernel logistic regression.

\myparagraph{Defending model exploratory attack} The most straightforward defense of this type of attack is to constrain the API, not returning confidences and not responding to probing queries. The minimization can be achieved in three approaches: \emph{Rounding confidence}, \emph{Differential privacy} and \emph{Ensemble methods}

\emph{Rounding confidence} is a type of defense that rounds confidence sources of an application to some fixed precisions \cite{fredrikson2015model}. Notably, some online ML service providers are already working on it. For instance, BigML and Amazon provide five decimal places and 16 significant digits for their confidence scores respectively when answering queries. Limiting the precision can decrease the success rate of attacks, this is because the outputs of the model are approximated. In equation-solving attacks, for example, if the output of an equation-system is rounded, it will increase the difficulty for attackers to guess the target function.

\emph{Differential privacy} is a class of mechanisms to protect, especially the privacy of training data \cite{vinterbo2012differentially}. A set of \emph{Differential privacy} methods has been applied to protect regressions \cite{chaudhuri2009privacy, zhang2012functional}, SVMs \cite{rubinstein2009learning}, decision trees \cite{jagannathan2009practical} and neural networks \cite{shokri2015privacy}.  The main idea of \emph{Differential privacy} is to avoid a query that allows an adversary to distinguish closely neighboring model parameters.

\emph{Ensemble methods} returns an aggregated output predicted by a set of models. This prevention method was mentioned in \cite{tramer2016stealing} and may have more resilience against the model exploratory attacks, compared with other methods. 
Bonawitz \cite{bonawitz2017practical} designed a new protocol to compute the sum of a large subset of models supporting a secure federated learning setting.

\subsubsection{Data poisoning attack}\label{sec:data_poison}

Unlike \emph{model exploratory attack}, an adversary performs the attacks during the model training phase. These attacks insert carefully constructed poison instances into the training dataset to manipulate the performance of a system.
We introduce types of data poisoning attacks both in traditional machine learning and deep learning.

\emph{Data poisoning attack in machine learning.}
\textit{(1) supervised learning.} A causative attack was proposed by Xiao et al. against SVMs which utilizes \emph{label flipping} to poison the training data \cite{xiao2012adversarial}.  A \emph{label flipping} attack attempts to add a noise label to the training data. These flipping labels are able to cause some malicious samples to be labeled as legal, or make legal samples appear to be malicious. To improve the efficiency of the attack, Biggio and Laskov \cite{biggio2012poisoning} utilized the gradient descent algorithm to find the best attack points to flip the labels.

\textit{(2) unsupervised learning.} The poisoning attack has been demonstrated against various clustering algorithms. The idea is to introduce carefully crafted data points to the training dataset to cause clusters to merge. In \cite{biggio2014poisoning, rieck2011automatic}, the authors assumed that the attacker has full knowledge of the clustering algorithm and then reduced the attack to an optimization problem. The evaluations show that the clustering algorithms are compromised  significantly with a very small percentage of poisoned input data.

\emph{Data poisoning attack in deep learning.} There are very few data poisoning attacks in neural networks. \cite{steinhardt2017certified} showed that a deep learning model lost 11\% accuracy after modifying 3\% training data. Moreover, if the attacks are focus on attacking the specific test instances, the successful rate, time consumption and required resources (the number of modified samples) can be reduced significantly \cite{gu2017badnets, chen2017targeted, steinhardt2017certified}.  \cite{shafahi2018poison, suciu2018does} targeted real-world scenarios where the labels are examined by human reviewers and malware detectors. The authors aimed to overfit the deep learning models by poisoning the training data. Thus, the target instants (trained models) would not perform well during inference time.

\myparagraph{Defending data poisoning attack} Compared with other systems, defending data poisoning attacks is more critical in machine learning systems, because training data coming from the outside world is very easy to poison. Steinhardt et al. \cite{steinhardt2017certified} developed an outlier detector for linear classifiers to generate approximate upper bounds of the training dataset. The data (poisoned) that exceeds the bounds will be removed from the training dataset, not changing the distribution of the clean data. The paper \cite{rubinstein2009antidote} aimed to develop a strategy to defend against attacks that attempts to poison the PCA based anomaly detection models. 
Since the PCA methods are less sensitive to outliers, an ideal approach against the variance injection caused by the perturbed data is to filter the poisoned data with predefined threshold. The paper also proposed two approaches \cite{rubinstein2009antidote} for selecting the threshold: the first uses covariance matrix, and the second is to find the maximal scale estimate of the data projection.

\subsubsection{Evasion attack}\label{evasion}
 With the explosive development of machine learning, evasion attacks are becoming the most prevalent type of attack in machine learning, attracting people's attention from both academia and industry. Fig. \ref{fig:attacker-defender} shows the arms race between the attacker and defender. It shows that the attacker attempts to confuse the defender with a crafted adversarial example, while the defender aims to strengthen its ability to filter out illegitimate input.

During both training and inference, the attacker can generate \emph{adversarial examples} by modifying the samples. The training phase modification is similar to \emph{data poisoning attacks} in that the decision boundary of the defender classifier is modified by insertion, modification or deletion of the training dataset. There are two approaches for generating adversarial samples, \emph{white-box} or \emph{black-box}.
In the \emph{white-box} setup, the adversarial samples are crafted based on the attacker who has access to both the training data and the targeted model.  Therefore, an adversary is able to obtain the boundaries of the targeted model by carefully modifying the training data. To be more explicit, as shown in Fig. \ref{fig:attacker-defender}, the defender aims to stop using the illegitimate input $X$ to train the itself. In order to fool the defender, the attacker attempts to learn the boundaries of the defender by adding the perturbations to $X$ and then performing the attack. This process is repeated until the adversarial samples break the boundaries. The most representative techniques \cite{papernot2016limitations,goodfellow2014explaining,kurakin2016adversarial} are based on the attacker who has knowledge of both target model and instance of data. In the \emph{black-box} setup, the attack introduced in \cite{papernot2017practical} is not aware of the training data and the targeted model. The only observation of the targeted model is the inputs and their labels given by the targeted model. Based on this, a local model is trained to replace the target DNN. 
84\% of the adversarial examples generated by the local substitute model are misclassified by the targeted DNN.

\input{figuresTex/fig-attacker-defender.tex}

\myparagraph{Defending evasion attack}
Crafting the adversarial samples is a complex optimization process as it is very hard to build a general tool for defending. Thus, \emph{adversarial training} which adds the adversarial examples to the training set is a straightforward way to increase the model robustness \cite{lyu2015unified,shaham2018understanding,tramer2017ensemble}.
Moreover, Defense-GAN \cite{samangouei2018defense} leverages a generative model to generate more samples similar to the training data, reducing the adversarial perturbation significantly. Apart from these general defending methods, some defenses use model hardening techniques.  \cite{xu2017feature} proposed a feature squeezing method to reduce the complexity of representing the data. Less important adversarial perturbations are filtered out afterwards. In the following section, we survey the attacks and the defenses in real IoT applications and then discuss the challenges of building a secure IoT application.

\subsubsection{The system challenges of building a secure ML-based IoT application}\label{syschallenges}
Most of the attacks and defenses reviewed in previous sections focus on developing the algorithms for \emph{functional tasks} such as computer version, natural language processing, audio speech processing etc. To build a secure IoT application, we must consider the security issues from the system perspective, as the \emph{functional tasks} cannot perform well when the system is under attack.
We discuss two system challenges to improve the security of ML-based IoT application.

\myparagraph{Developing new attacking and defending models}
ML has been widely used in IoT system developments including  network engineering  \cite{li2019learning, mao2017neural}, resource allocation \cite{li2018deep, xu2017deep}, system debugging \cite{sikder20176thsense,celik2018sensitive}, network intrusion
detection \cite{kolias2017termid, wang2010new, lopez2017conditional} and network operations\cite{hu2010qelar}. These systems can be exposed to the aforementioned attacks as well. The literature \cite{erpek2018deep,achleitner2017adversarial,han2019adversarial} has revealed successful examples of attacks and defenses. Further to the efforts on \emph{functional tasks}, more research and development (R\&D) work is required to improve the security of ML-based IoT applications.

\myparagraph{Developing new security platforms/frameworks}
 We have discussed the arms race game between the attacker and defender (see Fig.~\ref{fig:attacker-defender}). This can be utilized to ensure the resilience of IoT applications to various attacks. At a high level, an ideal platform would be able to launch various attacks via a predefined deployment pipeline to attack the experimental group. Meanwhile, the attack behaviors and system performance will be monitored to reinforce the capacity of the defender. To this end, three research questions need to be answered.
 1) \emph{How to automate the attacks.}  Unlike the traditional software deployment problem, deploying attacks is much more complicated. For example in an evasion attack, the proposed platform must be able to use various ML models to craft the adversarial examples. It is very difficult to automate this process. Due to the difference between the model inputs and outputs, the models may need to be retrained based on the observation of the real world to generate better adversarial examples.
 2) \emph{How to monitor the attacks.} As discussed in previous sections, attacks can happen in data collection, model training and model inference. Therefore, the traditional log system is not able to handle this complexity. In a data poisoning attack, for instance, the traditional log system is unable to capture the impact caused by fake data points injection into the system, thus the training of a defender is unfeasible.
 3) \emph{How to coordinate the attacker with the defender.} At the high level, the arms race game between the attacker and the defender is very logical. The challenge here is to continuously select the suitable attacks and thereby improve the defender's performance. This can be formalized as an optimization problem where one of the objective is to maximize the ability of a system in defending against certain types of attack.

\subsection{Fault Tolerance}
\label{sec:fault}

Distributed system fault tolerance has been studied for decades, many representative works have been proposed to handle the failures including system architecture \cite{randell1975system} and  algorithm design \cite{castro1999practical}. In the IoT environment, the probability of failure increases significantly, and many faults are very hard to detect. Our previous papers \cite{wen2017fog, garraghan2018emergent} reviewed the state-of-the-art research, and then discussed the key research directions. In this subsection, we will introduce some of the most common faults in ML applications.



%


\subsubsection{Faults in ML} \label{sec:generalization}
Generalisation is crucial for ML models, which measure the prediction capacity on unseen test data \cite{bousquet2003introduction}.
Generally, ML training can be regarded as an optimisation process. 
For example, the model can be trained by minimising a certain loss function.  
However, overfitting may occur when ML models are trained on less representative, noisy or small data, and in this case, trivial error patterns may be learned, causing lack of generalisation (i.e., faults) at the test stage.
There are many ways to reduce the overfitting effect, such as regularization (e.g., with regularization terms such as L1/L2 norm),  \textit{Stochastic gradient descent (SGD)}~\cite{bottou2010large} or dropout \cite{srivastava2014dropout} (for DL models), early stopping \cite{bengio2012practical} (stop training when validation error starts to increase), etc.

 Data imbalance, on the other hand, is very common in real-world scenarios and it may also cause overfitting. 
The model may mainly learn patterns from the majority classes while it may easily ignore the contributions from the minority classes (with limited training samples), yielding severe faults at the inference stage.
Various approaches have been proposed for mitigation including data augmentation (e.g.,\cite{wang2017effectiveness}), data upsampling (e.g., GAN-based data generation \cite{tang2019expression}, \cite{gen2018}), cost-sensitive learning (which will impose a larger penalty on training errors with minority classes),  transfer learning, etc.

In addition to overfitting effect, faults can also be attributed to the optimisation process. 
For example,  with very deep DL models or with RNN, gradient vanishing/explosion may occur during the optimisation process, causing representation learning to be challenging or even infeasible.
There are also several approaches to address this issue, e.g., \textit{Batch Normalization~(BN)}\cite{ioffe2015batch} (through normalizing the gradients in each layer), residual connection structure in DL (to preserve the gradient across many layers). 
For federated learning or distributed learning, RSA(i.e., Byzantine-Robust Stochastic Aggregation)\cite{li2019rsa} has also beenproposed to prevent the incorrect gradient aggregation.

\subsubsection{Fault tolerance in neural networks}
At a high level of abstraction, the neural network can be viewed as a distributed system.  Therefore, the failure can happen in \emph{neuron} or \emph{synapse}. In \cite{7967193}, Mhanmdi and Guerraoui proposed a general model to describe the fault model of neural networks. The \emph{neuron} may stop computing (Crash) or generate some abnormal outputs (Byzantine). Similarly, the failures of \emph{synapse} can be abstracted as Crash and Byzantine. Crash represents that the transmission has not succeeded, and Byzantine is that the incorrect messages are sent from the source \emph{neuron} to the destination \emph{neuron}.
Thus, we assume that a given neural network $\mathcal{N}$ performs an expected output $F_{\mathcal{N}}(X)$, and $F_{\mathcal{N}_{fault}}(X)$ is the output of the faulty network obtained from $\mathcal{N}$.
The distance $\epsilon$ between $F_{\mathcal{N}}(X)$ and $F_{\mathcal{N}_{fault}}(X)$ represents the fault tolerance of $\mathcal{N}$, when there are at most $n$ faulty components (including \emph{neuron} and \emph{synapse}):
\begin{equation}
\label{eq1}
	\parallel F_{\mathcal{N}}(X)-F_{\mathcal{N}_{fault}}(X) \parallel \leq \epsilon
\end{equation}


where $X$ is the training dataset, applied to both $\mathcal{N}$ and $\mathcal{N}_{fault}$. In order to guarantee the robustness of the neural model, the designer needs to ensure that the error (left hand in Equation~\ref{eq1}) is below a predefined threshold (right hand in Equation~\ref{eq1}). The threshold depends on the performance of the network and its intended application \cite{protzel1993performance, 7967193}.

Like traditional fault tolerance in distributed system, the fault tolerance in neural networks also has two types: \emph{Passive} and \emph{Active}.

In \textbf{passive} fault tolerance, no diagnostics, relearning, or reconfiguration is required thereby avoiding fault detection and location.
The most common \emph{passive} fault tolerance approach which is also one of the important features of neural network is inserting redundancy. Such methods learn a small network from the given input/output, and then add the replicated hidden neurons to share the load of the critical nodes, after the model has been trained.
Representative works \cite{chu1990fault, emmerson1993determining, chiu1993robustness} addressed the fault tolerance by adding extra links or nodes to the well trained neural network. The authors in \cite{chin1994training} proposed a solution that adds artificial faults to the network during the training time. Therefore, the network can tolerate the specific faults.
However, this approach requires that the neural network designers are aware of all the faulty scenarios while building the network. Also, adding redundancies makes the models very complex and huge, which brings the challenges of deploying them over lightweight and low-power IoT devices.

\textbf{Active} fault tolerance aims to recover the neural model from faults by resetting the neural network into a fault-free state.  However, it does not attract too much attention from research, a common strategy is to utilize high-performance computation resources to re-compute the lost work when the hardware fails \cite{abadi2016tensorflow, wei2015managed}. Notably, Qiao et al. proposed a checkpoint-based fault tolerance for deep learning in \cite{qiao2018fault}. This new method partially recovers the model from the checkpoints based on the priority of the checkpoints thereby significantly reducing the cost of recomputing.

\subsection{Performance Evaluation}\label{performanceeva}
In this section, we consider several performance criteria that need to be considered for evaluating the efficiency of the obtained ML models. The criteria is identified as \emph{two} main dimensions: \textit{model precision} and \textit{execution latency}.

\myparagraph{Model Precision}
In a typical IoT application, the software performance is assumed stable after deployment. However, this is not the case for ML application where precision degradation is always expected after deployment. Precision degradation can happen as various unexpected external changes  lead to shift in data distribution. Device location change, time and the weather are all important factors that may decrease the model performance. Therefore, it is critical that the model performance is monitored and new data is introduced continuously for retraining of the model. In ML, we define lifelong learning \cite{parisi2019continual} as continually acquiring data and extracting new information without catastrophic forgetting of past knowledge. Lifelong learning keeps the model precision at a steady level.

\myparagraph{Execution Latency}
Many IoT applications are latency-sensitive depending on their tasks. For example, in the aforementioned smart transportation system (see \S \ref{sec:smartcity}) where sensors monitor and detect car accidents, instant decisions have to be made to warn the drivers of potential hazards. Various factors, listed below, have to be evaluated to ensure seamless communication among the distributed components of a smart IoT application.

\textbf{Bandwidth Usage. }
In distributed IoT networks, large scale IoT sensors are generating a huge amount of data all the time. It is not possible to send all the data to the cloud for data analysis. Fog computing proposed to move the computing close to the sensors to reduce the data transmission over the IoT network.
However, the bandwidth of sensor network and edge network are still limited, some nodes may experiences high latency due to the network congestion. This may cause huge latency for the whole system as well.
We need to monitor and evaluate this network dynamic \cite{mao2016resource} in order to provide solutions to alleviate the congestion in the networks.

\textbf{Resource Consumption. }
Hardware in IoT applications varies in computing power, memory and storage capacity. For any resource-intensive tasks, for example those computation-heavy or memory-heavy ones, resource exhaustion in one node may lead to unacceptable latency for the whole application.
It is thus necessary to design efficient resource management systems \cite{yigitoglu2017foggy, moritz2018ray} to monitor and optimize task allocation for these physical devices,

\textbf{System Throughput. }
The ML-based IoT applications may be developed to serve millions of people, for example, the \emph{smart traffic routing application} mentioned in \ref{sec:smartcity}.
This massive number of users may send the requests simultaneously. Responding to these requests quickly without losing user satisfaction is still an unsolved problem in cloud computing. However, this issue is amplified in ML-based IoT applications, in which the queries may be performed on various devices and models. Some database optimization techniques such as caching frequent queries,  batching queries and approximate computing are applied \cite{crankshaw2017clipper, peng2018axnet}. There are remaining gaps in optimizing the query plans by considering heterogeneousness of the computing resources, uncertainty of the network, and diversity of ML models.

%

%% file: figuresTex/fig-attacker-defender.tex
\begin{figure}[ht]
    \centering
    \includegraphics[width=3.8in]{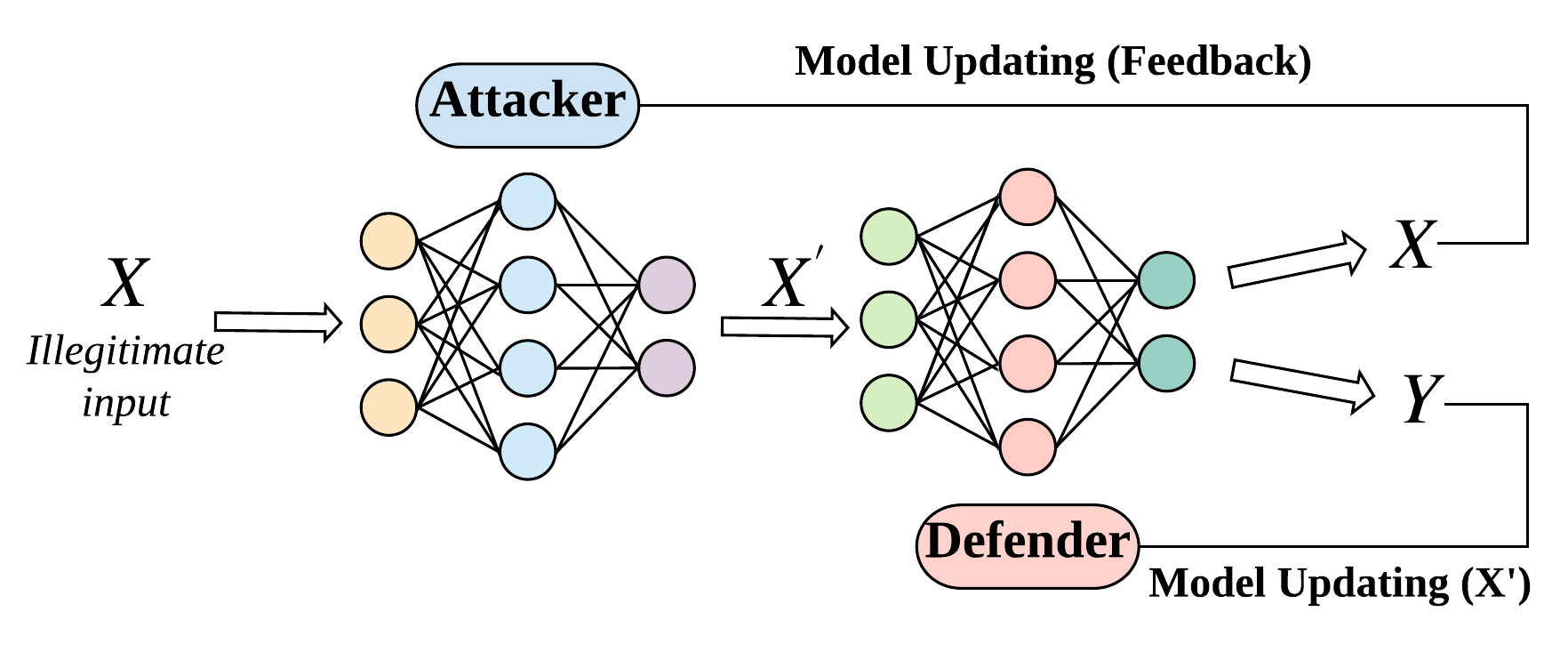}
    \vspace{-5mm}
     \caption{Arms race game between the attacker and the defender. The attacker takes an illegitimate example ($X$) as the input of his/her neural network and generates an adversarial example ($X'$). This example attempts to fool the defender which is a classification neural network. If the classifier recognizes the adversarial example as $Y$ which belongs to the legitimate input, it means the attacker wins the game. On the contrary, the defender wins the game if the adversarial example is classified as $X$. When the attacker fails, it will try to update its model to generate a stronger adversary example based on the feedback. Similarly, the defender will enhance its model based on the lesson learned from the successful attack.}
    \label{fig:attacker-defender}
    \vspace{-5mm}
    \end{figure}

%% file: collection.tex
\section{Data Acquisition}
\label{sec:collection}
Apart from ML models' actions and operations, acquisition of data plays a crucial role in Ml-based IoT applications. Data is one of the most important constituents in developing a ML model as the prediction accuracy of the model has a strong positive correlation with the quality of the input data \cite{mitchell1999machine}.
To provide high quality data for a ML-based IoT application, we orchestrate the data acquisition process into several steps.
In the whole data acquisition pipeline, we collect raw data from various data sources (\S \ref{datacollection}). 
With proper preprocessing techniques (\S \ref{datapreproc}) to remove redundant information and annotate the data, we are capable of performing several different ML tasks. 
While we have more data sources during the development process, we can also fuse (\S \ref{datafusion}) them to provide more consistent and useful information. 
The following subsections will focus on the mentioned steps and discuss how data acquisition can support  development of a robust ML-based IoT application.

\subsection{Data Collection}\label{datacollection}
The IoT data can be broadly categorized into \emph{Structured data} and \emph{Unstructured data} based on its representation.
\emph{Structured data} can be represented in a pre-defined format (rows and columns). The meaning of each field is explicit which eases the analysis and storage of the data. Examples of structured data include employee register information, visiting logs, etc. On the other hand, \emph{Unstructured data} lacks any specific structure or format. Varying from text, audio, video to mails and messages, it accounts for a large proportion of IoT data. These two types of data are generated in \emph{three} formats: \emph{signal  data}, \emph{log data} and \emph{packet data}. The \emph{signal data} collects the daily life signal through various hardware such as sensors, sound recorders, cctv cameras, etc. The \emph{log data} is usually used to capture the system status.  Finally, the \emph{packet data} is the data sent over the network and each unit transmitted consists of a header and the actual data.
To collect these data, \emph{three} important factors need to be considered: 1) \emph{Data exchange},  2) \emph{Resource consumption} and 3) \emph{Concept Drift}.

\myparagraph{Data exchange} 
The data is generated from an IoT device and sent to an edge (sink) node or 
other IoT devices and eventually the data will be collected and stored in the cloud. 
The computation power of gateways and edge nodes is improving, which brings an opportunity to remove data redundancy while saving the energy and bandwidth required for transferring data to downstream nodes \cite{wen2018approxiot}. This aggregation requires application of various data summarization  techniques \cite{cormode2011synopses} including sampling, sketching, histograms, wavelets and adaptation of these techniques to meet the constraints of the hardware and the time-varying channel conditions.  Henriette et al. \cite{roger2019comprehensive} investigated the state-of-the-art stream processing systems that can be used to implement these data summarization techniques and execute them in a parallel and elastic manner.  However, it still requires a lot of effort to develop new data summarization techniques and stream processing systems to handle the difficulty of processing high volumes data from various sources with multi-modality. 
%

\myparagraph{Resource consumption}
As mentioned earlier, IoT devices are very limited by resources such as processing capability, storage capacity, 
wireless bandwidth and battery power. Thus, it is very critical to optimize the resource utilization while processing, storing or transferring data to the edge device or cloud. To this end, we need to consider \emph{three} issues: \emph{resource allocation}, \emph{energy control} and \emph{task allocation}. 
\emph{Resource allocation} in the context of data collection is to assign computing, storage or bandwidth resources to the data generated by IoT devices before transferring to edge or cloud. Sending streaming data drains the battery at a faster rate while limited storage capacity does not enable large data storage.  
\emph{Energy control} focuses on optimizing the energy consumption when the IoT data is processed and transferred over the devices. \emph{Task allocation} aims to balance the  resources consumption in IoT devices while minimizing the overall latency. 
These \emph{three} factors are sometimes considered together and most of the available algorithms are based on market-enabled pricing schemes, which dynamically exchange the resources among the devices in IoT infrastructure by creating an artificial market \cite{jha2019multi,eswaran2012utility}. 
In ML-based IoT application, the ML models should be considered as the special tasks that are running on extremely heterogeneous computing resources in a distributed manner, and these tasks are usually  compute-intensive, data-intensive and network-intensive. As a result, it is crucial to develop new market models to describe these special resource consumption problems and new algorithms to solve the problems.

\myparagraph{Concept Drift}   
 Due to the dynamicity of the IoT environment, data distribution becomes very uncertain and changes frequently over time leading to \emph{concept drift} \cite{gama2014}. Changes can occur abruptly or gradually correlated with the occurrence of other events. Additionally, the change can be in different forms i.e., input data characteristics or relation between input data and target variables with single or multiple occurrence (constant or variable recurrence). For the successful execution of IoT applications, these drifts need to be predicted, distinguished from noise and handled properly. 
Numerous algorithms are proposed for managing concept drift. \cite{de2019overview,hu2019no} review the generic algorithms to handle the concept drift. There are two main detection methods, \emph{performance-based} and \emph{data distribution-based}. The former can work well if the data is labeled which 
may not be possible for all cases while the latter is able to detect only a subset of available drifts. Since IoT-based ML application data are not always labeled and high accuracy is desired, it is essential to develop new algorithms which are able to detect and manage the concept drift.

\subsection{Data Preprocessing}\label{datapreproc}
The real-world data collected from heterogeneous IoT devices usually contains outliers or is incomplete in nature, which makes it difficult to feed it into ML models directly.
Data preprocessing deals with these anomalies and improves the data quality and practicality. There are several things that need to be considered, namely \emph{data cleaning}, \emph{data annotation} and \emph{feature engineering}. We have discussed the details of \emph{feature engineering} in \S \ref{fs}, will not consider it in this section. 
	
\subsubsection{Data cleaning}\label{dataclean}
	Much data contains noise that is bound to confuse the ML models and reduce the accuracy of the prediction results. Data cleaning resolves this problem by completing several routine tasks such as \emph{filling missing values}, \emph{smoothing noise data} and \emph{removing outliers} \cite{alasadi2017review,blazquez2020review}. Empty records in the data set can be replaced manually by a specific value, for example the attribute mean or the most common attribute in the set. It can also be marked with "unknown" or just ignored if the dataset is large enough. Noisy data, though, can be smoothed by grouping first and then averaging over each group. Data outliers can also be detected during this process if the value exceeds a predefined threshold. 
	There are other common practices such as data normalization \cite{quackenbush2002microarray}, which is used to scale all dimensions of data to a specific range. This is a very efficient method for the case where there is high variation for different dimensions of the data.
	
\subsubsection{Data Annotation}\label{dataanno}
	As discussed in \S \ref{selection}, data annotation is necessary for supervised learning based ML models, in which 
	both the data and the corresponding target act as the input sample. The model is trained with the labeled data which is used to predict the target for new unseen data.
	This is usually costly and complex due to the requirement for a large volume of labeled data needed for the training. 
	The following investigates different annotation methods that can be applied according to the size of the data to be annotated and the cost of annotation per data.
	
	\myparagraph{Manual Annotation}
	At the initial stage of a ML project, quick prototyping of a workable model requires only few labeled data. 
	In this sense, the developers can manually annotate the collected data to create a small dataset. This is usually done by reviewing the data samples and attaching labels following the annotation guidelines. Manual annotation by the engineers is quick and precise without any
	professional training, and the data quality is usually great. the problem with this approach is the lack of scalability.

\myparagraph{Crowdsourcing Annotation}
	Crowdsourcing annotation is a scalable and cost-effective method. It is usually orchestrated by an online platform that provides access to a workforce of people to complete the annotation tasks. Famous crowdsourcing platforms include Amazon Mechanical Turk (MTurk). 
 Compared to  manual annotation, this approach can be scaled to large dataset labeling. However, the crowdsourcing method requires delicate design on quality control mechanisms to ensure the annotation quality, and the incentives or rewards for the crowds.
	
\myparagraph{Active Learning} Active learning \cite{settles2012active,gilyazev2018active} aims to design a system capable of choosing and learning from less training data while still achieving the same or even higher accuracy. An active learning system consists of two components: a \emph{learning module} that trains a model with the current training sample and a \emph{sample selection module} that selects the most informative samples from the unlabeled samples. The selected samples will then be annotated manually and added to the training set. The iterative process continues until the training converges. The key here is the sample selection module which can be approximately subdivided into \emph{five} categories, \textit{risk reduction}, \textit{uncertainty}, \textit{diversity}, \textit{density} and \textit{relevance} according to the selection criteria \cite{wang2011active}. These criteria can be used either single-handedly (e.g., risk reduction \cite{fawaz2019deep}, uncertainity \cite{joshi2009multi}, relevance \cite{ayache2007evaluation}) or in a combination. In \cite{wang2007interactive}, \textit{uncertainty}, \textit{diversity}, \textit{density} and \textit{relevance} are combined for multi-modality video annotation. Similarly, work in \cite{hoi2006batch} combines \textit{uncertainty}, \textit{diversity} and \textit{density} metrics and the evaluation proves the combination performs well on medical image classification tasks.

All the above explained methods work well for the case of static machine learning scenarios with batch data available beforehand, However, this may not be suitable for IoT-specific streaming data imminent 
with high concept drifts. In this case, the model needs to learn continuously with the upcoming data. Since the new data does not have any label, multiple delayed learning concepts \cite{plasse2016handling,gomes2017adaptive,grzenda2019delayed} are proposed to handle the non-negligible delay in data labeling. These methods are adequate for the scenario where labeling takes a constant time and latency is not a determining factor. 
For IoT data with variable constant drifts, cleaning and labeling may not take uniform time. Also, latency is one of the deciding factors for IoT-based ML applications. Thus, new sets of methods are essential for data labeling which considers the fluctuating IoT data with minimum possible delay.

\subsection{Data Fusion}\label{datafusion}

Data fusion aims to combine the data from multiple sources to provide more accurate and useful information. 
It offers numerous advantages for ML-based application by enhancing the data quality (finding the missing values), detecting any anomalies, conducting the prediction and finding any correlations among the distributed dataset \cite{yang2018novel,lau2019survey,bijarbooneh2016cloud}. However, there are multiple challenges in combining heterogeneous IoT data \cite{alam2017data} such as data frequency, data imperfection, data correlation, data alignment and dynamic iterative process.
To handle these challenges effectively, 
numerous data fusion methods are available in the literature. It is mainly categorized into \emph{three} groups as given below.

\subsubsection{Probabilistic Data Fusion Algorithms}\label{probdatafusion}
This group consists of the algorithms that use density function or probability distributions as a core method for data fusion. It includes Bayesian techniques \cite{biresaw2015tracker}, Markov models \cite{kumar2017coupled}, evidential reasoning \cite{xu2017data}
and other methods. These methods are simple and widely used in different applications to express the co-relation and dependency between numerous datasets. However, there are certain drawbacks with probabilistic data fusion methods highlighted in \cite{alam2017data}. First, it can not scale with the size and modality of the data. Second, uncertain and noisy data can not be handled properly. Finally,  prior probabilities and density functions are difficult to obtain.

\subsubsection{Knowledge-based Data Fusion Algorithms}\label{knowledagedatafusion}
To overcome the uncertainty of data and to increase the accuracy of fusion methods, knowledge-based data fusion methods are proposed. This method accumulates knowledge from the imprecise big data and apply over the fusion process. 
Different aggregation techniques  and ML methods are used for the data fusion process. 
For example, \cite{bigdeli2015fusion,merentitis2015automatic} (supervised learning method) and \cite{fuss2016fusion,zhang2016hierarchical} (unsupervised learning method) are used to discover the distribution of the complex datasets. However, the complexity of this type of method is higher than the probabilistic methods. This class of method, thus, may consume more computing resources and cost more time to process. 


	
\subsubsection{Evidence-based Data Fusion Algorithms}\label{evidancedatafusion}
This group of methods is based on Demster-Shafer Theorem (DST) and recursive operations. As compared to probabilistic methods, where there are only two states (happening or non-happening) of an event, DST includes an unknown state to capture real-world uncertainty. \cite{jamshidi2011fusion,saeidi2014fusion} are the applications of DST for data fusion. However, increasing the data evidence also increases the complexity of this method. Therefore, this method is not suitable for the applications running on less powerful computing resources.


\subsection{Discussion}
In this section, we reviewed core components in the data acquisition process and discussed how they can contribute to generation of high quality, ready-to-use data for IoT-ML application. 
There are \emph{multiple} research directions which can be considered to leverage others' efforts, thereby improving the performance of training a ML model.
First,  \emph{Data reuse}, with the scaling of the data volume, past data is stored and usually ignored after use. However, it can be reused and mined for more values. For example it can be used for boosting semi-supervised data annotation \cite{zhou2013active}, or it can be integrated with newly collected data for model training. Second, \emph{Data re-organization}, there exist datasets for different tasks in similar areas. They may not be the same, but can be re-organized to extract the common distributions. Proper identification and extraction can be explored to save effort on data collection. 

\emph{Feature evolution} is also an important trait of streaming data as a feature may appear and disappear over time. If a feature appears and is found to be relevant, it is necessary to incorporate that for the learning process. In this case, disappearance of a feature can be considered as a drift and the unavailability is treated as missing values. Ignoring this feature may lead to inaccurate prediction. Taking the relevancy of feature evolution for different problem domains.
Other challenges that related to the unbalanced data have been discussed in  \S \ref{sec:generalization} as well.

%% file: conclusion.tex
\section{Conclusion}
\label{sec:conclusion}
Growing numbers of internet-connected things (IoT) produce vast amounts of data, build applications and provide various services in domains such as smart cities, energy, mobility, and smart transportation.
ML is becoming a preliminary technique for analyzing IoT data. 
It produces high-level abstraction and insight that is fed to the IoT systems for fine-tuning and improvement of the services. 
In this survey, we reviewed the characteristics of the IoT development lifecycle and the role of ML for individual steps. 
Specifically, we divided the development lifecycles into different modules and presented a novel taxonomy to characterize and analyze various techniques used to build an ML-based IoT application. 
In summary, this survey seeks to provide systematic and insightful information for researchers. It assists the development of future orchestration solutions by providing a holistic view on the current status of ML-based IoT application development, deriving key open research issues that were identified based on our critical review. 

%% file: appendix.tex
\input{file/specification}
\input{file/model_selection}
\input{file/model_generation}
\input{file/model_opt}
\input{file/model_eva}

%% file: file/specification.tex
\section{Appendix A}
\begin{table}[h]
\begin{tabular}{|c|l|l|l|}
\hline
Software Specification   & \multicolumn{1}{c|}{\textbf{Cloud}}                                                                                                  & \multicolumn{1}{c|}{\textbf{Edge}}                                                                                                                  & \multicolumn{1}{c|}{\textbf{IoT devices}}                                                                                              \\ \hline
\multirow{6}{*}{Main OS} & \multirow{6}{*}{\begin{tabular}[c]{@{}l@{}}- Ubuntu\\ - CentOS\\ - Debian\\ - RHEL\\ - Windows Server\\ - Amazon Linux\end{tabular}} & \multirow{6}{*}{\begin{tabular}[c]{@{}l@{}}- Raspbian\\ - NOOBS\\ - Amazon FreeRTOS\\ - RIOT\\ - Google Fuchsia OS\\ - Windows 10 IoT\end{tabular}} & \multirow{6}{*}{\begin{tabular}[c]{@{}l@{}}- Amazon FreeRTOS\\ - Contiki\\ - TinyOS\\ - RIOT\\ - Ubuntu Core\\ - Mbed OS\end{tabular}} \\
                         &                                                                                                                                      &                                                                                                                                                     &                                                                                                                                        \\
                         &                                                                                                                                      &                                                                                                                                                     &                                                                                                                                        \\
                         &                                                                                                                                      &                                                                                                                                                     &                                                                                                                                        \\
                         &                                                                                                                                      &                                                                                                                                                     &                                                                                                                                        \\
                         &                                                                                                                                      &                                                                                                                                                     &                                                                                                                                        \\ \hline
Programming Language     & \begin{tabular}[c]{@{}l@{}}- Java\\ - ASP.NET\\ - Python\\ - PHP\\ - Ruby\end{tabular}                                               & \begin{tabular}[c]{@{}l@{}}- Java\\ - Python\\ - C\\ - C++\\ - JavaScript\end{tabular}                                                              & \begin{tabular}[c]{@{}l@{}}- C\\ - C++\\ - Java\\ - JavaScript\\ - Python\end{tabular}                                                 \\ \hline
Platforms                & \begin{tabular}[c]{@{}l@{}}- AWS\\ - Azure\\ - Google Cloud Platform\\ - IBM Cloud\\ - Oracle Cloud\end{tabular}                     & \begin{tabular}[c]{@{}l@{}}- Amazon Greengrass\\ - EdgeX\\ - Cisco IOx\\ - Akraino Edge Stack\\ - Eclipse ioFog\end{tabular}                        & \begin{tabular}[c]{@{}l@{}}- AWS IoT\\ - Azure IoT\\ - GCP IoT\\ - IBM Watson\\ - Cisco IoT cloud connect\end{tabular}                 \\ \hline
\end{tabular}
\caption{List of OS, programming language and platform in IoT layers}
\label{SoftwareList}
\end{table}

%% file: file/model_selection.tex
\section{Appendix B}

\subsection{Traditional Machine Learning~(TML) methods}
In this subsection, we give the details of several TML algorithms, as well as their IoT applications.

\paragraph{\textbf{Logistic Regression (LR)}}
Logistic regression is a linear classifier capable of performing binary or multi-class classification. It is among the simplest classification algorithms. 
In a binary classification setting, prediction target $y$ is usually formulated as $ y \in \{0,1\}$ and the prediction probability of the positive class given a d-dimensional input $\mathbf{x} = [x_1,x_2,...,x_d]\in \mathbb{R}^d$ can be calculated via:
\begin{equation}
p(y=1|\mathbf{x};\pmb{\theta})=\sigma(\mathbf{w}^T\mathbf{x}+b)=\sigma(\sum_{i=1}^d w_i x_i+b),
\end{equation}
In the equation $\pmb{\theta}=\{\mathbf{w}=[w_1, w_2, ..., w_d]\in\mathbb{R}^d, b\in\mathbb{R}\}$ denotes the model parameters, and
$\sigma(\cdot)$ is an activation function used to squash the linear output within the range $[0,1]$ for probabilistic interpretation.
The training of the classifier aims to learn suitable values for the parameters $\pmb{\theta} = \{\mathbf{w}, b\}$, starting from some random initialization, through minimizing of a loss or cost function $J(\pmb{\theta})$.
For LR, log (Cross-Entropy) loss  is used, i.e., $J(\pmb{\theta})=-\ln p(y|\mathbf{x};\pmb{\theta})$ for data point $\{\mathbf{x}, y\}$ to facilitate the calculation of the model gradient and to minimize the model loss. 
With the trained model, for any query data the classification decision can be made via thresholding the predicted probability.

Extend the binary LR to support $c$-class prediction scenarios, the target $y \in \{1, 2, ..., c\}$.
Softmax function is applied to the output layer to normalize the $c$ outputs into probabilities.
Different from binary LR, there are $c$ sets of parameters $\pmb{\theta}=\{\mathbf{W}=[\mathbf{w}_1,..., \mathbf{w}_c]^T, \mathbf{b}=[b_1,..., b_c]^T\}$
to be estimated by minimizing the cross-entropy loss (i.e., log loss).
With the trained model parameters, at the inference stage the class label will be assigned to the one with the largest classification probability.

\paragraph{\textbf{Decision Trees~(DT)}}\label{dt} Decision Tree is a tree-like model for classification or regression tasks.
 A decision tree is made up of nodes and edges where a node can be seen as a feature, and the edge represents a condition for classification.
The learning process of DT is to select the optimal feature that can best split the training examples in a recursive manner.
Based on certain criteria such as information gain~\cite{quinlan1986induction} or Gini impurity~\cite{du2002building}, the root node can be selected from the features, which will divide the whole training population into two or more homogeneous sets.
For each sub-population, sub-nodes will be selected in a similar manner and this process will repeat until all the subsets are pure (i.e., with the same class label for each subset).
These nodes and edges constitute the trained model and can be used for inference.

Different from most of the other classifiers (such as LR, SVM, ANN, DL) which require the input to be normalised to numerical values, the unique tree-structure of DT make it possible to take both numerical and nominal values, making it a highly interpretable tool for various classification tasks (e.g., medical records diagnosis).
However, DT suffers from the ``curse of dimensionality'', and it faces an overfitting problem when the input dimensionality is too high.
With unstructured high-dimensional data, feature engineering/extraction is one of the necessary steps to take (for dimensionality reduction), before DT is applied.


DT can be used as a main classifier (e.g., for low-dimensional structured data) or collaborative classifier with other machine learning algorithms on various IoT applications, such as Intrusion detection system~\cite{peng2018intrusion} in fog environment.

\paragraph{\textbf{Random Forest~(RF)}} As its name implies, a RF is an ensemble model with many DTs as base classifiers.
The individual DTs are constructed by random sampling the features and the training examples for diversity, boosting the performance of multiple classifier systems.
The random sampling process makes the individual DTs less correlated ---with different prediction errors--- and the aggregating function can smooth the large prediction variance, making RF a robust classifier with high generalization capabilities.
There are several key hyper-parameters for RF, and two main ones are: feature number for individual DTs, and number of DT classifiers.
For individual DTs, there is a trade-off between generalization and discrimination capabilities with respect to feature dimension, and one popular heuristic value is to use the square root of original dimension number (e.g., the default setting in scikit-learn).
For the RF though, the performance of the model usually correlates wit the number of DTs.
Yet the performance gain tends to become less significant since the diversity (among the DT classifiers, which are correlated to some extent) will decrease accordingly.
Since both the efficiency and storage are proportional (in a linear manner) to the classifier number, to find a number configuration that can result in optimal effectiveness, storage and efficiency performance is of vital importance.

RF is one of the most popular classifiers due to its great generalization capability.
For example, it has been used for intrusion detection~\cite{chen2018detection, meidan2017detection} and anomaly detection.
In a previous study~\cite{meidan2017detection}, the authors collected and manually labeled data from 17 distinct IoT devices and used RF algorithms to recognize IoT device categories from the white list. 

\begin{figure}
  \includegraphics[width=0.9\linewidth]{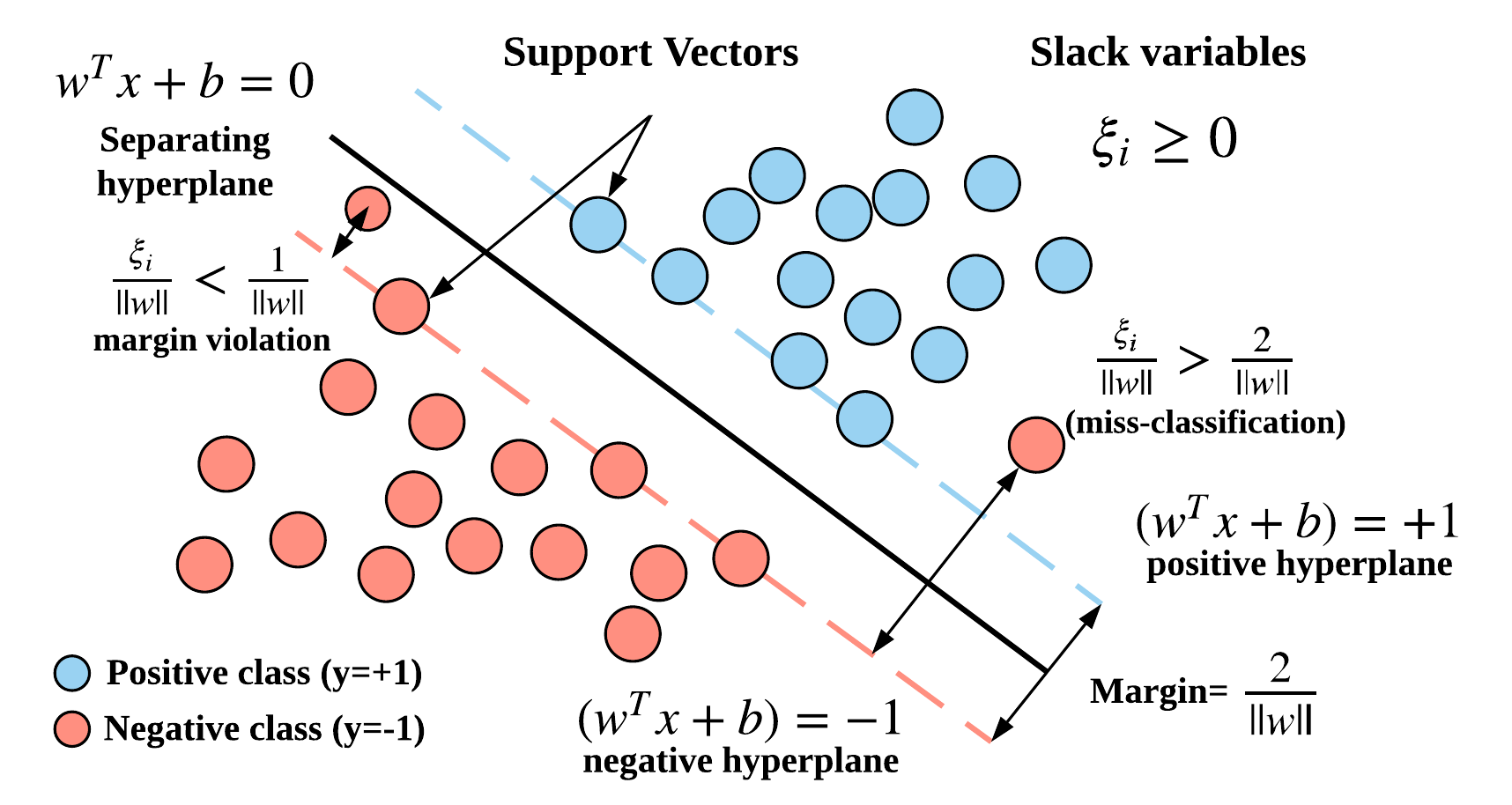}
  \caption{Basic concepts in SVM}
  \label{fig:svm}
\end{figure}

\paragraph{\textbf{Support Vector Machine~(SVM)}} Support Vector Machine~(SVM), also referred to as large margin classifier, aims to find a decision boundary (separating hyperplane) that can best separate (i.e., with the largest margin) positive/negative classes.
Fig. \ref{fig:svm} shows some basic concepts in SVM including support vectors, margin and slack variable $\xi$.
$\xi$ is a non-negative variable that is used to measure the misclassified instances and those within the margin (i.e., margin violation, as shown in Fig. \ref{fig:svm}). The objective function of (soft-margin) SVM can be constructed to maximize the margin while penalizing these instances:
\begin{equation} \label{eq:svm0}
\argmin_{\mathbf{w}, b, {\xi_i}} \frac{||\mathbf{w}||^2}{2}+C\sum_{i}^{N}\xi_i, \quad \text{subject to}\quad y_i(\mathbf{w}^T\mathbf{x}_i+b)\ge 1-\xi_i, \text{where} \quad \xi_i\ge0, \quad i=1,2,...,N.
\end{equation}
In Eq (\ref{eq:svm0}), $\{\mathbf{x}_i,y_i\}_{i=1}^N$ are the $N$ training sets with $y_i\in\{-1,1\}$, and $C$ is a regularization hyper-parameter that trades-off between the margin and errors (i.e., measured by $\sum_i^N\xi_i$).
It is obvious to see that if we set $C=0$, we can get a classifier with the large margin at a cost of potential high training errors.
On the other hand, if we set $C$ to a very large number (or $\infty$ in theory), errors are less likely to be 
tolerant, and we may end up with a classifier with narrow or hard margin.
It is worth noting that Eq.(\ref{eq:svm0}) can be further simplified into $\argmin_{\mathbf{w}, b} \frac{||\mathbf{w}||^2}{2}+C\sum_{i}^{N}\text{max}(0, 1-y_i(\mathbf{w}^T\mathbf{x}_i+b))$,
where the term $\text{max}(0, 1-y_i(\mathbf{w}^T\mathbf{x}_i+b))$ is also referred to as hinge loss.

\begin{figure}[!tbp]
  \centering
  \begin{minipage}[b]{0.62\textwidth}
    \includegraphics[width=\textwidth]{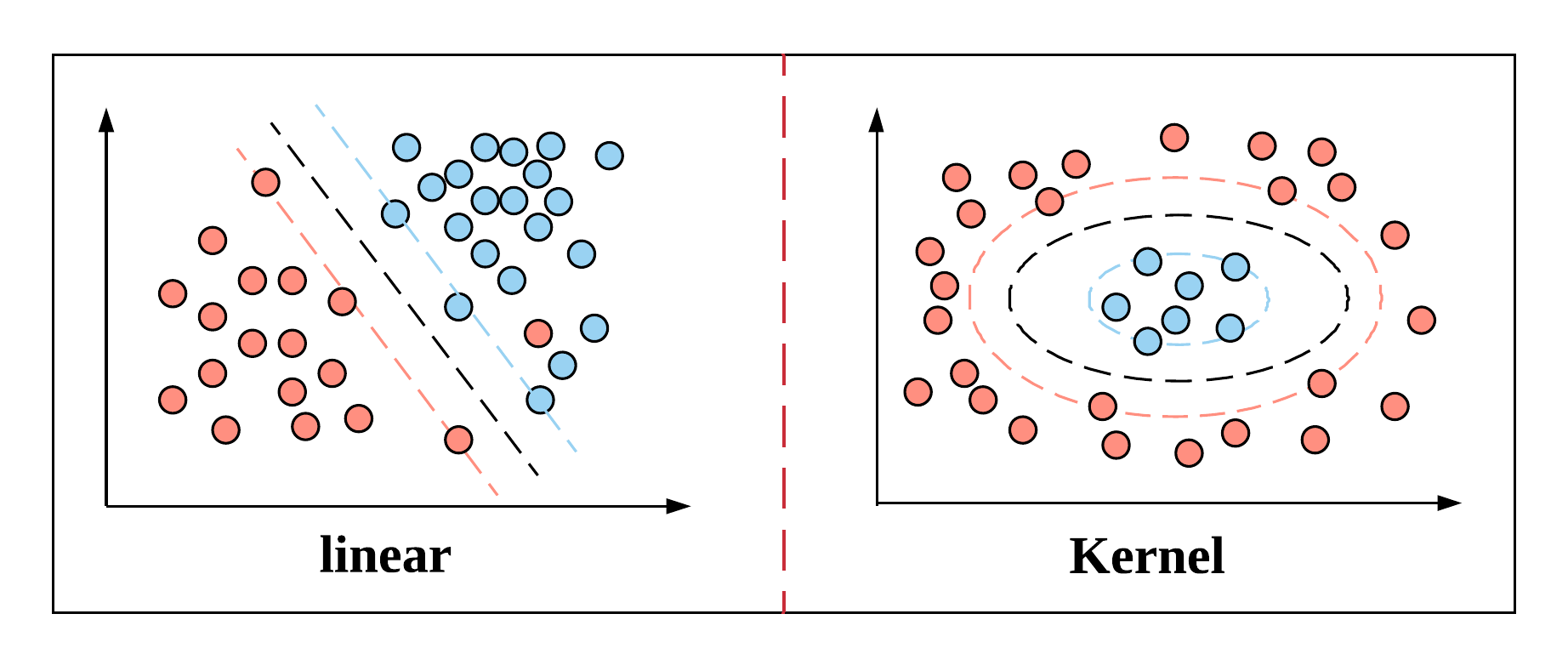}
    \caption{Margins for linear (left) and kernel (right) SVM}
    \label{fig:margin}
  \end{minipage}
  \hfill
  \begin{minipage}[b]{0.33\textwidth}
    \includegraphics[width=\textwidth]{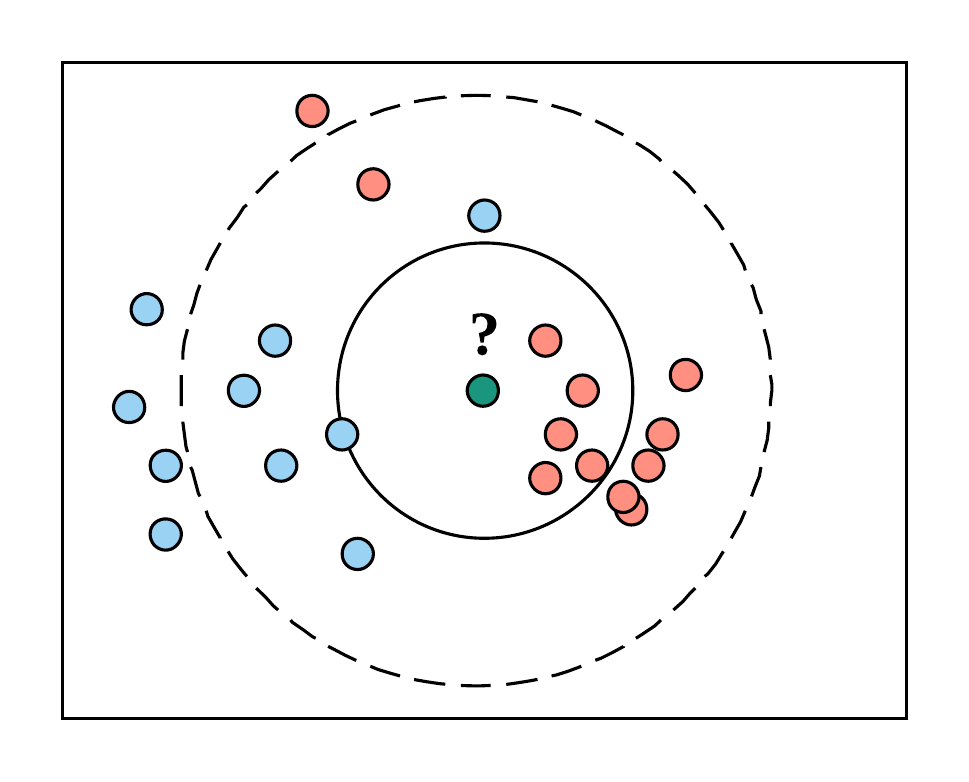}
    \caption{Majority voting process in KNN}
    \label{fig:knn}
  \end{minipage}
\end{figure}

On the other hand, for highly non-linearly separable data, instead of employing feature engineering/extraction processes, an elegant alternative---kernel SVM can be applied, and Radial Basis Function (RBF) is one of the most popular kernel functions.
Kernel SVM also has the aforementioned characteristics such as soft margin, and it tends to have great performance on small non-linearly separable data.
Fig. \ref{fig:margin} shows the margins on linearly separable (by linear SVM) and non-linearly separable (by kernel SVM) data, respectively.

SVM is notable for its generalization capability and is suitable for those small datasets with high-dimensional features \cite{vapnik2013nature},
and there are many IoT applications such as Android malware detection system \cite{sanjaa2013malware}, smart weather prediction \cite{rao2012efficient}, etc.


\paragraph{\textbf{$K$-Nearest Neighbour~(KNN)}} KNN is a nonparametric, instance-based, non-linear classifier.
Under the classification setting, given any query data, KNN essentially performs majority voting among the $K$ most similar training samples, as shown in Fig. \ref{fig:knn}.
The similarity can be measured by some distance metrics such as Minkowski distance $D(\mathbf{x}, \mathbf{y}) = (\sum_{j=1}^d|x_j-y_j|^p)^{\frac{1}{p}}$ (for $d$-dimensional vectors $\mathbf{x}, \mathbf{y}$).
It is worth noting that when $p=1$ and $p=2$, the Minkowski distance can be seen as Manhattan Distance and Euclidean distance respectively,
yet for different applications the optimal distance metric may vary from case to case.


One of the key properties is that KNN does not require any training process (i.e., lazy learning), and for any query data, the distance calculation has to be performed for each sample in the whole training set, which makes KNN a less-scalable approach for large datasets.
Another issue is the selection of hyper-parameter $K$. A small $K$ may make KNN sensitive to outliers in the training set while a large one may make KNN less discriminate. Nevertheless, KNN is a powerful non-linear classifier with low-dimensional small datasets, 
and there are many IoT applications such as network intrusion detection \cite{li2014new}, anomaly detection \cite{su2011real} and Urban noise identification \cite{alsouda2019iot}, etc.

\paragraph{\textbf{Naive Bayes~(NB)}}
NB is a probabilistic classifier which takes the class prior distribution into account, and assumes the features are conditionally independent.
Based on the Bayesian theory and the posterior probability, it can be presented as $p(y|\mathbf{x})=\frac{p(\mathbf{x}|y)p(y)}{p(\mathbf{x})}$, where $p(\mathbf{x}|y)$ is the likelihood; $p(y)$ is the prior; $p(\mathbf{x})$ is the evidence.
The classification process is to assign the label with the largest posterior probability, and in this case the term evidence $p(\mathbf{x})$ remains a constant which can be cancelled out, i.e., $p(y|\mathbf{x})\propto p(\mathbf{x}|y)p(y)$, which can be further written into $p(y|\mathbf{x})\propto \prod_{j=1}^dp(x_j|y)p(y)$ for d-dimensional input $\mathbf{x}$
due to the feature independence assumption.
In practice, logarithm is often used to avoid the problem of floating point underflow, and the class label $\hat{y}$ can be assigned via:
\begin{equation}
\hat{y} = \argmax_{y\in \{1, 2, ..., c\}}\ln p(y)+\sum_{j=1}^d \ln p(x_j|y), 
\end{equation}
NB is good at modelling both continuous and discrete data. For example the likelihood of a discrete feature can be calculated by frequency while the likelihood of a continuous feature can be calculated by density estimation (e.g., Gaussian).
It takes the prior of class distribution into consideration, which is helpful in data imbalanced problems.
However, it also suffers from the ``curse of dimensionality'' like DT, and normally can not be used directly on unstructured data before feature extraction/engineering approaches have been applied.
When the feature independence assumption is not significantly violated, 
it is normally served as an efficient and effective classifier.
There are many IoT applications, such as network traffic analysis for DoS attach detection \cite{hema2015attack}, animal health monitoring \cite{shinde2017iot}, etc.


 \begin{table}[t]
\centering
\caption{Summary of Traditional Machine Learning Models (Note: In time complexity, $m$ represents the number of training sample, $n$ represents the feature dimension, $k$ represents the selected $K$ value, $c$ represents the class number and $t$ indicates the tree number)}
\label{tab:traditional_ml}
\resizebox{\textwidth}{!}{%
\begin{tabular}{|l|l|l|l|c|l|l|}
\hline
\rowcolor[HTML]{C0C0C0}
\multicolumn{1}{|c|}{\cellcolor[HTML]{C0C0C0}Method} & \multicolumn{1}{c|}{\cellcolor[HTML]{C0C0C0}Learning model} & Category & \multicolumn{1}{c|}{\cellcolor[HTML]{C0C0C0}\begin{tabular}[c]{@{}c@{}}Typical \\ input data\end{tabular}} & \multicolumn{1}{c|}{\cellcolor[HTML]{C0C0C0}\begin{tabular}[c]{@{}c@{}}Time \\ Complexity\end{tabular}} &\multicolumn{1}{c|}{\cellcolor[HTML]{C0C0C0}Characteristics} & \multicolumn{1}{c|}{\cellcolor[HTML]{C0C0C0}IoT Application} \\ \hline \hline

DT &  &Discriminative& Various & $ O(m\cdot n^2)$ \cite{su2006fast} & \begin{tabular}[c]{@{}l@{}} \tabitem Dividing training samples to branches and leaves \\ \tabitem High interpretability method \\\tabitem Require large memory space due to the\\\enspace construction nature. \end{tabular} & \begin{tabular}[c]{@{}l@{}} \tabitem Intrusion detection~\cite{peddabachigari2004intrusion} \\ \tabitem Suspicious detection on \\ \enspace traffic sources~\cite{elovici2007applying} \\ \tabitem Future Heart Attack \\ \enspace Quantity prediction~\cite{kirmani2016prediction} \end{tabular} \\ \cline{1-1} \cline{3-7}

SVM &  & Discriminative & Various&$ O(m^2 \cdot n)$\cite{chu2007map} & \begin{tabular}[c]{@{}l@{}} \tabitem Good Generalization capability and suitability\\ \enspace for small dataset with large feature 
\\\tabitem Difficult on selection optimal kernel
\\ \tabitem High computation complexity on large dataset\\ \enspace with complex kernel \end{tabular} & \begin{tabular}[c]{@{}l@{}} \tabitem Malware detection~\cite{sanjaa2013malware} \\\tabitem Attack detection in
smart\\ \enspace grids~\cite{swetha2015smart}\\ \tabitem Smart weather prediction~\cite{rao2012efficient} \end{tabular}\\ \cline{1-1} \cline{3-7}

NB &  & Discriminative & Various & \begin{tabular}[c]{@{}c@{}}$ O(m\cdot n + n\cdot c)$ \\ \cite{chu2007map}\end{tabular}& \begin{tabular}[c]{@{}l@{}} \tabitem Ease to implement\\\tabitem Generalizes well to multi-class problem
\\ \tabitem Low dependence on large dataset and robustness\\ \tabitem Hard to capture relation information\\ \end{tabular} &  \begin{tabular}[c]{@{}l@{}}  \tabitem Detection of network\\
\enspace intrusion~\cite{mukherjee2012intrusion}\\ \tabitem Animal health monitoring~\cite{shinde2017iot} \end{tabular} \\ \cline{1-1} \cline{3-7}

KNN &  & Discriminative & Various &$O(m\cdot n\cdot k)$ \cite{zhao2012improved}&\begin{tabular}[c]{@{}l@{}} \tabitem Nonparametric, instance-based\\ \tabitem No training process, no model construction
\\ \tabitem Sensitive to the outlier \\ \tabitem Difficult on selection optimal K \end{tabular}  &  \begin{tabular}[c]{@{}l@{}}  \tabitem Anomalies detection~\cite{goldstein2016comparative}\\ \tabitem Urban noise identification~\cite{alsouda2019iot}  \end{tabular} \\ \cline{1-1} \cline{3-7}

RF &  \multirow{-14}{*}{Supvervised} & Discriminative & Various&\begin{tabular}[c]{@{}c@{}}$O(t\cdot n^2\cdot logn)$ \\  \cite{louppe2014understanding}\end{tabular} & \begin{tabular}[c]{@{}l@{}} \tabitem Robust to over-fitting\\ \tabitem Bypasses feature selection
\\ \tabitem Impractical in specific real-time application\end{tabular}  & \begin{tabular}[c]{@{}l@{}}  \tabitem DDoS attack detection~\cite{lakshminarasimman2017detecting}\\ \tabitem Unauthorized IoT devices \\\enspace detection~\cite{meidan2017detection}  \end{tabular} \\ \cline{1-1} \cline{3-7}  \hline

K-Means &  & Clustering & Various& $O(m \cdot n \cdot k)$\cite{chu2007map}& \begin{tabular}[c]{@{}l@{}} \tabitem Ease to use on Unlabelled data\\ \tabitem Produces tighter cluster than hierarchical clustering \\\tabitem Less effective than supervised learning method
\\ \tabitem Difficult on selection optimal K  \end{tabular} & \begin{tabular}[c]{@{}l@{}}  \tabitem Sensor fault detection~\cite{zhao2011fault}\\ \tabitem Sybil detection in
industrial \\ \enspace WSNs~\cite{yang2008detecting} \end{tabular} \\ \cline{1-1} \cline{3-7}

PCA & \multirow{-4}{*}{Unsupervised} & Dimension Reduction & Various&\begin{tabular}[c]{@{}c@{}}$O(m \cdot n^2 + n^3)$ \\ \cite{chu2007map} \end{tabular}& \begin{tabular}[c]{@{}l@{}}  \tabitem Used for dimensional reduction\\ \tabitem Consequently reduce the complexity of the model \\  \tabitem Should be used with other ML methods \end{tabular} & \begin{tabular}[c]{@{}l@{}}  \tabitem Real-time detection
systems \\\enspace in IoT
environments~\cite{elrawy2018intrusion}\\ \tabitem Traffic anomaly detection~\cite{ding2016pca} \end{tabular}  \\ \hline

\end{tabular}%
}
\end{table}



\paragraph{\textbf{K-Means}}
Different from the supervised classification models above, K-means is an unsupervised clustering algorithm without using class label information for training.
Given a number of data points, K-Means aims to find $K$ centroids (i.e., means), and the corresponding nearest samples to form the clusters.
Various distance metrics can be used for K-means algorithm, and the most common one is based on Euclidean distance, whose objective function is:
\begin{equation}\label{eq:kmeans}
\argmin \sum_{i=1}^N\sum_{k=1}^K\gamma_{ik} ||\mathbf{x}_i-\mu_k||^2,
\end{equation}
where $\mu_k$ is the centroid of the $k$th cluster and $\gamma_{ik}\in \{0,1\}$ denotes whether sample $\mathbf{x}_i$ belongs to the $k$th cluster ($1$) or not ($0$). 
K-means clustering is an heuristic process---starting from random values (of the centroids $\{\mu_k\}_{k=1}^K$), it will 1) assign each example to the nearest cluster and 2) update the $K$ centroids (by re-calculating the means of the corresponding samples for each cluster).
It is an iterative process and the updating will stop until Eq.(\ref{eq:kmeans}) is minimized (e.g., lower than a pre-defined threshold).

K-means is among the most popular clustering algorithms due to its simplicity.
However, using a Euclidean distance based method is limited to spherical datasets, and finding the most suitable distance metrics (for different applications/datasets) is one of the key issues for improving the performance of K-means algorithm.
K-means is widely used for IoT applications such as sensor fault detection ~\cite{zhao2011fault}, Sybil detection~\cite{yang2008detecting}, etc.



\paragraph{\textbf{Principal component analysis (PCA)}}
PCA is another unsupervised learning approach and it is normally used for dimensionality reduction or feature decorrelation.
Covariance matrix $\mathbf{S}$ which reflects the correlation of the features can be calculated via $\mathbf{S}=\frac{1}{N}\sum_{i=1}^N(\mathbf{x}_i-\mu)(\mathbf{x}_i-\mu)^T$ for $N$ $d$-dimensional training data points $\{\mathbf{x}_i\}_{i=1}^N$ with $\mu = \frac{1}{N}\sum_{i=1}^N\mathbf{x}_i$.
Eigenvalue decomposition can be performed on $\mathbf{S}$ such that the leading $s(<d)$ eigenvectors can be used as a transformation matrix for feature decorrelation or dimensionality reduction.
PCA is one of the most popular feature extraction tools due to its simplicity, 
and there are many IoT applications such as traffic anomaly detection~\cite{brauckhoff2009applying}.


Table \ref{tab:traditional_ml} summarizes the aforementioned TML, including their advantages, disadvantages and applications in IoT systems.

\subsection{Deep Learning~(DL) methods}


In this section, we detail several deep learning algorithms: Deep Neural Networks (DNN), Convolutional Neural Network~(CNN), Recurrent Neural Network~(RNN), etc.


\paragraph{\textbf{Deep Neural Networks~(DNN)}}
As previously mentioned, LR is one of the most popular linear classifiers, which means the decision boundary is a line or hyperplane in the feature space.
However, most of the data are more complex and are not linearly separable in real-world scenarios.
In this case, directly applying linear classifiers such as LR on the raw data would yield unsatisfied classification performance.
One could empirically use hand-crafted features followed by LR, yet this trial-and-error process can be time-consuming, and it may be challenging to define the high-order discriminant features where the data can be linearly separable.
One alternative is to use data-driven methods, and it is straightforward to extend LR to DNN (i.e., Multilayer Perceptron MLP).

DNN is built upon LR with a stack of hidden layers.
By applying activation functions like sigmoid, tanh, or ReLU on the hidden layers, raw features can be transformed in a non-linear manner, yielding discriminant (i.e., linearly separable) features before the output layer.
Similar to LR, the training process is to minimise the log loss.
However, the parameters of DNN are the layer-wise connections, which can be learned using the back-propagation approach.
DNN performs end-to-end learning, which means it learns the feature extractors and classifiers simultaneously.

DNN can be deemed as the simplest form of deep learning model which comprises one input layer, one output layer as well as multiple hidden layers for more complex feature extraction.
It is worth noting that Artificial Neural Network (ANN) is a special case of DNN with only one hidden layer.
The layers are organised in a hierarchical manner, with each layer being a function of the layer that preceded it.
For example, the second layer $\mathbf{h}^{(2)}$ (i.e., also the first hidden layer)  can be expressed as a function of the first (i.e., input) layer (e.g., with input vector $\mathbf{x}\in \mathbb{R}^d$):
\begin{equation}
\mathbf{h}^{(2)}=g^{(2)}(\mathbf{W}^{(2)T}\mathbf{x}+\mathbf{b}^{(2)}),
\end{equation}
where $\{\mathbf{W}^{(2)}, \mathbf{b}^{(2)}\}$ are the parameters and $g^{(2)}$ is the activation function (e.g., ReLU) for the second layer.
We also call these layers fully connected or dense layers.
Similarly, the $l^{th} (l>2)$ hidden layer can be written as
\begin{equation}
\mathbf{h}^{(l)}=g^{(l)}(\mathbf{W}^{(l)T}\mathbf{h}^{(l-1)}+\mathbf{b}^{(l)}).
\end{equation}
Assuming there are a total number of $L$ layers, then an output layer linearly transforms the previous hidden units $\mathbf{h}^{(L-1)}$, followed by a softmax function for probability scaling:
\begin{equation}\label{eq:softmax}
p(\mathbf{y}|\mathbf{h}^{(L-1)})=\mbox{softmax}(\mathbf{W}^{(L)T}\mathbf{h}^{(L-1)}+\mathbf{b}^{(L)}), 
\end{equation}
Note the model parameters include the layer-wise weight matrices and bias vectors, i.e., $\pmb{\theta}=\{\mathbf{W}^{(l)},\mathbf{b}^{(l)}\}_{l=2}^L$, which needed to be estimated by minimising the loss function (e.g., log loss).
DNN has been widely used for various IoT applications, including wearable-based activity recognition \cite{vacher2015speech}, traffic congestion prediction\cite{devi2017machine}, healthcare\cite{7821702}, etc.


\paragraph{\textbf{Convolutional Neural Networks~(CNNs)}}

\begin{figure}
  \includegraphics[width=0.9\linewidth]{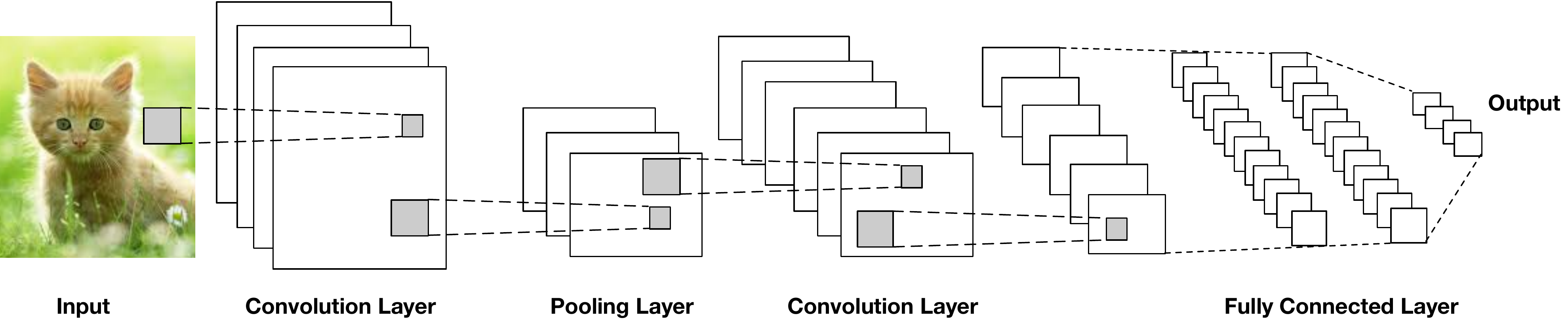}
  \caption{An architecture of CNN}
  \label{fig:cnn}
\end{figure}

Similar to the DNN, CNN also comprises multiple hidden layers. It learns to map from 2D data (e.g., images) at the input, output the class probabilities at the output.
DNN usually contains three types of hidden layers: convolution layer and pooling layer at each stage and FC/dense layer right before the output layer.
In a convolution layer, local patterns can be extracted by learning several convolutional kernels/filters (with a predefined size, such as $3\times3$), each of which has the same shared weights to be estimated (via back-propagation).
This weight-sharing scheme makes CNN effective in dealing with high-dimensional data (such as high-resolution images).
After convolution operation with a number of kernels/filters, the corresponding feature maps can be formed, and then a pooling layer can be employed for down-sampling these feature maps for more compact representation.
In CNN, normally there are multiple convolution/pooling layers, before FC/dense layers can be applied; Fig. \ref{fig:cnn} shows an example architecture of CNN.
Since cameras are one of the major parts of the IoT environment, CNN can be a very useful tool and has been widely used such as in traffic sign detection on autonomous driving \cite{shustanov2017cnn}.

\begin{figure}
  \includegraphics[width=0.8\linewidth]{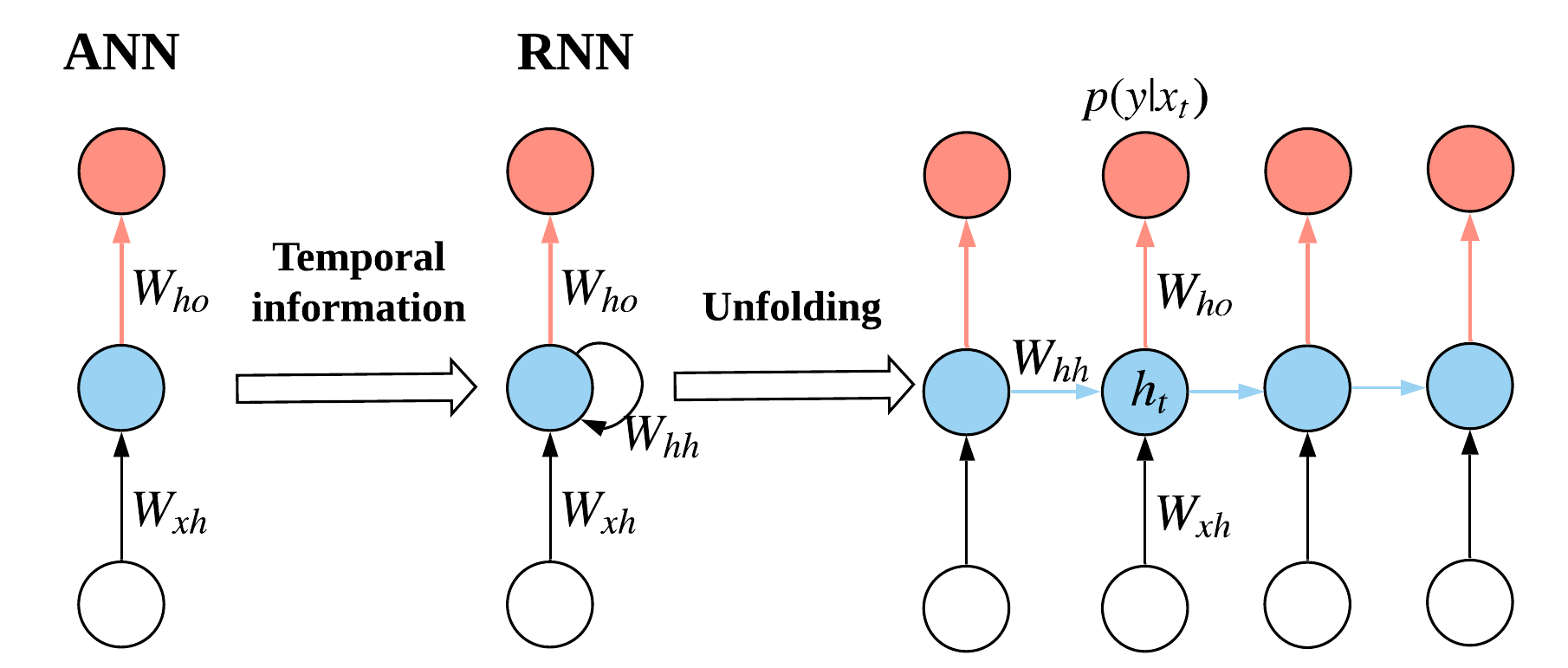}
  \caption{From ANN to RNN}
  \label{fig:RNN}
\end{figure}

\paragraph{\textbf{Recurrent Neural Networks~(RNNs)}}
RNN is a type of deep model and it is designed to model and recognise sequential data (e.g., time-series data).
Compared with DNN, a hidden unit of RNN can feed-forward on itself in the next timestamp and thus can memorise the temporal information for sequential inference.
Fig. \ref{fig:RNN} shows how to extend an ANN (DNN with one hidden layer) to a three-layer vanilla RNN by modelling the temporal information, we use an arrow to represent FC/dense layer for better visual effects. 
Compared with ANN, RNN has an additional set of parameters, i.e., the hidden-to-hidden transformation matrix $\mathbf{W}_{hh}$ to be estimated, and the full parameters can be expressed as:
$\Theta = \{\mathbf{W}_{xh}, \mathbf{W}_{hh}, \mathbf{W}_{ho}, \mathbf{b}_o, \mathbf{b}_h\}$.
For $\mathbf{x}_t$,  the input vector at the $t^{th}$ timestamp, 
the feed-forward pass can be written as:
\begin{equation}
\begin{split}
p(\mathbf{y}|\mathbf{x}_t)&=\text{softmax}(\mathbf{W}_{ho}^T\mathbf{h}_t+\mathbf{b}_o)\\
\text{where} \qquad \mathbf{h}_t &= \tanh(\begin{bmatrix} \mathbf{W}_{xh} \\ \mathbf{W}_{hh} \end{bmatrix}^T [\mathbf{x}_t, \mathbf{h}_{t-1}]+\mathbf{b}_h),
\end{split}
\end{equation}
It is clear that the current hidden state $\mathbf{h}_t$ is calculated based on the previous hidden state $\mathbf{h}_{t-1}$ and the current signal $\mathbf{x}_t$, while the output layer remains the same as DNN (see Eq.(\ref{eq:softmax})).
We can see RNN is an end-to-end method where both the sample-wise features $\mathbf{h}_t$ and predictions $p(\mathbf{y}|\mathbf{x}_t)$ can be learned simultaneously.

For complex time-series data analysis, we need a deeper RNN (i.e., larger window, multiple layers, etc.) to capture the high-level temporal/contextual information.
Yet vanilla RNN has a gradient explosion/vanishing problem owing to its numerical properties.
In the 1990s, a complex hidden unit named long short term memory (LSTM) \cite{hochreiter1997long} was proposed for large-scale recurrent neural network construction, which can preserve the error/gradient that can be back-propagated effectively through time and layers.
LSTM includes four different gates organised in a special internal structure, in contrast to tanh function for activation in vanilla RNN.
Accordingly, it also has four sets of specific gate parameters to be estimated.

RNNs achieve great performance in time-series applications, for example machine translation and speech recognition.
In IoT applications, 
it has shown its great performance in sensor-based power station condition prediction \cite{zhang2018lstm}, wearable-based activity recognition \cite{guan2017ensembles}, etc.

\begin{figure}
  \includegraphics[width=0.7\linewidth]{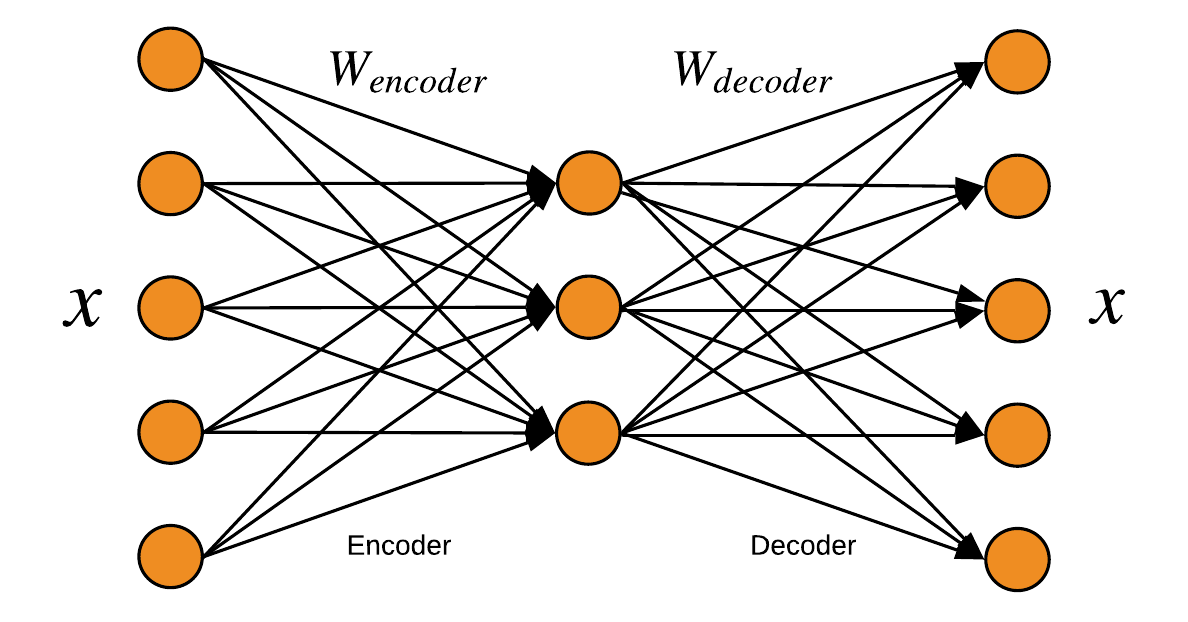}
  \caption{The encoder-decoder structure of a three-layer Auto-encoder}
  \label{fig:AE}
\end{figure}

\paragraph{\textbf{Auto-encoder~(AE)}}
An Auto-encoder(AE) is a type of unsupervised neural network, and it can transform the data into latent code/representation/feature (e.g., in lower-dimensionality), from which the original data can be reconstructed. 
AE is widely used for data compression and feature extraction, and a typical AE includes two parts: encoder, and decoder. 
Fig. \ref{fig:AE} shows a three-layer encoder-decoder structure for AE, and the $d$-dimensional input vector $\mathbf{x}$ can be transformed into the latent representation $\mathbf{h}$ via a linear transformation followed by an activation function $g(.)$, i.e., $\mathbf{h}=g(\mathbf{W}_\text{encoder}^T\mathbf{x}+\mathbf{b}_0)$.
Similarly, decoder can be used to reconstruct the data such that $\mathbf{x}' = g(\mathbf{W}_\text{decoder}^T\mathbf{h}+\mathbf{b}_1)$.
Note $\mathbf{\Theta} = \{\mathbf{W}_\text{encoder}, \mathbf{b}_0, \mathbf{W}_\text{decoder}, \mathbf{b}_1 \}$ are the model parameters that can be estimated via minimising a reconstruction loss 
\begin{equation}\label{eq:AE}
\argmin_{\mathbf{\Theta}} L(\mathbf{x}-\mathbf{x}').
\end{equation}
Note in Eq. (\ref{eq:AE}), $L(.)$ can be very flexible, and some popular ones include mean squared error, log loss (for binary input), etc. 
It is worth noting that with small dataset, a popular constraint is to set $\mathbf{W}_\text{encoder} = \mathbf{W}^T_\text{decoder} $, which can limit the degree of freedom of this model with better generalisation capabilities.
AE can be used as a powerful tool for dimensionality reduction. 
Compared with PCA, the non-linear transformation may extract more discriminant information for classification. 
It can also be used for anomaly detection, and the intuition is that the trained AE (based on normal class) cannot reconstruct the abnormal class well, yielding large reconstruction errors. 
There are many IoT applications for AE, such as fault diagnosis in hardware devices~\cite{chopra2015fault}, and anomaly detection in the performance of assembly lines~\cite{oh2018residual}.

\paragraph{\textbf{Generative Adversarial Network~(GAN)}}
Generative Adversarial Networks (GAN) is a generative model with adversarial architecture. Fig. \ref{fig:gan} shows the basic network architecture of a GAN, which contains a generator and a discriminator.

\input{figuresTex/fig-gan-arch}
The generator aims at generating indistinguishable samples compared to the real data while the discriminator works adversarially to distinguish the generated fake samples from the real data. 
It is an iterative competition process that will eventually lead to a state where the generated samples are indistinguishable from the real data. 
The generative model aims to learn the data distribution and generate data samples for those domains that lack data samples.

The decent idea of applying adversarial structure on deep neural networks brought impressive results related to content 
generated, and realistic visual content. Moreover, it can be applied to many applications such as image restoration, style transfer, sample generation. However, there are three main drawbacks on GAN frameworks. The first one is that the network is hard to train and the training process is destabilized ~(i.e., ``Non-convergence''). The second drawback is that the generator will tend to produce limited varieties of samples at the end of a training period~(i.e., ``Mode collapse''). The last drawback of GAN is the gradient diminishing problem that occurs when the balance between generator and discriminator were broken.

GANs have been recently implemented in the IoT environment, mainly on IoT security. For example, a previous study~\cite{zhao2018research} proposed a GAN based framework for improving open-categorical classification on individual identity authentication application. Moreover, GAN has been used as a tool to generate large datasets that do not need manual annotation. For example, a study\cite{zhao2018application} explored the feasibility of using GAN to generate illegal Unmanned Aerial Vehicles (UAVs) dataset and obtain a better classification model with better accuracy. As the data generated from sensors may be unlabelled, GANs may have more potential applications in the IoT environment.

\begin{table}[t]
\centering
\caption{Summary of Deep Machine Learning Models}
\label{tab:my-table}
\resizebox{\textwidth}{!}{%
\begin{tabular}{|l|l|l|l|l|l|}
\hline
\rowcolor[HTML]{C0C0C0}
\multicolumn{1}{|c|}{\cellcolor[HTML]{C0C0C0}Method} & \multicolumn{1}{c|}{\cellcolor[HTML]{C0C0C0}Learning Type} & Category & \multicolumn{1}{c|}{\cellcolor[HTML]{C0C0C0}\begin{tabular}[c]{@{}c@{}}Input data\\ Type\end{tabular}} & \multicolumn{1}{c|}{\cellcolor[HTML]{C0C0C0}Characteristics} & \multicolumn{1}{c|}{\cellcolor[HTML]{C0C0C0}IoT Application} \\ \hline \hline

CNNs & \multirow{5}
{*}{Supvervised} & Discriminative &  2-D (image, sound, etc.) &\begin{tabular}[c]{@{}l@{}}  \tabitem Mainly used on image processing \\ \tabitem Less connection compared
to DNNs.\\ \tabitem Require large training samples. \end{tabular}  & \begin{tabular}[c]{@{}l@{}}  \tabitem Traffic sign detection\\ \tabitem Plant disease detection \\ \tabitem Bridge crack detection\end{tabular} \\ \cline{1-1} \cline{3-6}

RNNs &  &Discriminative  & Sequential data & \begin{tabular}[c]{@{}l@{}} \tabitem Mainly used to analyze sequential data \\\tabitem Useful in IoT applications with\\\enspace time-dependent data \end{tabular} & \begin{tabular}[c]{@{}l@{}} \tabitem Identify movement pattern \\\tabitem Behavior detection \\\tabitem Human activity recognition\\ \tabitem Mobility prediction \end{tabular}\\ \cline{1-1} \hline

%

AEs &  & Generative & Various & \begin{tabular}[c]{@{}l@{}} \tabitem Mainly used for feature extraction,\\ \enspace
and dimensionality reduction \\ 
\tabitem Optimized by reconstructs
input data\\\tabitem Can be used on unlabeled data \end{tabular} & \begin{tabular}[c]{@{}l@{}} \tabitem Emotion recognition\\ \tabitem Machinery fault diagnosis \\\tabitem Intrusion detection\\ \tabitem Failure detection\end{tabular} \\ \cline{1-1} \cline{3-6}

GANs & \multirow{-5}{*}{Unsupervised} &Generative  & Various & \begin{tabular}[c]{@{}l@{}}  \tabitem Learn data discribution\\  \tabitem Can be used as a data generation tool\\ \tabitem Two part networks:
\\\enspace a generator and a discriminator \end{tabular} & \begin{tabular}[c]{@{}l@{}}  \tabitem Localization and way finding\\ \tabitem Image to text \end{tabular} \\ \hline
\end{tabular}%
}
\end{table}
\subsection{Reinforcement Learning (RL) methods}

The goal of a reinforcement learning agent is to find an optimal policy to maximize the expected sum of future rewards $J(\theta)$ parameterized by $\theta$. At each time step $t$, reward $r_{t} = r(a_{t},s_{t})$ is given when an agent takes an action $a_{t} \in A$ at state $s_{t} \in S$.
\begin{equation} \label{equa:reward}
	\mathop{\arg\max}_{\theta} \ \ J(\theta) = \mathbb{E}_{t \sim p_{\theta}(t)} [\sum_{t} r(s_{t},a_{t}) ]
\end{equation}
In the equation, $p_{\theta}(t)$ represents the interactions between the RL agent and outer environments. The trajectory depends on two factors: the agent policy and the environment dynamic. Agent takes actions based on its policy $\pi_{\theta(a_{t}|s_{t})}$. The dynamic transition for the environment $s_{t} \times a_{t} \rightarrow s_{t+1}$ can be expressed as $p(s_{t+1}| s_{t},a_{t})$ and is usually unknown. Overall, the whole trajectory $p_{\theta}(t)$ can be represented as:
\begin{equation}
	P_{\theta}(s_{1}, a_{1} ... s_{T}, a_{T}) = p(s_{1})\prod_{t=1}^{T}\pi_{\theta(a_{t}|s_{t})}p(s_{t+1}|s_{t},a_{t})
\end{equation}
It is clear from the above equation that, if we are smart on the agent policy $\pi_{\theta(a_{t}|s_{t})}$ or the transition dynamic $p(s_{t+1}| s_{t},a_{t})$, we can find the best trajectory $p_{\theta}(t)$ which maximizes reward at each time step. The RL community has formalized these two approaches as \textit{Model-Free} and \textit{Model-Based} approaches based on the fact that the later one directly models the system dynamic transitions.

Fig. \ref{fig:rl-categorisation} gives a broad view of the reinforcement learning world. The reinforcement learning method can be divided into \textit{Model-free} methods and \textit{Model-based} methods with correlations combining the features of both methods. The model-free approach can be subdivided into \textit{Policy based}, \textit{Value based} and \textit{Actor-Critic} approaches according to the ways that the best policy is generated. In this section, we will detail both approaches with simple mathematical expressions and list the most famous algorithms. We will also cover some works that take advantage of both methods.

\myparagraph{Model-Free Methods} They learn a policy and decide the best action to take given a certain state. It can be categorized as \emph{policy-gradient}, \emph{value-based} and \emph{actor-critic} methods, based on how the policy is generated.

\emph{Value-Based} methods learn a value function to estimate the ``goodness''  $V(s_{t})$ for reaching a certain state $s_{t}$, or the ``goodness'' $Q(s_{t},a_{t})$ for taking certain action $a_{t}$ given the state $s_{t}$. Hereby the ``goodness'' function estimates the sum of future rewards from current state $s_{t}$ till the end $s_{T}$ (given a finite trajectory). At each step, the agent chooses the action with the highest score based on the estimated value function. \emph{Value-based} approach is deterministic, and may not be sufficient to solve complex problems. A list of the most prevalent algorithms includes Q-Learning \cite{watkins1992q}, DQN~\cite{mnih2015human}, Prioritised Experience Replay~\cite{schaul2015prioritized}, Dueling DQN~\cite{wang2015dueling}, Double DQN~\cite{van2016deep} and Retrace~\cite{munos2016safe}.

\emph{Policy-Gradient} methods provide an attractive paradigm by directly maximising $J(\theta)$ (Equa \ref{equa:reward}) with respect to the parameters $\theta$ of the policy $\pi_{\theta}(a|s)$. The gradient with respect to the parameters $\theta$ can be derived as
 \begin{equation}
 \label{eqn:gradient}
 	\nabla_{\theta}J = \mathbb{E}_{\theta} [\sum_{t} \nabla_{\theta} log \pi_{\theta}(a_{t}|s_{t})(R_{t}-b_{t})]
 \end{equation}
 $R_{t} = \sum_{t' = t}\gamma^{t'-t}r(s_{t'},a_{t'})$ with a discounting factor $\gamma$ 
 emphasizing the agent more  on recent rewards.
 To reduce the policy variance given different datasets, a baseline $b_{t}$ that does not depend on future states or actions is subtracted. In practice, the expected future return is sampled and aggregated within a trajectory. Several works have been proposed based on this paradigm to either reduce the policy variance \cite{schulman2015trust}, increase scalability \cite{heess2017emergence} or reduce sample complexity \cite{schulman2017proximal}.

\emph{Actor-Critic} algorithms \cite{konda2000actor} are similar to the policy-gradient approach in updating the policy, while an estimated value function $V(s_{t})$ is leveraged in place of the constant baseline $b$ in the original equation \ref{eqn:gradient}.
The term $(R_{t} - b_{t})$ is thus an estimate of the \textit{Advantage} defined as $A(a_{t},s_{t}) = Q(a_{t},s_{t}) - V (s_{t})$ where $a_{t}$ is the action and $s_{t}$ denotes the current state, with $R_{t}$ estimating $Q(a_{t},s_{t})$. Actor-critic methods experience much lower variance with the policy $\pi$ and value function $V$ seen as actors and critics respectively. Works utilising this architecture benefit from both the policy-gradient and value-based methods, with DDPG \cite{lillicrap2015continuous} combining deep Q-learning for continuous action space, A3C \cite{mnih2016asynchronous} for concurrent training. Other prevalent works include TD3 \cite{fujimoto2018addressing}, Soft-Critic SAC \cite{haarnoja2018soft}.

\myparagraph{Model-based Methods} The model-based algorithms differ from the model-free methods in that the latter  cares less about the environment's inner working and the rewards are estimated through sampling. On the contrary, a model-based approach focuses on the model to predict the next state at each time step. Model-based RL achieves good sample efficiency by learning the transition dynamics of the environments directly. During learning, sample trajectories are collected and trained with supervised learning. Several different approaches have been applied to study the dynamics, Gaussian Process approaches \cite{deisenroth2011pilco}\cite{ko2009gp}\cite{boedecker2014approximate}, Time-varying linear models approaches \cite{levine2014learning}\cite{lioutikov2014sample} or Deep networks \cite{nagabandi2018neural}\cite{gal2016improving}\cite{chua2018deep}.

%% file: figuresTex/fig-gan-arch.tex
\begin{figure}[ht]
\centering
\includegraphics[width=5.0in]{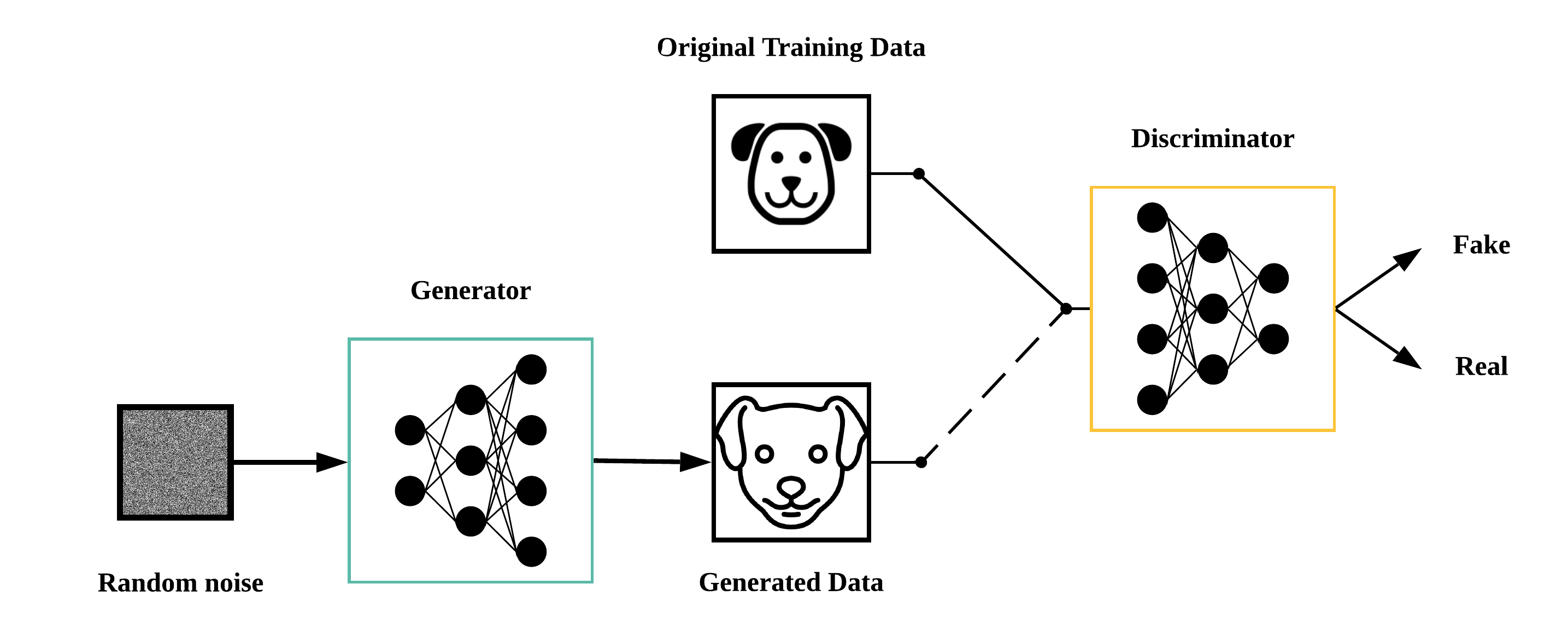}
 \caption{Generative Adversarial Network Architecture}
\label{fig:gan}
\end{figure}

%% file: file/model_eva.tex
\section{Appendix C}

\begin{table}[H]
\centering
\begin{tabular}{|c|c|c|c|}
\hline
Metrics                                                       & Formula                                                         & Type                                                  & Evaluation Focus                                                                                                                                                        \\ \hline
\multirow{2}{*}{Accuracy(acc)}                                & $\frac{tp+tn}{tp+fp+tn+fn}$                                     & Binary                                                & \begin{tabular}[c]{@{}c@{}}This metric measures the correct percentage of\\ the total samples\end{tabular}                                 \\ \cline{2-4} 
                                                              & $\frac{\sum^{l}_{i=1}\frac{tp_i+tn_i}{tp_i+tn_i+fp_i+tn_i}}{l}$ & \begin{tabular}[c]{@{}c@{}}Multi\\ class\end{tabular} & Average accuracy for all classes                                                                                                                                     \\ \hline
\multirow{2}{*}{Error Rate(err)}                              & $\frac{fp+fn}{tp+fp+tn+fn}$                                     & Binary                                                & \begin{tabular}[c]{@{}c@{}}This metric measures the miss-classification\\ percentage over evaluated samples\end{tabular}                                      \\ \cline{2-4} 
                                                              & $\frac{\sum^{l}_{i=1}\frac{fp_i+fn_i}{tp_i+tn_i+fp_i+tn_i}}{l}$ & \begin{tabular}[c]{@{}c@{}}Multi\\ class\end{tabular} & Average Error Rate of all classes                                                                                                                                   \\ \hline
\multirow{2}{*}{Precision(p)}                                 & $\frac{tp}{tp+fp}$                                              & Binary                                                & \begin{tabular}[c]{@{}c@{}}Precision measures correct classified\\ positive samples in a positive class\end{tabular} \\ \cline{2-4} 
                                                              & $\frac{\sum^{l}_{i=1}\frac{tp_i}{tp_i+fp_i}}{l}$                & \begin{tabular}[c]{@{}c@{}}Multi\\ class\end{tabular} &  Average of precision on each class                                                                                                                                 \\ \hline
\multirow{2}{*}{Recall(r)}                                    & $\frac{tp}{tp+fn}$                                              & Binary                                                & \begin{tabular}[c]{@{}c@{}}Recall measures the fraction of correct classified \\positive samples\end{tabular}                                    \\ \cline{2-4} 
                                                              & $\frac{\sum^{l}_{i=1}\frac{tp_i}{tp_i+tn_i}}{l}$                & \begin{tabular}[c]{@{}c@{}}Multi\\ class\end{tabular} & The average of recall for each class                                                                                                                                    \\ \hline
\multirow{2}{*}{F1-Score(FS)}                                 & $\frac{2*p*r}{p+r}$                                             & Binary                                                & \begin{tabular}[c]{@{}c@{}}This metric measures the harmonic mean \\ of recall and precision\end{tabular}                                                 \\ \cline{2-4} 
                                                              & $\frac{2*p_M*r_M}{p_M+r_M}$                                     & \begin{tabular}[c]{@{}c@{}}Multi\\ class\end{tabular} & The average F1-Score                                                                                                                                       \\ \hline
\begin{tabular}[c]{@{}c@{}}Geometric\\ Mean(GM)\end{tabular} & $\sqrt{tp*tn}$                                                  & Binary                                                & \begin{tabular}[c]{@{}c@{}}This metric is similar to the F1-Score but aims to\\ maximize the $tp$ rate and $tn$ rate\end{tabular}    \\ \hline
\end{tabular}
\caption{Evaluation Metrics for Classification Problem. \\Note: $i$ indicates the class $C_i$, and $M$ donates the macro-averaging.} 
\label{threshold}
\end{table}
\begin{table}[H]
\centering
\begin{tabular}{|c|c|c|}
\hline
Metric & Formula & Evaluation Focus \\ \hline
Mean Squared Error & $\frac{1}{N}\sum^{N}_{i=1}(y_i-\hat{y})^2$ & \begin{tabular}[c]{@{}c@{}}MSE calculates the square residual for every data point,\\contribute more to the outlier  \end{tabular} \\ \hline
Mean Absolute Error & $\frac{1}{N}\sum^{N}_{i=1}|y_i-\hat{y}|$ & \begin{tabular}[c]{@{}c@{}}MAE calculates the absolute residual for every data point,\\so that negative and positive residuals do not cancel out\end{tabular}  \\ \hline
Mean Absolute Percentage Error & $\frac{1}{N}\sum^{N}_{i=1}|\frac{y_i-\hat{y}}{y}|$ & Percentage equivalent of MAE  \\ \hline
Mean Percentage Error & $\frac{1}{N}\sum^{N}_{i=1}(\frac{y_i-\hat{y}}{y})$ & \begin{tabular}[c]{@{}c@{}}MPE can help to check whether\\ the model underestimates or overestimates\end{tabular} \\ \hline
\end{tabular}
\caption{Evaluation Metrics for Classification Problem. Note: $y_i$ indicates the predicted value for $i$th sample, and $\hat{y}$ denotes the ground truth value. $N$ indicates the total number of samples.}
\label{tab:regression_eva}
\end{table}